\documentclass[10pt,journal,compsoc]{IEEEtran}

\ifCLASSOPTIONcompsoc
  \usepackage[nocompress]{cite}
\else
  \usepackage{cite}
\fi

\ifCLASSINFOpdf

\else

\fi

\usepackage{microtype}
\usepackage{graphicx}
\usepackage{subfigure}
\usepackage{booktabs} 
\usepackage{hyperref}
\usepackage{url,amsmath}
\usepackage{booktabs,stfloats}       
\usepackage{amsfonts}       
\usepackage{nicefrac}       
\usepackage{microtype}      
\usepackage{xcolor}         
\usepackage{lipsum}		
\usepackage{doi}
\usepackage{multirow}
\usepackage{amsfonts,amssymb}
\usepackage{bm} 
\usepackage{amsmath}
\usepackage{mathtools}
\usepackage{array}
\usepackage{booktabs}
\usepackage{algorithm}
\usepackage{setspace}
\usepackage{float}
\usepackage{color}
\usepackage{graphicx} 
\usepackage{epstopdf}
\usepackage{epsfig}
\usepackage{subfigure}
\usepackage{algorithmic}
\usepackage{algorithm}
\usepackage{enumerate}
\usepackage{makecell}
\usepackage{wrapfig}
\usepackage{booktabs}
\usepackage{mathrsfs}

\usepackage{hyperref}

\usepackage{multicol}

\newtheorem{mythm}{Theorem}
\newtheorem{mydef}{Definition}

\newtheorem{lemma}{Lemma}
\newtheorem{remark}{Remark}
\newtheorem{example}{Example}

\begin{document}

\title{Global Nash Equilibrium in Non-convex Multi-player Game: Theory and Algorithms}

\author{Guanpu Chen, Gehui Xu, Fengxiang He, Yiguang Hong,~\IEEEmembership{Fellow,~IEEE,} Leszek Rutkowski,~\IEEEmembership{Fellow,~IEEE,} Dacheng Tao,~\IEEEmembership{Fellow,~IEEE}
\IEEEcompsocitemizethanks{\IEEEcompsocthanksitem G. Chen, F. He, and D. Tao are with JD Explore Academy, JD.com Inc, Beijing 100176, China.\protect\\
E-mail: chengp@amss.ac.cn, fengxiang.f.he@gmail.com, and\protect\\
dacheng.tao@gmail.com.
\IEEEcompsocthanksitem G. Xu is with Academy of Mathematics and Systems Science, Chinese Academy of Sciences, Beijing 100190, China.\protect\\
E-mail: xghapple@amss.ac.cn.
\IEEEcompsocthanksitem Y. Hong is with Research Institute for Intelligent Autonomous Systems, Tongji University, Shanghai 210201, China.\protect\\
E-mail: yghong@iss.ac.cn,
\IEEEcompsocthanksitem L. Rutkowski is with Systems Research Institute, Polish Academy of Sciences, 01-447 Warsaw, Poland and AGH University of Science and Technology, 30-059 Krak\'ow, Poland.\protect\\
E-mail: leszek.rutkowski@ibspan.waw.pl.
\IEEEcompsocthanksitem G. Chen and G. Xu contributed equally. Correspondence to F. He.
}
\thanks{Manuscript received xxx; revised xxx.}}


\IEEEtitleabstractindextext{\begin{abstract}
Wide machine learning tasks can be formulated as non-convex multi-player games, where Nash equilibrium (NE) is an acceptable solution to all players, since no one can benefit from changing its strategy unilaterally. Attributed to the non-convexity, obtaining the existence condition of global NE is challenging, let alone designing theoretically guaranteed realization algorithms. This paper takes conjugate transformation to the formulation of non-convex multi-player games, and casts the complementary problem into a variational inequality (VI) problem with a continuous pseudo-gradient mapping. We then prove the existence condition of global NE: the solution to the VI problem satisfies a duality relation. Based on this VI formulation, we design a conjugate-based ordinary differential equation (ODE) to approach global NE, which is proved to have an exponential convergence rate. To make the dynamics more implementable, we further derive a discretized algorithm. We apply our algorithm to two typical scenarios: multi-player generalized monotone game and multi-player potential game. In the two settings, we prove that the step-size setting is required to be $\mathcal{O}(1/k)$ and $\mathcal{O}(1/\sqrt k)$ to yield the convergence rates of $\mathcal{O}(1/ k)$ and $\mathcal{O}(1/\sqrt k)$, respectively. Extensive experiments in robust neural network training and sensor localization are in full agreement with our theory.

\end{abstract}

}

\maketitle

\IEEEdisplaynontitleabstractindextext

\IEEEpeerreviewmaketitle

\IEEEraisesectionheading{\section{Introduction}
\label{Intro}}

%
 

\IEEEPARstart{M}{any}
advanced learning approaches in artificial intelligence are developed toward  multi-agent ways, distributed manners, or federated frameworks \cite{yu2019multi,li2019interaction,fan2021fault,zhu2022topology}.  For instance, as one of the most popular schemes,  adversarial learning is gradually generalized to multiple {agents } \cite{song2018multi,zhao2020improving,9613799}, not restricted to classic models with one generator and one discriminator.
Also, most complex systems involve the interaction and interference of the multiple participants therein, such as smart grids \cite{saad2012game}, intelligent transportation \cite{saharan2020dynamic}, and cloud computing \cite{pang2008distributed}.  The common core ideology is
to sufficiently utilize the autonomy and evolvability of individual computational units in large-scale tasks. General optimization frameworks  or min-max adversarial protocols will be no longer evergreen, and proper models should be established for describing these multi-agent systems as well as the accompanied solvers.

Game theory exploits the advantages to the full in such multi-player scenarios. Actually, game theory has been playing an essential role in the leading edge of machine learning  nowadays  such as adversarial training   and reinforcement learning \cite{busoniu2008comprehensive,lanctot2017unified, dai2018sbeed,6330964}. The Nash equilibrium (NE) therein \cite{nash1951non} becomes a popular concept in various fields like applied mathematics, computer sciences, and engineering, in addition to economy. {When all players' strategy profile reaches an NE, no one can benefit from changing its strategy
unilaterally.}
This paper focuses on a typical class of non-convex multi-player games. 
Player $i$ minimizes its own payoff function $J_i(x_i,\boldsymbol{x}_{-i}):\mathbb{R}^{Nn}\rightarrow \mathbb{R} $, which is influenced by both the player's own decision $x_i\in \mathbb{R}^n$ and others' decisions $\boldsymbol{x}_{-i} \in \mathbb{R}^{(N-1)n}$.  
Specifically, the non-convex structure in players' payoff is endowed with 
\begin{equation}\label{fda}
	J_{i}(x_{i},\boldsymbol{x}_{-i})=\Psi_{i}(\Lambda_{i}(x_{i},\boldsymbol{x}_{-i})).\nonumber
\end{equation}
{Here, $ \Lambda_{i}: \mathbb{R}^{Nn} \rightarrow \mathbb{R}^{q_{i}} $ is a vector-valued 
	nonlinear operator, where $ \Lambda_{i}=(\Lambda_{i,1},\cdots,\Lambda_{i,q_{i}})^{T} $ and for $ k \in\{1,\cdots,q_{i}\}$, $ \Lambda_{i,k}: \mathbb{R}^{Nn} \rightarrow \mathbb{R} $ is a quadratic function 
	{in $x_i$}.
}
Besides, $ \Psi_{i}: \mathbb{R}^{q_i} \rightarrow \mathbb{R}$ is a 
canonical function \cite{gao2017canonical}, whose gradient $\nabla\Psi_{i}$ is a one-to-one mapping from the primal space into the dual space.

{This setting has been widely investigated in machine learning applications, 
such as robust network training and sensor network localization.
For example, 
in sensor localization \cite{ke2017distributed,yang2018df},  $x_i$ is the sensor node, $\Lambda_{i,k}$ represents the estimated distance between $x_i$ and $\boldsymbol{x}_{-i}$, while
$\Psi_{i}$ is reified as Euclidean  norms to  measure the error of the true distance and the estimated distance. 
In robust neural network training \cite{nouiehed2019solving,deng2021local}, $x_i$ denotes the model parameter, $\Lambda_{i,k}$ serves as  the  output of training data,  while  $\Psi_{i}$ represents the cross-entropy function. Moreover,
this setting may also inspire solutions to resource allocation problems in unmanned vehicles \cite{yang2019energy} and secure transmission \cite{ruby2015centralized}, where $x_i$ stands for the transmit resources, $\Psi_{i}$ denotes the transmission cost together with $\Lambda_{i,k}$, which is a   logarithmic-posynomial function.
}

{Given the above formulation, it is natural and essential from both game-theoretic and machine-learning perspectives to seek its global NE, which characterizes a global optimum solution since  no one will deviate from its strategy unilaterally with others' given strategies.
However, it is the status quo that finding the global optimum or equilibrium in non-convex settings is still an open problem \cite{1177151,8643982,9398583}. That is not only owing to the lack of powerful
tools compared with convex categories, but also due to the diversity of non-convex structures, which
may not be solved by a common methodology.  Despite many efficient tools within convex conditions leading to fruitful achievements of multi-player game models \cite{yi2019operator,chen2021distributed,facchinei2010penalty}, they may be far from enough when encountering the non-convexity, and may be stuck in local NE or approximations when tracking along the pseudo-gradients, rather than reaching a global NE.
On the other hand, although
some inspiring breakthroughs have been
made for solving non-convex two-player min-max games in different situations, such as  Polyak-{\L}ojasiewicz cases \cite{nouiehed2019solving,fiez2021global},   concave cases \cite{lin2020gradient,rafique2021weakly}, 
they may not provide available help in multi-player  settings, because the global stationary conditions are mutually
coupled and cannot be handled individually by each player. Thus, we need novel processes to explore the existence of global NE and design algorithms to realize it.}

To this end, we first employ the canonical duality theory \cite{gao2017canonical} and
obtain a one-to-one duality relation within a conjugate transformation \cite{rockafellar1974conjugate}, in order to deal with the non-convexity in payoff functions.
By generalizing to continuous vector fields,
we compactly formulate the coupled stationary conditions of the transformed problem as a continuous mapping.
Thus, seeking all players' NE profile can be accomplished by  verifying a fixed point of this continuous mapping. 
We then cast the fixed point seeking 
into solving a variational inequality (VI) problem \cite{facchinei2003finite}. By the above procedures, we can transform the global NE of a non-convex multi-player game into the solution to a VI problem, which is much easier to be solved since all players' coupled stationary conditions are regarded from an entire perspective.  
So far, we obtain the existence condition of the global NE in such a non-convex multi-player game: the solution to the VI problem is required to satisfy the {duality relation}.

Based on this transformation, we then propose a conjugate-based ordinary differential equation (ODE) for solving the VI problem.
{No longer flowing in the  primal spaces,} the ODE evolves in the dual spaces of both decision variables and canonical variables.
Then a mapping from this dual space to the primal space is enforced via  the gradient information of differentiable Legendre conjugate functions \cite{diakonikolas2019approximate}.
We prove that the equilibrium of this conjugate-based ODE is the global NE of this non-convex multi-player game if the aforementioned existence condition is verified. 
Besides, we provide rigorous convergence analysis on the continuous dynamics, as well as
 prove that the ODE has an exponential convergence rate.

For practical implementation, we further derive a discrete algorithm associated with the proposed conjugate-based ODE. We analyze the step-size settings for some desired convergence rates in two typical  non-convex game models. Specifically, with a step size as $\mathcal{O}(1/k)$, the convergence rate achieves $ \mathcal{O}(1/k) $ in a class of multi-player generalized monotone games \cite{facchinei2010penalty,koshal2016distributed}; while with another step size as $\mathcal{O}(1/\sqrt k)$, the convergence rate achieves $ \mathcal{O}(1/\sqrt{k}) $ in a class of multi-player potential games \cite{ke2017distributed,yang2018df}.

We conduct extensive experiments in robust neural network training and sensor localization. 
All these experimental results show that our algorithm converges to the global NE of non-convex games, instead of being stuck in local NE or approximate NE. Moreover, we compare
our algorithm  against several popular  methods on these multi-player settings and  demonstrate that our  algorithm outperforms all other techniques. 

To our best knowledge, this is the first paper on the existence condition and realization algorithms of the global Nash equilibria in a non-convex multi-player game. Our contributions are summarised below:
\begin{itemize}
    \item \textbf{Existence condition.}  We employ canonical duality theory to transform the non-convex multi-player game  into a complementary dual problem, and  cast  solving all players’ stationary point profile  into solving 
    a VI problem. Then  we provide the existence condition of global NE: the solution to the VI problem is required to satisfy a duality relation.
    \item \textbf{Conjugate-based ODE.} We propose a conjugate-based ODE for solving the VI problem, which evolves in the dual spaces of both decision variables and canonical variables.
The equilibrium of the ODE is the global NE of this non-convex multi-player game if the existence condition is verified.
The convergence analysis of the ODE  and its exponential convergence rate are provided. 

    \item \textbf{Discrete algorithm.} We derive a discrete algorithm based on the continuous ODE for practical implementation in two typical  non-convex scenarios --  a convergence rate of $ \mathcal{O}(1/k) $ in a class of multi-player generalized monotone games  with a step size of $\mathcal{O}(1/k)$, and  another convergence rate of $ \mathcal{O}(1/\sqrt{k}) $ in a class of multi-player potential games with a step size of $\mathcal{O}(1/\sqrt k)$.
\end{itemize}

{
The rest of this paper is organized as follows: Section  \ref{relatedwork} introduces the related work. Section  \ref{gen_inst} formulates a non-convex multi-player game, while Section \ref{headings} investigates the existence of global NE. Section \ref{ODE} proposes an ODE to seek the global NE   and gives the exponential convergence analysis. Section \ref{d5} derives a discretized algorithm from the ODE and presents the step-size settings and convergence rates in two typical scenarios.  
Section \ref{e11} examines the  effectiveness of the proposed approach with several experiments. {Section \ref{discuss} gives some discussions on the
results of theory and algorithms obtained in this paper. } 
 Finally,   Section \ref{conclusion} concludes the paper and gives some future directions.
}

\section{Related Work}\label{relatedwork}




\textbf{Convex multi-player games.} {Many} theoretical results in multi-player games have been built on fundamental convexity assumptions 
\cite{facchinei2010penalty,yi2019operator,chen2021distributed}.
Under the frame of convexity, NE seeking has been extensively studied in  many typical multi-player game models, including
aggregative  games \cite{koshal2016distributed,xu2022efficient}, potential games \cite{lei2020asynchronous,yang2018df},  and hierarchical games \cite{kim2001hierarchical}. {For example, \cite{facchinei2010penalty,lei2020asynchronous} directly required convex payoffs on each player's decision variable, while \cite{koshal2016distributed,chen2021distributed} needs strongly/strictly monotone pseudo gradients to display the interaction on all players’ actions.}
Despite many efficient tools within convex conditions leading to fruitful achievements, they may be far from enough
when encountering non-convexity in practical circumstances.

\textbf{Non-convex two-player min-max problems.}
Some inspiring breakthroughs have been achieved
for solving  non-convex two-player min-max problems, including
Polyak-{\L}ojasiewicz cases \cite{nouiehed2019solving,fiez2021global}, strongly-concave cases \cite{lin2020gradient,rafique2021weakly}, and  general non-convex non-concave cases \cite{heusel2017gans,daskalakis2018limit}.
Such popular researches owe to the  success of GAN  and its  variants
\cite{goodfellow2014generative,daskalakis2018training}. { For instance, \cite{heusel2017gans} proposed a two time-scale
update rule  with stochastic gradient descent to find a local NE in GANs, while \cite{daskalakis2018limit} designed an optimistic mirror descent  algorithm  to explore NE in GANs with a theoretical guarantee.
However, it is not straightforward and realistic to directly generalize the above two-player approaches to solve multi-player settings.  It is because the global stationary conditions are mutually coupled among multiple players and cannot be handled individually, which is  unlike  two-player situations.}

\textbf{Non-convex multi-player games with local NE or approximations.} Initial efforts have been made for solving non-convex multi-player games.
\cite{pang2011nonconvex} proposed a best-response scheme for Nash stationary points  of  a class of
non-convex  games  in signal processing, and then \cite{hao2020piecewise} extended this method in  multi-player  bilevel  games with non-convex constraints.   
Moreover,  \cite{raghunathan2019game} introduced a gradient-based Nikaido-Isoda  function  to find Nash stationary points in a reformulated non-convex game,
while \cite{liu2020approximate} 
designed a  gradient-proximal algorithm for approximate NE in a class of non-convex aggregative games. The algorithms within these works lead to  local NE or Nash stationary points dependent on the initial points.  How to guarantee the existence of global NE and how to design algorithms for seeking  global NE deserve further investigation in non-convex multi-player game models.

\textbf{Similar non-convex structures in optimization.} 
Related results exist in solving such non-convex problems where the objectives or payoffs are composited with canonical functions and quadratic operators, which are however somewhat premature. \cite{zhu2012approximate} considered an  approximate optimization to relax such non-convex constraints and provided the optimality conditions for the simplified problem, while \cite{latorre2016canonical} proposed similar sufficient conditions  and discussed the existence of the global optimum via canonical duality theory. On this basis, \cite{ren2021distributed} investigated the global optimal solution of such a non-convex optimization in a distributed manner over multi-agent networks, while \cite{liang2019topology} focused on  approximating the solutions of discrete variable topology problems with multiple constraints. 
We notice that regarded as so important a class of non-convex problems, most of the existing work concentrated on  the optimization perspective. Taking into account the interference and interaction among multiple players, the global stationary conditions are mutually coupled and can not be handled individually by each player. Therefore, 
the above optimization methods cannot completely match our problems.
Serving as a widely accepted equilibrium concept, the global NE therein needs to be further discussed with respect to its existence conditions and seeking algorithms.
The consequent results in this paper will help demystify the complicated interactions among
players and provide trustworthy insights for large-scale problems afterward.

\section{Preliminaries on Game Theory}
\label{gen_inst}
We begin our study of the non-convex games   with multiple players indexed by $\mathcal{I} = \{1,\cdots,N\}$.
For $i \in \mathcal{I}$, the $ i $th player has an action variable $ x_{i} $ in an action  set $  \Omega_{i}\subseteq \mathbb{R}^{n} $,  where $\Omega_{i}$ is compact and convex, and $   \boldsymbol{\Omega} =\prod_{i=1}^{N}\Omega_{i} $. Let $ \boldsymbol{x} = \operatorname{col}\{x_{1}, . . . ,x_{N}\} \in \mathbb{R}^{nN} $ be the profile of all players' actions, while $ \boldsymbol{x}_{-i} $ be the profile of all players' actions except for the $ i $th player's. Moreover,
	the $  i  $th player has a payoff function $ J_{i}(x_{i},\boldsymbol{x}_{-i}) : \boldsymbol{\Omega}\rightarrow \mathbb{R} $, which is dependent on both $ x_{i} $ and $ \boldsymbol{x}_{-i} $, and
	twice continuously differentiable in $  x_{i} $. 
Given $ \boldsymbol{x}_{-i} $, the $ i $th player intends to solve the following problem
\begin{equation}\label{f1}
\min \limits_{x_{i} } J_{i}\left(x_{i}, \boldsymbol{x}_{-i}\right), \quad \text{s.t. } x_{i} \in \Omega_{i}.
\end{equation}
In this paper, we focus on a typical class of non-convex multi-player games, in which the $i$th player's payoff function  is endowed with the following structure
\begin{equation}\label{fda}
J_{i}(x_{i},\boldsymbol{x}_{-i})=\Psi_{i}(\Lambda_{i}(x_{i},\boldsymbol{x}_{-i})).
\end{equation}
{Here $ \Lambda_{i}: \mathbb{R}^{Nn} \rightarrow \Theta_{i} \subseteq \mathbb{R}^{q_{i}} $ is a vector-valued 
	nonlinear operator
	with $ \Lambda_{i}=(\Lambda_{i,1},\cdots,\Lambda_{i,q_{i}})^{T} $.
	For $ k \in\{1,\cdots,q_{i}\}$, each $ \Lambda_{i,k}: \mathbb{R}^{Nn} \rightarrow \mathbb{R} $ is quadratic 
	{in $x_i$}, whose second-order partial derivative in $x_i$ is both
	$x_i$-free and $\boldsymbol{x}_{-i}$-free,
	e.g., $\Lambda_{i,k}=x_{i}^{T}A_{i,k}x_{i}+\sum_{i\neq j}x_{i}^{T}B_{i,k}x_{j}$.
}
Moreover, $ \Psi_{i}: \Theta_{i} \rightarrow \mathbb{R}$ is a convex differential
canonical function \cite{gao2017canonical}, whose gradient $\nabla\Psi_{i} : \Theta_{i}\rightarrow \Theta^{*}_{i}$ is a one-to-one mapping. 
Such non-convex structures composited by canonical functions and quadratic operators emerge in broad applications, including robust network training \cite{nouiehed2019solving}, sensor localization \cite{yang2018df}, and GAN \cite{gidel2018variational}.
We provide specific examples in the following for  intuition about the above non-convex model. 

\begin{example}[Euclidian distance function]
\begin{equation}\label{j1}
 \sum\nolimits_{j\in\mathcal{N}_{i}}( \left\|x_{i}-{x}_{j}\right\|^{2}-d_{i,j })^{2},
\end{equation}
{where $\Psi_{i}=\sum\nolimits_{j=1}\nolimits^{\mathcal{N}_{i}} \Lambda_{i,j}^{T}\Lambda_{i,j}$ and $\Lambda_{i,j}=\left\|x_{i}-{x}_{j}\right\|^{2}-d_{i,j }$.}
(\ref{j1}) usually serves as the payoffs in sensor localization
\cite{jia2013distributed,ke2017distributed,yang2018df},
where $x_{i}\in \Omega_i$ is a sensor node, $\mathcal{N}_{i}$ is  the neighbors of node $i$, and $d_{i,j}$ is the distance parameter. 
\end{example}

\begin{example}[Log-sum-exp function]
\begin{equation}\label{j2}
\beta_{1} \log [1+ \exp 
(- x_{i}^{T}x_{i}-\sum\nolimits_{j=1}^{N} x_{i}^{{T}} x_{j}-\beta_{2}^{{T}} x_{i}	)],
\end{equation}
{where $\Psi_{i}=\beta_{1}\log [1+ \exp \Lambda_{i}]$ and $\Lambda_{i}=-x_{i}^{T}x_{i}-\sum\nolimits_{j=1}^{N} x_{i}^{{T}} x_{j}-\beta_{2}^{{T}} x_{i}$.}
(\ref{j2}) 
usually {appears in}    the tasks like robust neural network training \cite{nouiehed2019solving,deng2021local}, where
$x_i$ is  the neural network parameter,  $x_{j}$ is the perturbation, and $\beta_{1}, \beta_{2}$ 
are  training data.  
\end{example}

\begin{example}[Log-posynomial function]
\begin{equation}\label{e345}
\log(  x_{i}^{T} \mathbf{C}_{i} x_{i}+x_{i}^{T}\mathbf{D}_{i}\boldsymbol{x}_{-i})^{-1},
\end{equation}
{where $\Psi_{i}=\log(\Lambda_{i})^{-1}$ and $\Lambda_{i}=x_{i}^{T} \mathbf{C}_{i} x_{i}+x_{i}^{T}\mathbf{D}_{i}\boldsymbol{x}_{-i}$.}
(\ref{e345})
usually {occurs} in  resource allocation \cite{yang2019energy,ruby2015centralized,chiang2007power},
where $x_{i}$ stands for transmitting resources,  and matrices $\mathbf{C}_{i}$ and $\mathbf{D}_{i}$ represent the correlation coefficients.
\end{example}

\begin{figure}[t]
	\centering	
			\includegraphics[width=0.9\columnwidth]{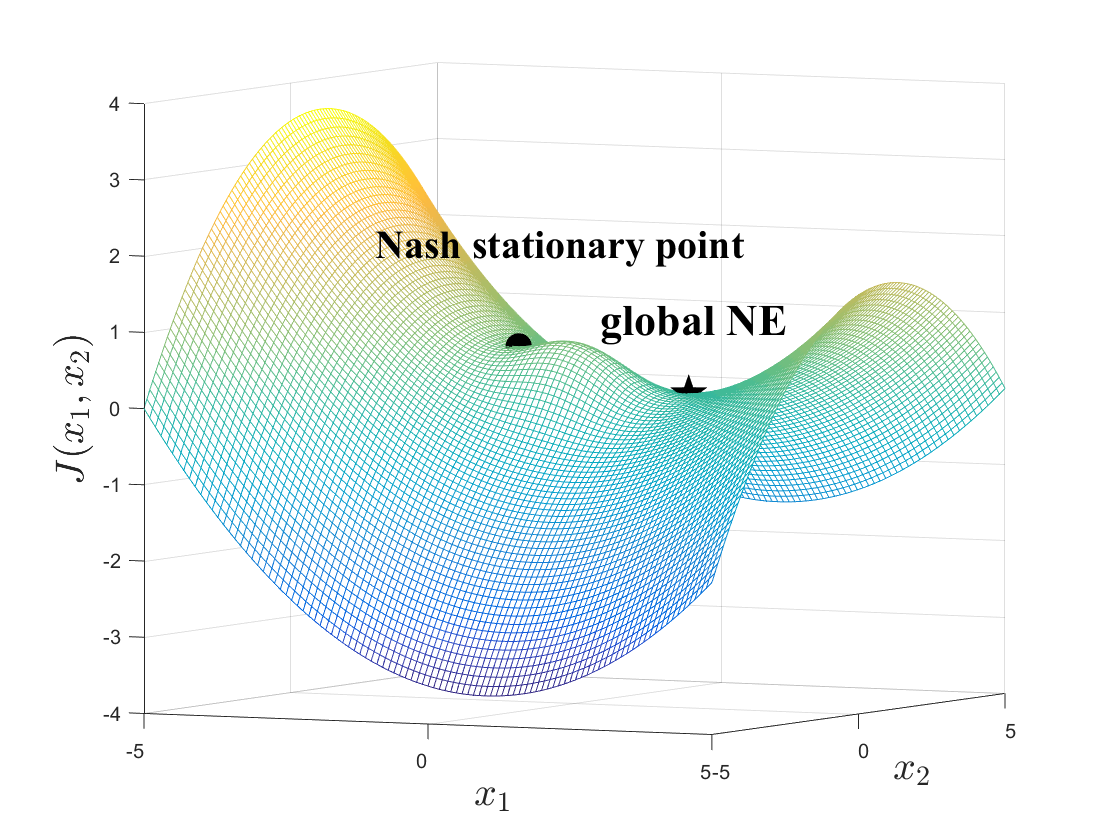}
\\
	\caption{A non-convex two-player demo with log-sum-exp payoffs in (\ref{j2}).
  } 
	\label{fig11}
\end{figure}

We introduce the following important concept for solving the non-convex multi-player game (\ref{f1}).
\begin{mydef}[global Nash equilibrium]
	A strategy profile $ \boldsymbol{x}^{\Diamond}\in\boldsymbol{\Omega} $ is said to be a
	{global Nash equilibrium} (NE)
	of (\ref{f1}), 
	if for all $ i\in \mathcal{I}$,
	\begin{equation}\label{ee}
	J_{i}(x_{i}^{\Diamond}, \boldsymbol{x}_{-i}^{\Diamond})\leq J_{i}(x_{i}, \boldsymbol{x}_{-i}^{\Diamond}), \quad \forall x_{i}\in \Omega_i. 
	\end{equation}
\end{mydef}		
The global NE above characterizes a strategy profile that  each player adopts its globally optimal strategy. That is, given others' actions, no player can benefit from changing  her/his action unilaterally.  Actually, the conception of global NE here is indeed the concept of NE \cite{nash1951non}, and we emphasize \textit{global} in the non-convex formulation to tell the difference from  \textit{local} NE \cite{pang2011nonconvex,nouiehed2019solving,heusel2017gans}.
Also, we consider another mild but well-known concept 
to help characterize the solutions to (\ref{f1}).
\begin{mydef}[Nash stationary  point]	
	A strategy profile $ \boldsymbol{x}^{\Diamond} $ is said to be a {Nash stationary point}  of  (\ref{f1}) if for all $ i\in \mathcal{I}$, 
	\begin{equation}\label{e234}
	\begin{aligned}
	&\mathbf{0}_{n} \in \nabla_{x_{i}} J_{i}(x_{i}^{\Diamond}, \boldsymbol{x}_{-i}^{\Diamond})+\mathcal{N}_{\Omega_{i}}({x_{i}^{\Diamond}}). \\
	\end{aligned}
	\end{equation}
\end{mydef}
It is not difficult to reveal that if $ \boldsymbol{x}^{\Diamond} $ is a global NE, then it must be a NE stationary point, but not vice versa.
For instance, in Fig. \ref{fig11}, the global NE distinguishes from Nash stationary points, {shown} on the	surface plot of one player's non-convex payoff.

Actually, as for convex games, most
existing research computes global NE   via investigating Nash stationary points \cite{facchinei2010penalty,koshal2016distributed,chen2021distributed}.
However,   
considering the bumpy geometric structure of the non-convex payoff function,    one cannot expect to find a global NE of (\ref{f1}) merely via the Nash stationary conditions in (\ref{e234}).  
To this end,  we aim at obtaining a global NE of  such a non-convex multi-player model (\ref{f1}) and begin the exploration in the sequel.

\section{Existence Condition of Global NE }
\label{headings}

In this section,  we  primarily explore 
the  existence of global NE with the following
 procedures, that is, 
\begin{enumerate}[i)]
    \item We employ canonical duality theory to transform the original game (\ref{f1}) into a complementary dual problem, and investigate the relationship between  the stationary points of the dual problem and the Nash stationary points of game (\ref{f1}); 
    \item  We adopt a sufficient feasible domain for the introduced conjugate variable to  investigate the global optimality of the stationary points;
    \item  We cast  solving all players’ stationary point profile of the dual problem into solving 
    a variational inequality (VI) problem with a continuous pseudo-gradient
mapping;   
\item We provide the existence condition of global NE of the non-convex multi-player game: the solution to the VI problem is required to satisfy a duality relation. 
\end{enumerate}


 
 
\noindent\textbf{Step 1: Complementary dual problem} 

We first take $\xi_{i}=\Lambda_{i}(x_{i},\boldsymbol{x}_{-i}) \in \Theta_{i}$ in the payoff function of (\ref{fda}),  which is  called a canonical measure. This follows the definition of canonical functions, {for $i\in\mathcal{I}$}. Since  $ \Psi_{i}(\xi_{i}) $  is a convex	canonical function, the one-to-one duality relation $\sigma_{i}=\nabla \Psi_{i}\left(\xi_{i}\right) : \Theta_{i}\rightarrow \Theta^{*}_{i}$ implies the existence of
the conjugate function $ \Psi^{*}_{i}  : \Theta^{*}_{i}\rightarrow \mathbb{R}$, which
can be uniquely described by the Legendre transformation \cite{rockafellar1974conjugate,8770111,gao2017canonical}, that is, 
\begin{align*}
\Psi_{i}^{*}\left(\sigma_{i}\right)=\xi_{i}^{T} \sigma_{i}-\Psi_{i}\left(\xi_{i}\right),
\end{align*}
where $\sigma_{i}\in \Theta_{i}^{*} $
is a canonical dual variable.
Thus, denote
$\boldsymbol{\sigma}=\operatorname{col}\{\sigma_1,\cdots,\sigma_N\}$
and $\boldsymbol{\Theta}^{*} =\prod_{i=1}^{N}\Theta^{*}_i \subseteq \mathbb{R}^{q} $  with $q=\sum_{i=1}^{N} q_{i}$.
Then
the  complementary
function $\Gamma_{i}: \boldsymbol{\Omega}\times \Theta_{i}^{*}\rightarrow \mathbb{R} $ referring to the canonical duality
theory can be  defined as 
\begin{align}\label{ee2}
\Gamma_{i}(x_{i},\sigma_{i},\boldsymbol{x}_{-i})=&\xi_{i}^{T} \sigma_{i}-\Psi_{i}^{*}(\sigma_{i})\nonumber\\
=&
\sigma_{i}^{T} \Lambda_{i}\left(x_{i},\boldsymbol{x}_{-i}\right)-\Psi_{i}^{*}\left(\sigma_{i}\right).
\end{align}

\begin{lemma}\label{t1}
	There exists a profile $ \boldsymbol{x}^{\Diamond} $ as a Nash stationary point of (\ref{f1}) 
	if $\boldsymbol{\sigma}^{\Diamond}\in \boldsymbol{\Theta}^{*}$ and for $ i\in \mathcal{I}$, 
	$ ({x_{i}^{\Diamond}},{\sigma_{i}^{\Diamond}})$
	is a stationary point of  complementarity function  $ \Gamma_{i} (x_{i},\sigma_{i},\boldsymbol{x}_{-i}^{\Diamond})$.
\end{lemma}
Lemma \ref{t1} reveals the equivalency relationship of stationary points between (\ref{ee2}) and  (\ref{f1}). This
means that we can close the duality gap between the non-convex original game and its canonical dual problem with the canonical  transformation. {More proof details of Lemma \ref{t1} can be found in 
the Supplementary Materials due to space limitations.} 

\noindent\textbf{Step 2: Sufficient feasible domain}

For $i\in\mathcal{I}$, we define the second-order partial derivative of  $\Gamma_i(x_i, {\sigma}_i, \boldsymbol{x}_{-i} )$ in $x_i$ is defined as follows. 
\begin{align*}
P_{i}(\sigma_{i})= \nabla^{2}_{x_{i}}\Gamma_i=\sum\nolimits_{k=1}\nolimits^{q_i} [\sigma_{i}]_{k} \nabla^{2}_{x_{i}} \Lambda_{i, k}(x_{i},\boldsymbol{x}_{-i}).
\end{align*}
Recalling that 
$\Lambda_{i}: \boldsymbol{\Omega} \rightarrow \Theta_{i} $ is a quadratic
operator and   $\nabla^{2}_{x_{i}} \Lambda_{i}$ is both $x_i$-free and $\boldsymbol{x_{-i}}$-free  (see the cases  in (\ref{j1})-(\ref{e345})), we can easily check that $P_{i}(\sigma_{i})$ is indeed   a linear combination of $[\sigma_i]_{k}$.
On this basis, we introduce the   following set of $\sigma_i$  for $ i\in\mathcal{I}$. 
\begin{equation}\label{s23}
\mathscr{E}_{ i}^{+}= \Theta_{i}^{*}\cap \{\sigma_{i} : 
P_{i}(\sigma_i) \succeq
\kappa_{x}\boldsymbol{I}_{n}
\}, 
\end{equation} 
where the constant $\kappa_{x}>0$ and the further notation $$ \boldsymbol{\mathscr{E}}^{+}=\mathscr{E}_{ 1}^{+} \times \cdots \times \mathscr{E}_{ N}^{+}.$$
When $ \sigma_{i}\in\mathscr{E}_{ i}^{+}  $,  the positive definiteness of $P_{i}(\sigma_{i})$ 
implies that $\Gamma_i(x_i,{\sigma}_i, \boldsymbol{x}_{-i} )$ is convex with respect to $x_i$. Besides, the convexity of $\Psi_{i}(\xi_i)$ derives that its Legendre conjugate $\Psi_{i}^{*}(\sigma_i)$ is also convex.
Hence, the  complementary function  $\Gamma_i(x_i, {\sigma}_i, \boldsymbol{x}_{-i} )$ is 
concave in $\sigma_{i}$. {This convex-concave property of $\Gamma_i$ enables us to further investigate the optimality of  the stationary points of   (\ref{ee2}), that is,  the optimality of  the Nash stationary point of (\ref{f1})}. 

\begin{remark}
	The computation of
$\mathscr{E}_{ i}^{+}  $ is actually not so hard in most practical cases. For example, 	take the payoff function in (\ref{j1}) with  $i=1,2$ and $n=q_i=1$.  	The complementary function is
$\Gamma_{i}(x_{i},  \sigma_{i},x_{3-i})=\sigma_{i}((x_{i}-x_{3-i})^{2}-d_{i,3-i}) - \sigma_{i}^{2}/4$, 
where $x_i\in \Omega_i= [a,b]$ and $\sigma_i\in \Theta_{i}^{*}=[-2d_{i,3-i},2(b-a)^2-2d_{i,3-i}]$. Thus, 
the subset $\mathscr{E}_{ i}^{+}=\left\{\sigma_{i}: 2\sigma_{i}  \geq \kappa_x\right\}\cap   \Theta_{i}^{*} =[\kappa_x/2, 2(b-a)^2-2d_{i,3-i}]$, which can serve as the feasible constraint set for the dual variable $\sigma_i$.  
\end{remark}

\noindent\textbf{Step 3:  Variational inequality} 

With the interference of  $\boldsymbol{x}_{-i}$,  the transformed problem actually reflects a cluster of $\Gamma_{i}$ with a mutual coupling of stationary conditions,  
rather than a deterministic one. Therefore, different from classic optimization works \cite{zhu2012approximate,latorre2016canonical,gao2017canonical},    the  stationary points for  player $i$ cannot be calculated independently. {We should consider the computation of all players' stationary point profile and discuss its  optimality in an entire perspective.

	To this end,  variational inequalities (VI) help us  to carry forward \cite{facchinei2003finite}.  
	Specifically,
	denote  $\boldsymbol{z}=\operatorname{col}\{\boldsymbol{x},\boldsymbol{\sigma}\}$ and $\boldsymbol{\Xi} = \boldsymbol{\Omega} \times \mathscr{E}^{+}$. 
	Take the following continuous mapping as the pseudo-gradient of (\ref{ee2}).
	\begin{align*}
	F(\boldsymbol{z})=\operatorname{col}\Big\{\operatorname{col}\{
	\sum\nolimits_{k=1}\nolimits^{q_i}[\sigma_{i}]_{k} \nabla_{x_{i}} \Lambda_{i, k}(x_{i},\boldsymbol{x}_{-i}) \}&_{i=1}^{N}
	, \\
	\operatorname{col}\{
	-\Lambda_{i}(x_{i},\boldsymbol{x}_{-i})+\nabla \Psi_{i}^{*}(\sigma_{i})
	\}&_{i=1}^{N}  \Big\}.
	\end{align*}
 	Note that the interaction on all players' variables is displayed in mapping $F$, which is a joint function  of the partial derivatives of all players'  complementary functions (\ref{ee2}).
	Then (\ref{ee2}) can be cast as a VI  problem   $\operatorname{VI}(\boldsymbol{\Xi},F)$ to solve, i.e.,  finding    $\boldsymbol{z}^{\Diamond}\in  \boldsymbol{\Xi}$ such that
	\begin{equation}\label{e2e}
	(\boldsymbol{z}-\boldsymbol{z}^{\Diamond})^{T}F(\boldsymbol{z}^{\Diamond}) \geq 0, \quad \forall \boldsymbol{z}\in \boldsymbol{\Xi}.   
	\end{equation}
 
\noindent\textbf{Step 4: Existence condition} 

Based on the above steps,
 we have the following existence condition for identifying the global NE of 
	(\ref{f1}).
	\begin{mythm}\label{t2}
				There exists $\boldsymbol{x}^{\Diamond}$ as the global NE		of the non-convex multi-player game (\ref{f1}) if 
			$(\boldsymbol{x}^{\Diamond},\boldsymbol{\sigma}^{\Diamond})$ is a solution to $\operatorname{VI}(\boldsymbol{\Xi},F)$ with  $\sigma_{i}^{\Diamond}=\nabla \Psi_{i}\left(\xi_{i}\right) \mid_{\xi_{i}=\Lambda_{i}\left(x_{i}^{\Diamond}, \boldsymbol{x}_{-i}^{\Diamond}\right)}$ for $i\in\mathcal{I}$.
\end{mythm}
{The proof sketch can be summarized as below.
	Under Assumption 1, if 
	there exists $\boldsymbol{\sigma}^{\Diamond}\in \mathscr{E}^{+}$ such that $\boldsymbol{z}^{\Diamond}=\operatorname{col}\{\boldsymbol{x}^{\Diamond},\boldsymbol{\sigma}^{\Diamond}\}$ is a solution to $\operatorname{VI}(\boldsymbol{\Xi},F)$, 
	then it satisfies the first-order condition of the VI. 
Together with $\sigma_{i}^{\Diamond}=\nabla \Psi_{i}\left(\xi_{i}\right) \mid_{\xi_{i}=\Lambda_{i}\left(x_{i}^{\Diamond}, \boldsymbol{x}_{-i}^{\Diamond}\right)}$, we claim that  
	the canonical duality relation holds over $\Theta_i\times\mathscr{E}_{ i}^{+}$ for $i\in\mathcal{I}$. It follows from Lemma \ref{t1} that  the solution to $\operatorname{VI}(\boldsymbol{\Xi},F)$ is a stationary point profile of (\ref{ee2}) on $\Theta_i\times\Theta_i^{*}$.
	We can further verify that the total complementary function  $\Gamma_{i}(x_{i},\sigma_{i},\boldsymbol{x}_{-i})$ is  concave in  dual  variable $\sigma_{i}$ and convex in $x_{i}$.	
	In this light, we obtain the  globally optimality of $(\boldsymbol{x}^{\Diamond}, \boldsymbol{\sigma}^{\Diamond})$ on $\boldsymbol{\Omega}\times \mathscr{E}^{+} $, that is, for $x_{i}\in \Omega_i$ and $\sigma_{i}\in \mathscr{E}_{i}^{+}$, 
	\begin{equation*}
	\Gamma_{i}(x_{i}^{\Diamond},  \sigma_i,\boldsymbol{x}_{-i}^{\Diamond})\leq\Gamma_{i}(x_{i}^{\Diamond},  \sigma_i^{\Diamond},\boldsymbol{x}_{-i}^{\Diamond}) \leq \Gamma_{i}(x_{i},  \sigma_i^{\Diamond},\boldsymbol{x}_{-i}^{\Diamond}). 
	\end{equation*}
	This confirms that $\boldsymbol{x}^{\Diamond}$ is the global NE of (\ref{f1}). 
	The whole proof of Theorem \ref{t2} can be found in the Supplementary Materials due to space limitations.
}

	The result in Theorem \ref{t2}  reveals that  
once the solution of $\operatorname{VI}(\boldsymbol{\Xi},F)$ is obtained, we can check whether the duality  relation $\sigma_{i}^{\Diamond}=\nabla \Psi_{i}\left(\xi_{i}\right) \mid_{\xi_{i}=\Lambda_{i}\left(x_{i}^{\Diamond}, \boldsymbol{x}_{-i}^{\Diamond}\right)}$
holds, so as to identify whether the solution of $\operatorname{VI}(\boldsymbol{\Xi},F)$  is a global NE.
Based on the above conclusion, we are inspired  to solve  $\operatorname{VI}(\boldsymbol{\Xi},F)$ via its first-order conditions and employ the duality relation as a criterion for identifying the global NE. 

\begin{remark}
The foundation  to realize the above idea  is the  nonempty set $ \mathscr{E}_{ i}^{+} $. 
It is possible to obtain an empty $\mathscr{E}_{ i}^{+}$ in reality, provided by $P_{i}(\sigma_{i}) \succeq \kappa_{x}\boldsymbol{I}_{n}$  has no intersection with  $\Theta_i^*$, and these situations  
make the above  duality theory approach  unavailable. 	Thus,  $\mathscr{E}_{ i}^{+}$
should be effectively checked  once  the  problem is formulated.  Such a process has also been similarly employed in some classic optimization works to solve non-convex problems \cite{zhu2012approximate,liang2019topology,ren2021distributed,zheng2012zero,latorre2016canonical}.  In addition, this is why we cannot directly employ the standard Lagrange multiplier method and the associated KKT theory, because we need to confirm a feasible d domain of multiplier $\sigma_i$ by utilizing canonical duality information (referring to $\Theta_i^*$).
\end{remark}
\begin{remark}
By generalizing the coupled stationary conditions to continuous vector fields, we compactly formulate these stationary conditions of the dual problem (\ref{ee2}) as a continuous mapping in a VI problem $\operatorname{VI}(\boldsymbol{\Xi},F)$. 
Thus, seeking all players' stationary point profile (or Nash stationary point) can be accomplished by  verifying a fixed point of 
	this continuous mapping. 
	The seed of employing the VI idea in  game problems dates back to \cite{harker1990finite}, and has since found wide applications in various game models, for a survey, see {\cite{giannessi2006equilibrium}} and the references therein.
\end{remark}

\section{Approaching Global NE via Conjugate-based ODE}\label{ODE}


{In this section,  we propose an ODE to seek the solutions to $\operatorname{VI}(\boldsymbol{\Xi},F)$  (\ref{e2e}) with the assisted complementary information (the   Legendre conjugate of $\Psi_i$  and the canonical dual variable $\sigma_i$).}
In fact,  
an ODE provides continuously evolved dynamics, which help reveal how the primal variables and the canonical dual ones influence  each other via conjugate gradient information.
Meanwhile, the analysis techniques in modern calculus and nonlinear systems for theoretical guarantees of ODEs may lead to comprehensive results with mild assumptions.  

\subsection{ODE Design}

 {Consider that the local set constraints of variables, like $\Omega_i$ and $ \mathscr{E}_{i}^{+} $ of (\ref{e2e}), are  usually equipped with specific structures in various tasks. We intend to employ 
 conjugate properties of the generating functions within Bregman divergence to design ODE flows.}
Take   $ \phi_{i}(x_i) $ and $ \varphi_{i}(\sigma_i) $ as 
two {generating functions}, 
{where $ \phi_{i}(x_i) $ is 
	$\mu_{x}$-strongly convex and $L_{x}$-smooth  on $ \Omega_{i} $, and $ \varphi_{i}(\sigma_i) $ is
	$\mu_{\sigma}$-strongly  convex and $L_{\sigma}$-smooth on $ \mathscr{E}_{ i}^{+} $.}
It follows from  the Fenchel inequality \cite{diakonikolas2019approximate} that the Legendre conjugate $\phi_i^{*}$ and $\varphi_i^{*}$ are   convex and differentiable,
where for $y_i\in \mathbb{R}^{n}$,
$$\phi_i^{*}(y_i)\triangleq\operatorname{min}_{x_i \in \Omega_i}\{-x_i^{T} y_i+\phi_i(x_i)\},$$  and  for $\nu_i\in  \mathbb{R}^{q_{i}}$, 
$$\varphi_{i}^{*}(\nu_i) \triangleq \operatorname{min}_{\sigma_i \in \mathscr{E}_{ i}^{+}}\{-\sigma_i^{T} \nu_i+\varphi_i(\sigma_i)\}.$$ 
Accordingly, their conjugate gradients
satisfy
\begin{equation}\label{c1}
\nabla \phi_i^{*}(y_i) = \operatorname{argmin}_{x_i \in \Omega_i}\left\{-x_i^{T} y_i+\phi_i(x_i)\right\}, 
\end{equation}
\begin{equation}\label{e222}
\nabla \varphi_{i}^{*}(\nu_i) = \operatorname{argmin}_{\sigma_i \in \mathscr{E}_{ i}^{+}}\{-\sigma_i^{T} \nu_i+\varphi_i(\sigma_i)\}.
\end{equation}

\begin{table*}[htb]
	\centering
	
	\small
	\caption{Closed-form conjugate gradients with different generating functions.}
	\setlength\tabcolsep{10pt}
	\renewcommand\arraystretch{2.3}
	\begin{tabular}{llll}
		\hline
		\hline
		& Feasible set  &  Generating function  &  Conjugate gradient \\ \hline
		
		General convex set&	$\Omega$  &$\!\frac{1}{2}\|x\|^2_2$ & $\!{\operatorname{argmin}_{x \in \Omega}} \frac{1}{2}\|x\!-\!y\|^{2}$  \\ 
		\makecell[l]{Non-negative orthant }&$ \mathbb{R}^n_+$ &$\!\!\sum\nolimits_{l=1}\nolimits^{n} \!x^{l} \log (x^{l})\!-\!x^{l}$ & ${\exp (y)}$  \\
		Unit square $ [a,b]^n$ 
		&
		$ \!\!\{x^l\!\in\! \mathbb{R}\!\!: \!a\!\leq\! x^l\!\leq \!b \}$
		&$\makecell[l]{\!\sum\nolimits_{l=1}\nolimits^{n} \!(x^{l} \!- \!a)\! \log (x^{l} \!- \!a)\!\\\;\;+\!(b\!-\!x^{l}) \!\log (b\!-\!x^{l})}$ & $\operatorname{col}\{\frac{a+b\exp (y^{l})}{ \exp (y^{l})+1}\}_{l=1}^{n}$  \\
		Simplex $\Delta^{n}$
		&$ \!\!\{x\!\in\! \mathbb{R}^n_+\!:\!\sum\nolimits^{n}\nolimits_{l=1}\!x^l=1 \}$ &$\!\sum\nolimits_{l=1}\nolimits^{n} \!x^{l} \log (x^{l})$ & $\operatorname{col}\{\frac{\exp (y^{l})}{\sum_{j=1}^{n} \exp (y^{j})}\}_{l=1}^{n}$  \\ 
		Euclidean sphere $\textbf{B}^{n}_{\rho}(w)$&$\!\!\{x\!\in\! \mathbb{R}^n\!:\!\!\|x\!-\!w \|^2_2\!\leq\! p\}$ &$\!\!-\!\sqrt{p^2\!-\!\|x\!-\!w \|^2_2}$ & $py[\sqrt{1\!+\!\|y\|^2_2} ]^{-1}\!-\!w  $  \\
		\hline\hline
	\end{tabular}
	\label{tab1}
\end{table*}


On this basis, for each player $i\in\mathcal{I}$, the conjugate-based ODE for seeking a global NE of the non-convex multi-player game (\ref{f1}) is presented by
\begin{equation}\label{e21}
\left\{\begin{array}{l}
\vspace{0.16cm}
\dot y_i=- \sigma_{i}^{T}\nabla_{x_{i}} \Lambda_{i}\left(x_{i},\boldsymbol{x}_{-i}\right)
+\nabla \phi_i(x_i)-y_i,  \\\vspace{0.16cm}
\dot{\nu}_{i} = \Lambda_{i}\left(x_{i},\boldsymbol{x}_{-i}\right)\!-\!\nabla \Psi_{i}^{*}\left(\sigma_{i}\right)\!+\!\nabla \varphi_i(\sigma_i)\!-\!\nu_i,    \\\vspace{0.16cm}		
x_{i}=\nabla \phi_{i}^{*}
\left(y_{i}\right),\\\vspace{0.16cm}
\sigma_{i}=\nabla \varphi_{i}^{*}
\left(\nu_{i}\right),
\end{array}\right.
\end{equation}
where the initial condition is $y_{i}(0)=y_{i0}\in\mathbb{R}^{n}$, $\nu_{i}(0)=\nu_{i0}\in\mathbb{R}^{q_i}$, $x_{i}(0)=\nabla \phi_{i}^{*}(y_{i0})$, and $\sigma_{i}(0)=\nabla \varphi_{i}^{*}(\nu_{i0})$. Here, $t$ represents continuous time, and we drop  $t$
in the dynamics  for a concise expression. 

We give an explanation of two important operations for designing the conjugate-based ODE (\ref{e21}).
On the one hand, we design the dynamics for $y_i(t)$ and $\nu_i(t)$ in dual spaces via the stationary conditions in (\ref{e2e}).
The terms about $- \sigma_{i}^{T}\nabla_{x_{i}} \Lambda_{i}(x_{i},\boldsymbol{x}_{-i})$  and $\Lambda_{i}(x_{i},\boldsymbol{x}_{-i})-\nabla \Psi_{i}^{*}(\sigma_{i})$ represent the directions of gradient descent and ascent according to $\Gamma_i$ in (\ref{ee2}).
Besides, $\nabla \phi_i(x_i)$ and $\nabla \varphi_i(\sigma_i)$ are regarded as damping terms in ODE to avoid $y_i$ and $\nu_i$  going to infinity \cite{nemirovskij1983problem,krichene2015accelerated}.
On the other hand, 
 the mapping from dual spaces  back to primal spaces is implemented to update $x_i(t)$ and $\sigma_i(t)$ by virtue of conjugate gradients $ \nabla \phi_{i}^{*} $ and $ \nabla \varphi_{i}^{*} $, serving as the output feedback in system updating. 
%

Actually, the above conjugate-based idea not only helps solve the inherent non-convexity in such a class of multi-player games, but also brings the convenience of dealing with local set constraints with specific structures.  We would like to remark that the mappings via conjugate gradients $ \nabla \phi_{i}^{*} $ and $ \nabla \varphi_{i}^{*} $ are established based on valid generating functions rather than a conventional Euclidean norm,  which yields explicit map relations between dual spaces and primal spaces to flexibly deal with diverse constraint conditions. 
\begin{remark}
{Without subscript $i$ in $\Omega$, we give some practical examples to show different  closed-form conjugate gradients.
%
When  $\Omega$ is an $n$-dimensional unit simplex 
of soft-max output layers in GAN \cite{daskalakis2018training}, i.e., $ \Delta^{n}=\!\!\{x\!\in\! \mathbb{R}^n_+\!:\!\sum\nolimits^{n}\nolimits_{l=1}\!x^l=1 \}$, a widely
used generating function is the (negative) Gibbs–Shannon entropy 
$\phi(x)=\!\sum\nolimits_{l=1}\nolimits^{n} \!x^{l} \log (x^{l})$, which yields the closed-form conjugate gradient  $\nabla \phi^{*}(y)=\operatorname{col}\{\frac{\exp (y^{l})}{\sum_{j=1}^{n} \exp (y^{j})}\}_{l=1}^{n}$. When $\Omega$
 is a Euclidean sphere of parameter perturbation in  adversarial training \cite{deng2021local}, i.e., $\textbf{B}^{n}_{\rho}(w)=\!\!\{x\!\in\! \mathbb{R}^n\!:\!\!\|x\!-\!w \|^2_2\!\leq\! p\} $, the generating function can be chosen as $\phi(x)=\!\!-\!\sqrt{p^2\!-\!\|x\!-\!w \|^2_2}$, which  explicitly yields $\nabla \phi^{*}(y)=py[\sqrt{1\!+\!\|y\|^2_2} ]^{-1}\!-\!w   $.  
 Moreover, recalling  sensor localization tasks \cite{jia2013distributed},
we can take $\phi(x)=(x-a) \log (x-a)+(b-x)\log(b-x)$ since   $\Omega=[a,b]$ has a unit-square form, which brings the closed-form $\nabla \phi^{*}(y)=\operatorname{col}\{\frac{a+b\exp (y^{l})}{ \exp (y^{l})+1}\}_{l=1}^{n}$. 
Readers can check Table  \ref{tab1}
for more cases.}
\end{remark}


\subsection{Convergence Analysis}

Next, we investigate the variables' trajectories of (\ref{e21}) and reveal the convergence. Similarly to  $\boldsymbol{x}$ and $\boldsymbol{\sigma}$,  compactly denote $\!\boldsymbol{y}\!\in\!\mathbb{R}^{nN}\!$ and $\boldsymbol{\nu}\!\in\!\mathbb{R}^{q}\!$. 
 Then, ODE (\ref{e21}) can be compactly presented  by
\begin{equation}\label{e1}
	\left\{\begin{array}{l}\vspace{0.15cm}
		\dot{\boldsymbol{y}}=-G(\boldsymbol{x},\boldsymbol{\sigma}) +\nabla\boldsymbol{\phi}(\boldsymbol{x})-\boldsymbol{y},\\\vspace{0.15cm}
		\dot{\boldsymbol{\nu}}={\Lambda}\left(\boldsymbol{x}\right)-\nabla \Psi^{*}\left(\boldsymbol{\sigma}\right)+\nabla \boldsymbol{\varphi}(\boldsymbol{\sigma})-\boldsymbol{\nu},\\\vspace{0.15cm}
		\boldsymbol{x}=\nabla \boldsymbol{\phi}^{*}(\boldsymbol{y}),\\\vspace{0.15cm}
		\boldsymbol{\sigma}=\nabla \boldsymbol{\varphi}^{*}\left(\boldsymbol{\nu}\right).
	\end{array}\right.
\end{equation}

\begin{lemma}\label{l2}
	Suppose that  
	$(\boldsymbol{y}^{\Diamond},\boldsymbol{\nu}^{\Diamond},\boldsymbol{x}^{\Diamond},\boldsymbol{\sigma}^{\Diamond})$ is an equilibrium point of ODE (\ref{e21}). If $\sigma_{i}^{\Diamond}=\nabla \Psi_{i}\left(\xi_{i}\right) \mid_{\xi_{i}=\Lambda_{i}\left(x_{i}^{\Diamond}, \boldsymbol{x}_{-i}^{\Diamond}\right)}$ for $i\in\mathcal{I}$,
	then $ \boldsymbol{x}^{\Diamond} $ is {the} global NE of  (\ref{f1}). 
\end{lemma}
 Lemma 2 establishes the relationship between the equilibrium of (\ref{e21}) and {the} global NE of (\ref{f1}). {More proof details of Lemma \ref{l2} can be found in 
 %
 the Supplementary Materials due to space limitations.
 } 
 
 In fact, the conjugate-based ODE (\ref{e21}) is designed around solving $\operatorname{VI}(\boldsymbol{\Xi},F)$ (\ref{e2e}), i.e, the equilibrium point of   (\ref{e21}) corresponds to the solution to  (\ref{e2e}).  {Recalling the existence condition  in Theorem \ref{t2}, when the duality relation $\sigma_{i}^{\Diamond}=\nabla \Psi_{i}\left(\xi_{i}\right) \mid_{\xi_{i}=\Lambda_{i}\left(x_{i}^{\Diamond}, \boldsymbol{x}_{-i}^{\Diamond}\right)}$ is satisfied, we can derive that the equilibrium of  ODE (\ref{e21}) realizes the global NE of  non-convex multi-player game (\ref{f1}). }
The subsequent theorems presents the main convergence results of ODE   (\ref{e21}), which implies that global NE can be found along  the trajectory of ODE (\ref{e21}).

\begin{mythm}\label{l1}
	If $\mathscr{E}_{ i}^{+}$ is nonempty for $i\in\mathcal{I}$,
	then ODE (\ref{e21}) is bounded and convergent.
	Moreover, if 
	the convergent point $(\boldsymbol{y}^{\Diamond},\boldsymbol{\nu}^{\Diamond},\boldsymbol{x}^{\Diamond},\boldsymbol{\sigma}^{\Diamond})$  satisfies 
	$\sigma_{i}^{\Diamond}=\nabla \Psi_{i}\left(\xi_{i}\right) \mid_{\xi_{i}=\Lambda_{i}\left(x_{i}^{\Diamond}, \boldsymbol{x}_{-i}^{\Diamond}\right)}$ for $i\in\mathcal{I}$, 
	then $\boldsymbol{x}^{\Diamond}$ is the 
	global NE of  (\ref{f1}).
\end{mythm}
Furthermore, the convergence rate of (\ref{e21}) can be obtained.
\begin{mythm}\label{t4}
	If $\mathscr{E}_{ i}^{+}$ is nonempty and $ \Psi_{i}\left(\cdot\right)$ is $\frac{1}{\kappa_{\sigma}}$-smooth	for $i\in\mathcal{I}$,  
	then 
	(\ref{e21}) converges at an exponential rate, i.e.,
	\begin{align*}
	\|\boldsymbol{z}(t)-\boldsymbol{z}^{\Diamond}\|\leq \sqrt{\frac{{\tau}}{{\mu}}}\|\boldsymbol{z}(0)\|\operatorname{exp}(- \frac{{\kappa}}{2{\tau}}\,t),
	\end{align*}
	where  
	$\mu=\min\{{\mu_{x}}/{2},{\mu_{\sigma}}/{2} \}$, 
	${\kappa}=\min\{\kappa_{\sigma}, \kappa_{x} \}$, ${\tau}=\max\{{L_{x}}/{2\mu_{x}},{L_{\sigma}}/{2\mu_{\sigma}} \}$.
\end{mythm} 
{We provide a proof sketch here for Theorems \ref{l1} and \ref{t4}. We first prove that the trajectory $( \boldsymbol{y}(t)$, $\boldsymbol{x}(t)$,  $\boldsymbol{\nu}(t) $, $\boldsymbol{\sigma}(t) )$ is bounded along ODE (\ref{e21}).  
	To this end, we construct a Lyapunov candidate function with the Bregman divergence as below
	\begin{equation*}
	\begin{aligned}
	V_{1}=& \sum_{i=1}^{N} D_{\phi_{i}^{*}}(y_{i},y_{i}^{\Diamond})+D_{\varphi_{i}^{*}}(\nu_{i},\nu_{i}^{\Diamond}).
	\end{aligned}
	\end{equation*}
We can carefully verify that
	\begin{equation*}
	V_{1} \geq \frac{\mu_{x}}{2}\left\|\boldsymbol{x}-\boldsymbol{x}^{\Diamond}\right\|^{2}+ \frac{\mu_{\sigma}}{2}\left\|\boldsymbol{\sigma}-\boldsymbol{\sigma}^{\Diamond}\right\|^{2}.
	\end{equation*}
	This means that $V_{1}$ is positive semi-definite, and is  radially unbounded in $\boldsymbol{x}(t)$ and $\boldsymbol{\sigma}(t)$. So far, the utilized technique tools include the convexity of generating functions, the canonical dual relations, and the optimal conditions of VI problem, though the derivations are not straightforward.
	We next investigate the derivative of $V_1$ along the ODE. We can cautiously obtain that $	d{V}_{1}/dt\leq 0$, which yields that the trajectories of $\boldsymbol{x}(t)$ and $\boldsymbol{\sigma}(t)$ are bounded along  ODE (\ref{e21}). Similarly, we can get  that $\boldsymbol{y}(t)$ and $\boldsymbol{\nu}(t)$ are bounded too. The above completes most parts in the proof of Theorem \ref{l1}, except for some discussions on the invariant sets of the ODE. 
	For Theorem \ref{t4}, we can get $
	d{V}_{1}/dt\leq -\frac{\kappa}{\tau} {V}_{1}$ more than $d{V}_{1}/dt\leq 0$ thanks to the smoothness of function $ \Psi_{i}$. This eventually results in the exponential rate of the ODE. More detailed proofs of Theorems \ref{l1} and \ref{t4} can be found in the Supplementary Materials due to space limitations.

}
\begin{remark}\label{roadmap}
We summarize the road map for seeking global NE in this non-convex game for friendly comprehension.
First, we  
should  check whether   $\mathscr{E}_{ i}^{+}$ is nonempty  once the problem is defined and formulated.
Next, we should seek the solution to $\operatorname{VI}(\boldsymbol{\Xi},F)$ via  ODE flows, wherein the variable $\sigma_i$ is restricted  on the nonempty $\mathscr{E}^{+}_{i}$ and the implementation of the ODE is guaranteed. 
Finally, after obtaining the solution to $\operatorname{VI}(\boldsymbol{\Xi},F)$ by convergence, we should identify whether the convergent point satisfies the duality relation $\sigma_{i}^{\Diamond}=\nabla \Psi_{i}\left(\xi_{i}\right) \mid_{\xi_{i}=\Lambda_{i}\left(x_{i}^{\Diamond}, \boldsymbol{x}_{-i}^{\Diamond}\right)}$. If so,
the  convergent point is a global NE.   
\end{remark}

\section{Discrete Algorithm}\label{d5}


{In this section, we consider deriving the discretization from the conjugate-based ODE, so as  to make the dynamics implementable
in practice. Notice that each step in  discrete algorithms can directly compute the minimum of a sub-problem, rather than walks along some trajectories of  conjugate functions with explicit expressions  in the continuous ODE.  Tapping into this advantage,
the corresponding   discrete algorithm  is not obligated to resort to conjugate information like variables $y_i$ and $\nu_i$, which results in simplifying the algorithm iteration.}
\subsection{Discrete Algorithm Design}

We redefine an operator
generated by $\Psi_i$ on $\Theta_i$   as 
\begin{align*}
\Pi_{\Theta_i}^{\Psi_i}(\sigma_i)=\operatorname{argmin}_{\xi_i \in \Theta_i}\{-\sigma_i^{T} \xi_i+\Psi_i(\xi_i)\}.
\end{align*}
This operator avoids computing
the conjugate information of $\Psi_i^*$. Similarly, redefine  operators $\Pi_{\Omega_{i}}^{\phi_i}(\cdot)=\nabla \phi_i^{*}(\cdot)$ in (\ref{c1})
and $\Pi_{\mathscr{E}_{ i}^{+}}^{\varphi_i}(\cdot)=\nabla \varphi_i^{*}(\cdot)$ in (\ref{e222}). 
With  a step size $\alpha_k$ at discrete time $k$,
we  discretize the conjugate-based ODE (\ref{e21}) in the following.
\begin{algorithm}[htb]
	\caption{ }
	\label{nplayer}
	\renewcommand{\algorithmicrequire}{\textbf{Input:}}
	\renewcommand{\algorithmicensure}{\textbf{Initialize:}}
	\begin{algorithmic}[1]
		\REQUIRE 
		Step size  $ \{\alpha_{k} \} $,
		proper generating functions $\phi_{i}$  on $\Omega_{i}$ and  $\varphi_{i}$ on $\mathscr{E}_{ i}^{+}$.  
		\ENSURE $x_{i}^{0}\in \Omega_{i},   \sigma_{i}^{0}\in \mathscr{E}_{ i}^{+},\,i\in\{1,\dots,N\}$. 
		\FOR{$k = 1,2,\cdots$}
		\FOR{player $i\in\{1,\dots,N\}$}
		\STATE compute the conjugate of $\Psi_i$:\\
		\;$\xi_i^{k}=\Pi_{\Theta_i}^{\Psi_{i}}(\sigma_i^{k})  $\\\vspace{0.05cm}
		\STATE update the decision variable :\\\vspace{0.05cm}
		\;$	x_{i}^{k+1} \!=\! \Pi_{\Omega_{i}}^{\phi_i}\!(-\alpha_k \!\sigma_{i}^{kT}\!\nabla_{x_{i}} \Lambda_{i}(x_{i}^{k},\boldsymbol{x}_{-i}^{k})\!+\!\nabla\! \phi_{i}(x_i^{k})  )
		$\\
			\STATE update  the canonical dual variable:\\\vspace{0.05cm}
		\;$\sigma_{i}^{k+1} \!= \!\Pi_{\mathscr{E}_{ i}^{+}}^{\varphi_i}(\nabla \varphi_{i}(\sigma_i^{k})\!+\!\alpha_k(\Lambda_{i}\left(x_{i}^{k},\boldsymbol{x}_{-i}^{k}\right)-\xi_i^{k}))
		$
		\ENDFOR
		\ENDFOR
	\end{algorithmic}
\end{algorithm}

The update of $x_{i}^{k+1}$ in Algorithm 1  can be equivalently expressed as
\begin{align*}
{\operatorname{argmin}_{x \in \Omega_{i}}}\{\langle x, \sigma_{i}^{kT}\nabla_{x_{i}} \Lambda_{i}(x_{i}^{k},\boldsymbol{x}_{-i}^{k})\rangle+\frac{1}{\alpha_k} D_{\phi_{i}}(x, x_{i}^{k})\},
\vspace{-0.5pt}
\end{align*}
where $D_{\phi_{i}}(x, x_{i}^{k})$ is the {Bregman divergence} with generating function $\phi_{i}$.
A similar equivalent scheme can be found in $\sigma_i^{k+1}$. 
These equivalent iteration schemes reveal that, parts of the idea in Algorithm 1 derived from the conjugate-based ODE (\ref{e21}) actually  coincide with the \textit{mirror descent} method \cite{nemirovskij1983problem}.  
Therefore, after computing the conjugate of $\Psi_i$ and plugging it into the update of $\sigma_i$ referring to properties of  canonical functions and VIs, readers may also regard Algorithm 1 from the {mirror descent} perspective.
\subsection{{Step Size } and Convergence Rate}
Hereinafter,  we provide the step-size  settings and the corresponding convergence rates of Algorithm 1 in two typical non-convex multi-player games.

\noindent\textbf{Multi-player generalized monotone games}
Monotone games stand for a broad category in game models, where the pseudo-gradients of all players' payoffs satisfy the properties of monotonicity \cite{facchinei2010penalty,koshal2016distributed,chen2021distributed}. 
The monotone property  yields the equivalence between the weak and the strong solutions to VI problems \cite{minty1962monotone}, which makes most convex games solvable by the first conditions in VI. 
Analogously, we consider Algorithm 1 under a class of generalized monotone conditions \cite{giannessi2006equilibrium}, referring to  the canonical complementary function (\ref{ee2}),
and are  rewarded by the following results.

\begin{mythm}\label{t6}
	If $\mathscr{E}_{ i}^{+}$ is nonempty and 	 $\Pi_{\Theta_{ i}}^{\Psi_i}(\cdot)$ is $\kappa_{\sigma}$-strongly monotone, 	then Algorithm \ref{nplayer} converges at a rate of $\mathcal{O}({1}/{k})$ with step size $\alpha_k=\frac{2}{\kappa(k+1)}$, i.e.,
	\begin{align*}
		\|\boldsymbol{x}^k-\boldsymbol{x}^{\Diamond} \|^2+\|\boldsymbol{\sigma}^k-\boldsymbol{\sigma}^{\Diamond} \|^2
	\leq 
	\frac{1}{k+1}\frac{{M}_1}{{\mu}^2{\kappa}^2},
	\end{align*}
		where  ${\mu}=\min\{\frac{\mu_{x}}{2},\frac{\mu_{\sigma}}{2} \}$,  ${\kappa}=\min\{\kappa_{\sigma}, \kappa_{x} \}$, and ${M}_1$ is a positive constant.
\end{mythm}
{The proof sketch is presented below.  Take a collection of the Bregman divergence  as
\begin{equation*}
\Delta(\boldsymbol{z}^{\Diamond}\!,\boldsymbol{z}^{k+1})\!\triangleq\!\sum_{i=1}^{N}\! D_{\phi_{i}}(x_{i}^{\Diamond},x_{i}^{k+1})\!+\!D_{\varphi_{i}}(\sigma_{i}^{\Diamond},\sigma_{i}^{k+1}).
\end{equation*}
Here we employ the three-point identity, the Fenchel's inequality, the strong monotonicity of $F(\boldsymbol{z})$, and the optimality of VI solution to process  the above formula. Then we get
\begin{equation*}
\Delta(\boldsymbol{z}^{\Diamond}\!,\boldsymbol{z}^{k+1})\!\triangleq
\leq \!\Delta(\boldsymbol{z}^{\Diamond}\!,\boldsymbol{z}^{k}\!)\!-\! \alpha_{k}\kappa \|\boldsymbol{z}^{k}\!-\!\boldsymbol{z}^{\Diamond}\|^2 \!\!+\!\! \frac{\alpha_{k}^2}{4\mu}\|F(\boldsymbol{z}^{k}) \|^2.
\end{equation*}
Moreover, by substituting above with $ \eta_k= \kappa\alpha_k $, we take the sum of these  inequalities over $k, \cdots , 1$ and obtain
\begin{equation*}
\Delta(\boldsymbol{z}^{\Diamond},\boldsymbol{z}^{k+1})\leq\eta_{k}^{2} (k+1) \frac{{M}_1}{4\kappa^2\mu}.
\end{equation*}
Owing to the step-size setting   $ \eta_k= \kappa\alpha_k= {2}/{(k+1)}$, we finally reach the conclusion. The whole proof of Theorem \ref{t6} can be found 
%
in the  Supplementary Materials due to space limitations.
}

\noindent\textbf{Multi-player potential games}
Potential games also have a wide spectrum of applications such as 
power allocation
\cite{yang2018df},
congestion control 
\cite{lei2020asynchronous},
and multi-target tracking \cite{soto2009distributed}. In  a potential game, there exists a unified potential function for all players such that the change in each player’s payoff is equivalent to the change in the potential function.
Hence, the deviation in the payoff of each player in (\ref{fda}) can be concretely mapped to a uniformed canonical  potential function, that is, 
\begin{equation}\label{p2}
J_{i}\left(x_{i}^{\prime}, \boldsymbol{x}_{-i}\right)-J_{i}(\boldsymbol{x})=\Psi(\Lambda(x_i^{\prime},\boldsymbol{x}_{-i}))
-\Psi(\Lambda(\boldsymbol{x})).
\end{equation}
The  complementary function is thereby obtained with a common canonical dual variable  $\sigma$  as
\begin{equation}\label{p11}
\Gamma_i(x_{i},\sigma,\boldsymbol{x}_{-i})=\Gamma(\boldsymbol{x},\sigma)=\sigma^{T} \Lambda\left(\boldsymbol{x}\right)-\Psi^{*}\left(\sigma\right).
\end{equation}
Also,  the  set  $\mathscr{E}^{+}$ of  $\sigma$ is in a unified form similar to
(\ref{s23}).
Considering the weighted averages $\hat{\boldsymbol{x}}^{k}$ and $\hat{{\sigma}}^{k}$ in the course of $k$ iterations,
we give the convergence rate of Algorithm 1 in the result below.

\begin{mythm}\label{t9}
	If $\mathscr{E}^{+}$ is nonempty and players' payoffs are subject to the potential function in
	(\ref{p2}), then Algorithm 1 converges at a rate of $ \mathcal{O}(1/\sqrt{k}) $  with  step size $\alpha_k=\frac{2{\mu{d}}}{{M}_2\sqrt{k}}$,  i.e.,
	\begin{align*}
	\Gamma(\hat{\boldsymbol{x}}^{k},\sigma^{\Diamond})-
	\Gamma(\boldsymbol{x}^{\Diamond},\hat{\sigma}^{k})
	\leq 
	\frac{1}{\sqrt{k}} \sqrt{\frac{{d}}{\mu}}{M}_2,
	\end{align*}
	where ${\mu}=\min\{\frac{\mu_{x}}{2},\frac{\mu_{\sigma}}{2} \}$,  
	and ${d}$, ${M}_2$ are two positive constants.
\end{mythm}
{The proof sketch can be summarized as below. 
	Take another collection of the Bregman divergence  as
	\begin{equation*}
	\widetilde{\Delta}(\boldsymbol{z}^{\Diamond},\boldsymbol{z})\triangleq D_{\varphi}(\sigma^{\Diamond},\sigma)+\sum_{i=1}^{N} D_{\phi_{i}}(x_{i}^{\Diamond},x_{i}).
	\end{equation*}
	By the three-point identity  and $\sigma\in\mathscr{E}^{+}$ in (\ref{s23}), the VI with respect to $\boldsymbol{z}^{k}$ can be bounded by the duality gap of the complementary function, that is,
	\begin{equation*}
	\begin{aligned}
	\left\langle F(\boldsymbol{z}^{k}), \boldsymbol{z}^{\Diamond}-\boldsymbol{z}^{k}\right\rangle \leq \Gamma(\boldsymbol{x}^{\Diamond},\sigma^{k})-\Gamma(\boldsymbol{x}^{k},\sigma^{\Diamond}).
	\end{aligned}
	\end{equation*} 
	Then we can further derive over $1, \cdots , k$ that
	\begin{align*}
	\sum_{j=1}^{k} \alpha_{j}\left(
	\Gamma(\boldsymbol{x}^{j},\sigma^{\Diamond})-
	\Gamma(\boldsymbol{x}^{\Diamond},\sigma^{j})
	\right) \leq \widetilde{\Delta}(\boldsymbol{z}^{\Diamond},\boldsymbol{z}^{1})+\frac{\sum^{k}_{j=1}\alpha_{j}^2 {M}^2_{2}}{4\mu}.
	\end{align*}
	Owing to the Jensen's inequality and the step-size setting   $\alpha_k=2\sqrt{\mu{d}}/{M}_2\sqrt{k}$, we finally reach the conclusion. The whole proof of Theorem \ref{t9} can be found in 
 the Supplementary Materials due to space limitations.
}
\begin{remark}
{
In the aforementioned two cases, our algorithm performs with the best-known convergence rates, and lines up with the results under convex circumstances. With strongly monotone conditions, 
our algorithm for non-convex settings achieves the same convergence rate of $ \mathcal{O}(1/{k})$ as convex cases\cite{yousefian2015self,pavel2019distributed}. The proof is established on the scales from variational theory and the measurement of Bregman divergence, which reflects the convergence rate with respect to the equilibrium point. As for potential games, we sufficiently utilize the unified potential function to derive the same convergence rate of $ \mathcal{O}(1/\sqrt{k}) $ as convex cases \cite{mertikopoulos2019learning,freund1999adaptive}. The convergence result is described by the duality gap within the potential function correspondingly.
}
\end{remark}

\begin{figure*}[t!]
	\centering
 \subfigure[$\boldsymbol{x}^{0}=\{2,-2.5\}$]{
\begin{minipage}[l]{0.3\linewidth}
\centering
\includegraphics[width=6.5cm]{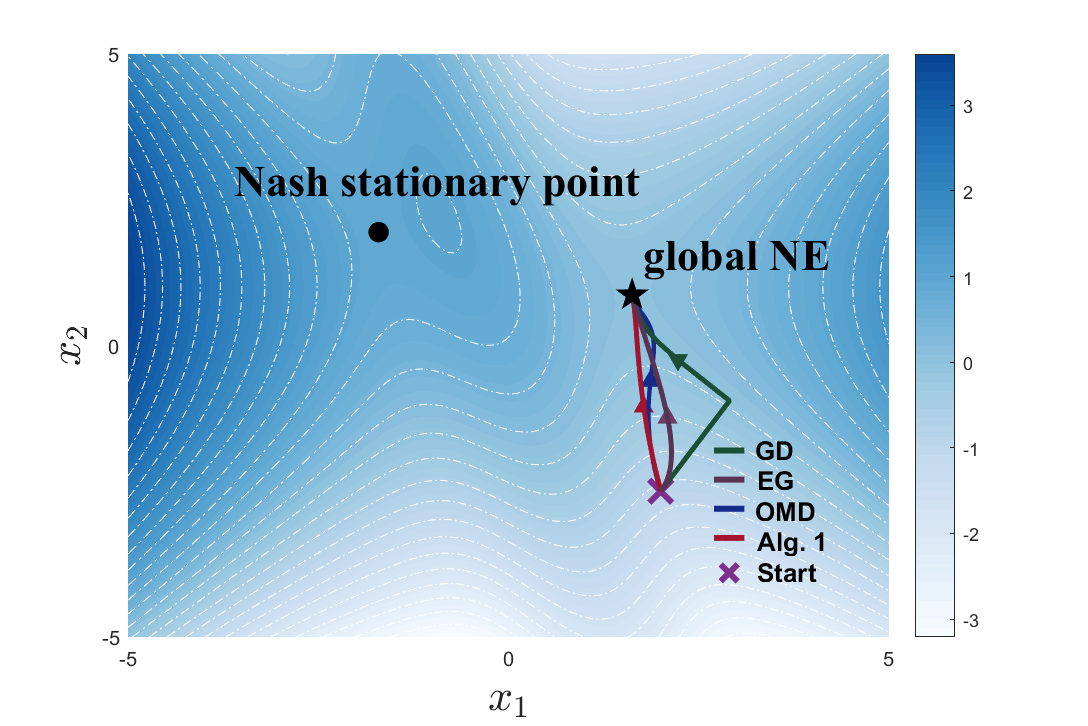}
\end{minipage}
}
 \subfigure[$\boldsymbol{x}^{0}=\{3.5,4.25\}$]{
\begin{minipage}[l]{0.3\linewidth}
\centering
\includegraphics[width=6.5cm]{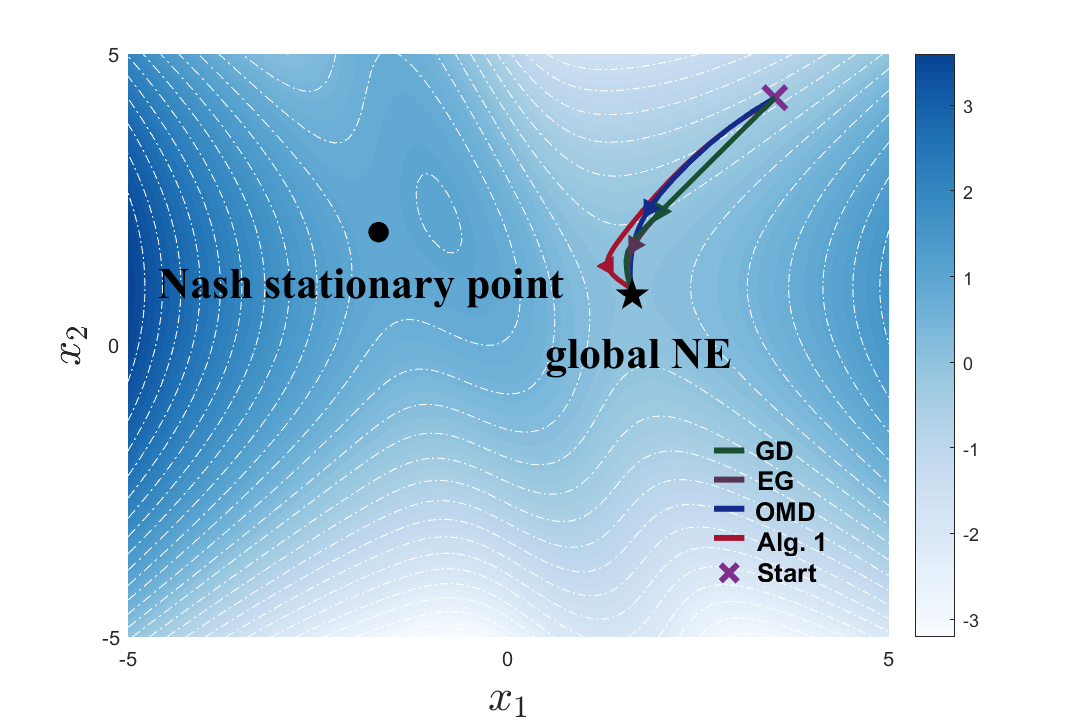}
\end{minipage}
}
 \subfigure[$\boldsymbol{x}^{0}=\{-4,-4\}$]{
\begin{minipage}[l]{0.3\linewidth}
\centering
\includegraphics[width=6.5cm]{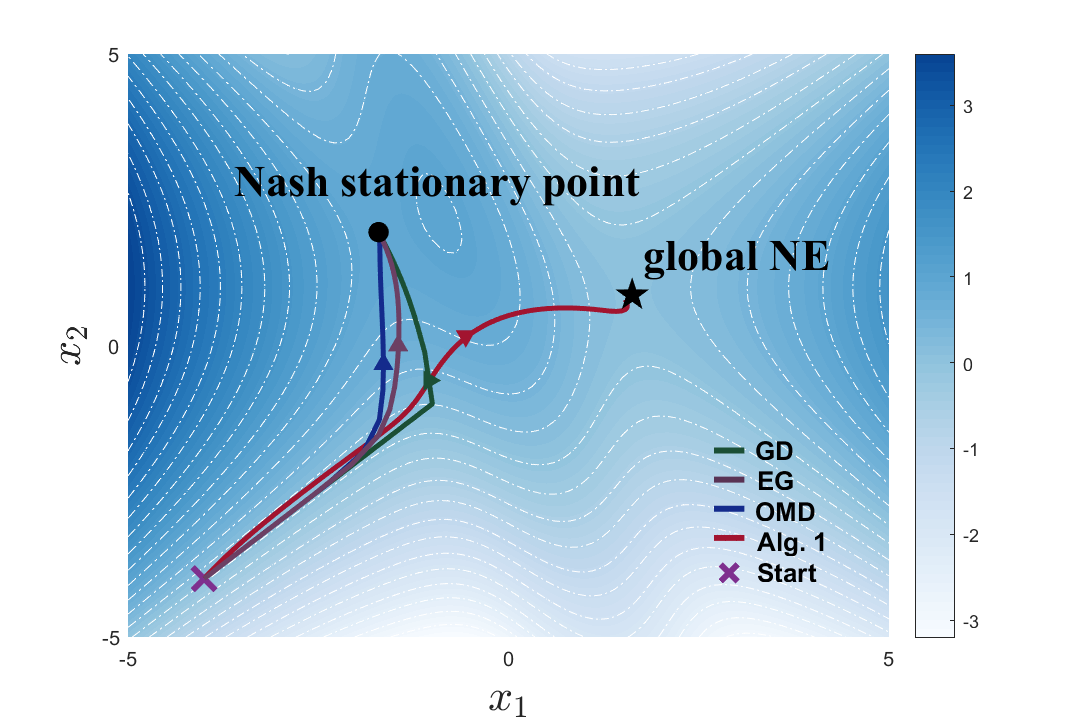}
\end{minipage}
}
\\
 \subfigure[$\boldsymbol{x}^{0}=\{0,-3\}$]{
	\begin{minipage}[l]{0.3\linewidth}
		\centering
		\includegraphics[width=6.5cm]{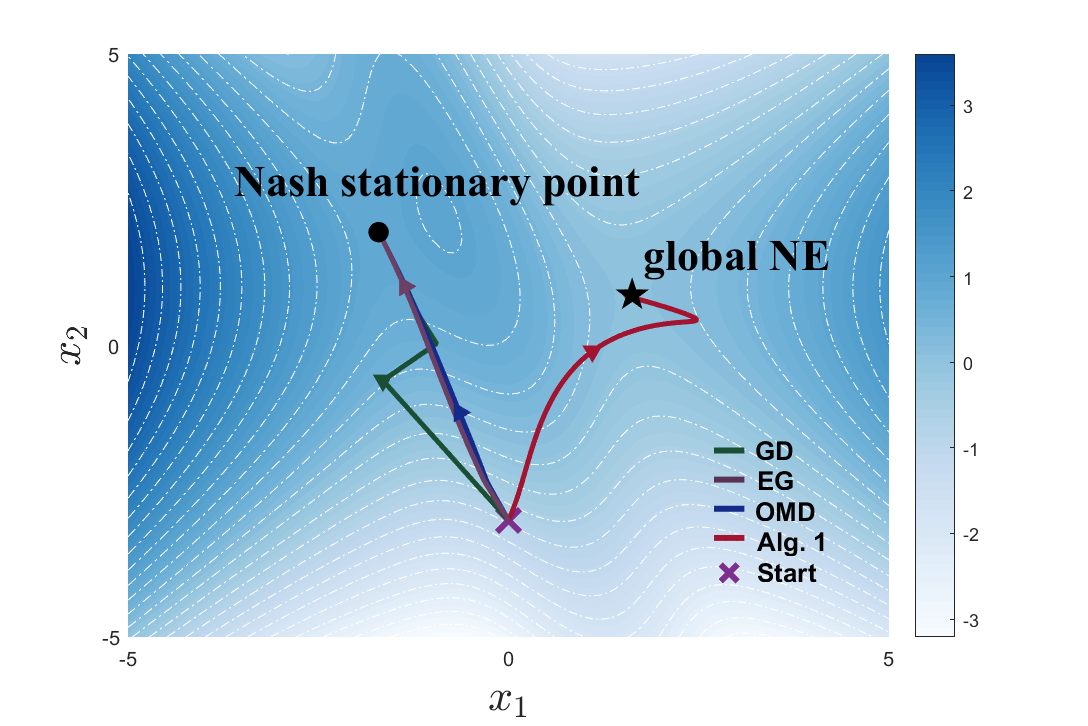}
	\end{minipage}
}
\subfigure[$\boldsymbol{x}^{0}=\{-1.5,3.5\}$]{
	\begin{minipage}[l]{0.3\linewidth}
		\centering
		\includegraphics[width=6.5cm]{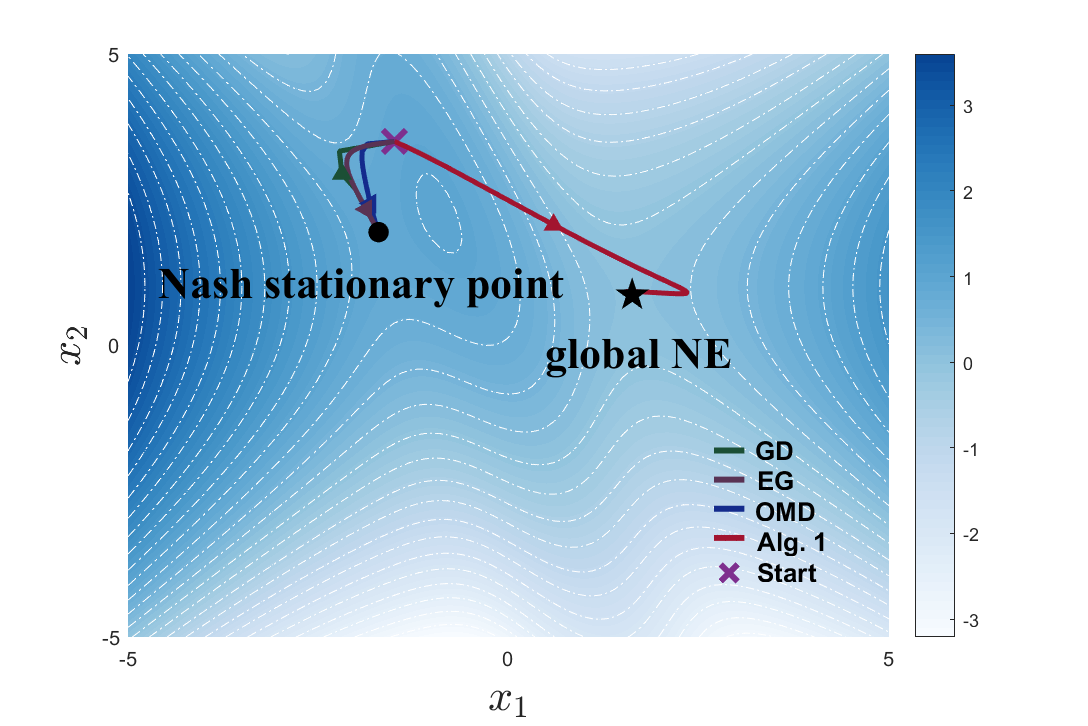}
	\end{minipage}
}
\subfigure[$\boldsymbol{x}^{0}=\{-4,-0.5\}$]{
	\begin{minipage}[l]{0.3\linewidth}
		\centering
		\includegraphics[width=6.5cm]{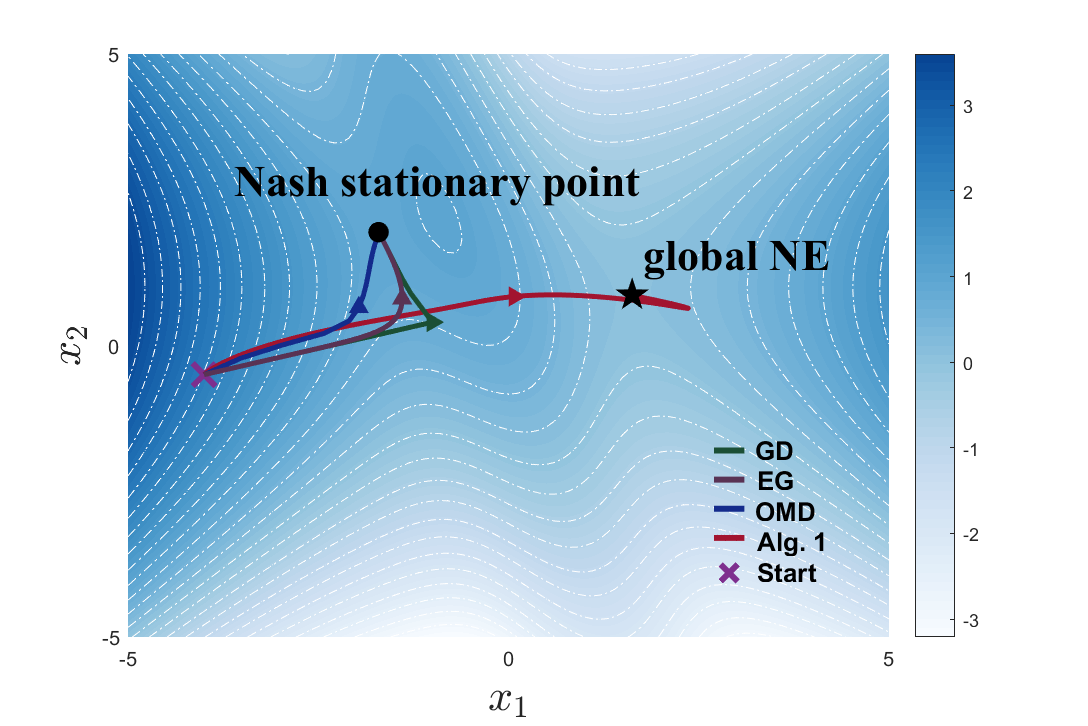}
	\end{minipage}
}

	\caption{Performance of different methods with different initial points.}
	\label{fig6}
\end{figure*}

\section{Experiments}\label{e11}

We examine the  effectiveness of our approach for seeking global NE in the  tasks of robust neural network training and sensor localization. The code will be released publicly.

\subsection{Robust neural network training}

In this part,  we show the convergence performance of our approach under a  robust neural network problem.

\subsubsection{Experiment implementation}


\noindent\textbf{Model and data  pre-processing.} Consider a two-player 
setting in adversarial training \cite{deng2021local,nouiehed2019solving}. The payoff for two players can be described as below 
  \begin{equation}\label{rnn-2}
\min _{x_1} \max _{x_2 } \ell( 
g_{x_1}( \beta_1, x_2),\beta_2)+\frac{\lambda_1}{2}\|x_1\|^{2}-\frac{\lambda_2}{2}\|x_2\|^{2},
\end{equation}
  where function $\ell(\cdot)$  represents the cross-entropy loss on training data $(\beta_1,\beta_2)$, who is perturbed by variable $x_2$. Function $g_{x_1}(\cdot)$ is a 2-layer DNN model, within which variable $x_1$ denotes neural network parameters. The input data  is drown
from a Gaussian distribution $\beta_1\sim\mathcal{N}(\mu,\zeta)$, where $\mu$ is randomly generated from $\mu\sim\mathcal{N}(0,0.25)$ and  $\zeta =1$.
The action set $\Omega_1=\{x_1\in \mathbb{R}: x_{\operatorname{min}}\leq x_{1}\leq x_{\operatorname{max}}\}$ is endowed with a  unit square
form, while set $\Omega_2=\{x_2\in \mathbb{R}: \|x_2\|^{2}\leq \epsilon \}$ is endowed with a Euclidean sphere 
form.

\noindent\textbf{Methods.} We employ Algorithm 1 to solve this problem. Assign $\alpha_k=\frac{2}{k+1}$ as the step size. { Take $\phi_1={(x_{1} \!- \!x_{\operatorname{min}}) \log (x_{1} \!- \!x_{\operatorname{min}})+\!(x_{\operatorname{max}}\!-\!x_{1}) \log (x_{\operatorname{max}}\!-\!x_{1})}$, $\phi_2=-\!\sqrt{\epsilon^2\!-\|x\|^2_2}$, $\varphi_1=\frac{1}{2}\|\sigma_1\|^2$ and $\varphi_2=\frac{1}{2}\|\sigma_2\|^2$.
}
To show the convergence performance of our  approach, 
we compare Algorithm 1 with several 
familiar methods based on stationary information,
such as the classic gradient descent (GD), 
the optimistic mirror descent (OMD) \cite{daskalakis2018training}, and the extra-gradient method (EG) \cite{korpelevich1976extragradient}. Set tolerance $t_{\operatorname{tol}} = 10^{-3}$ and the terminal criterion $\|\boldsymbol{x}^{k+1}-\boldsymbol{x}^{k}\|\leq t_{\operatorname{tol}}$.



\subsubsection{Experiment results}

We verify the existence of global NE in  this non-convex task 
via checking the nonempty set $\mathscr{E}_{ i}^{+}$  and the existence condition in Theorem \ref{t2}. Then we calculate the global NE by employing Algorithm 1, and confirm that the convergent point of Algorithm 1 is the global NE. 
The plot of two opponents' payoffs is shown in  Fig. \ref{fig6}. It can be seen that  
this non-convex game setting has a Nash stationary point and a global NE. Here we show $\boldsymbol{x}^{\dagger}=[-1.69,1.90]$ as the Nash stationary point, while
$\boldsymbol{x}^{\Diamond}=[1.63,0.86]$ as the global NE. 

Moreover, we compare Algorithm 1 with other methods. We randomly initialize these methods. The results are shown in Fig. \ref{fig6} (a)-(f).  Interestingly, the initialization is seen to have little impact on
the convergence of our algorithm, however, changed drastically for
that of others.
{The evolution of $\boldsymbol{x}(t)$ with initial value $\boldsymbol{x}^{0}=[2,-2.5]$ and $\boldsymbol{x}^{0}=[3.5,4.25]$
are respectively shown in Fig. \ref{fig6} (a) and (b). In this case,  all methods  find the global NE.  The evolution of $\boldsymbol{x}(t)$ with initial value $\boldsymbol{x}^{0}=[-4,-4]$ and $\boldsymbol{x}^{0}=[0,-3]$ are respectively shown in Fig. \ref{fig6} (c) and (d). In this case, only Algorithm 1 still achieves the target, while other methods are stuck into a Nash stationary point instead.  The evolution of $\boldsymbol{x}(t)$ with initial value $\boldsymbol{x}^{0}=[-1.5,3.5]$ and $\boldsymbol{x}^{0}=[-4,-0.5]$ are respectively shown in Fig. \ref{fig6} (e) and (f). In this case,  although these initial values are close to the Nash stationary point $\boldsymbol{x}^{\dagger}$  respectively, Algorithm 1 still converges to the global NE. }
These phenomenons clearly imply that Algorithm 1 is an effective algorithm to seek  global NE wherever the initial point lies,  while other methods are susceptible to varied initial points. This fully supports our theory.

\begin{figure}[t]
 \centering
	\includegraphics[width=0.9\columnwidth]{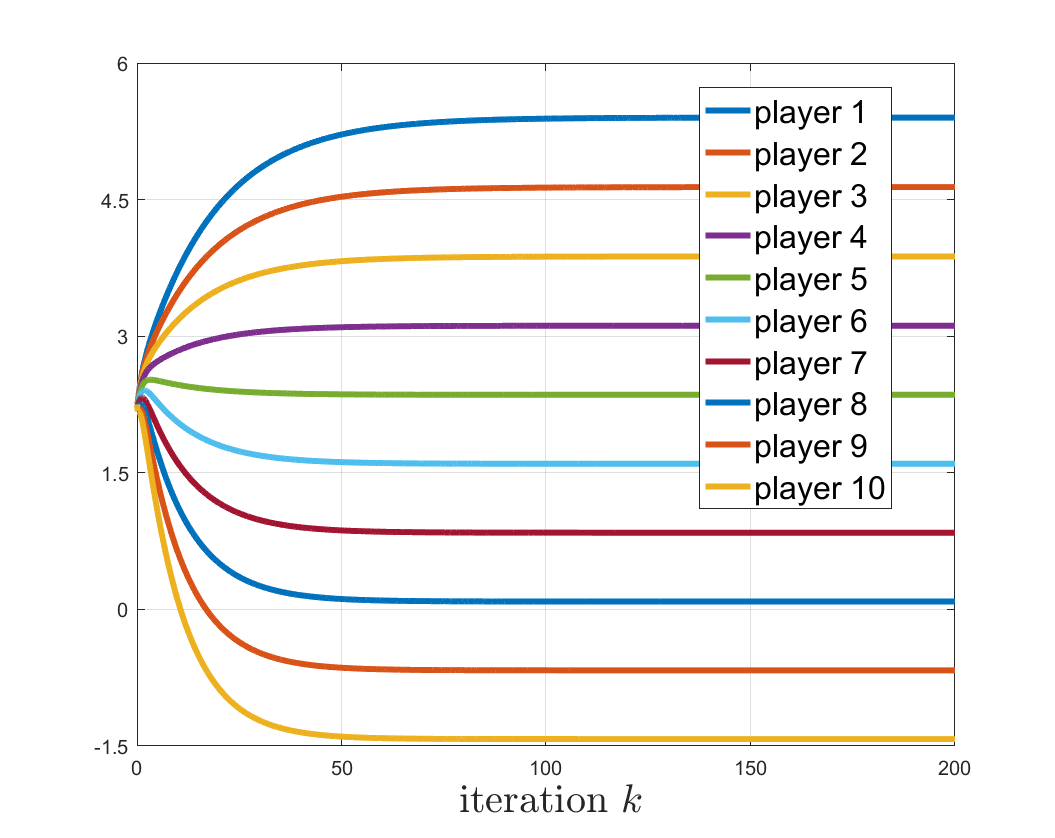}
	\caption{ Convergence of all players’ decisions in Algorithm 1.}
	\label{fg2}
\end{figure}

\begin{figure}[t]
 \centering
	\includegraphics[width=0.9\columnwidth]{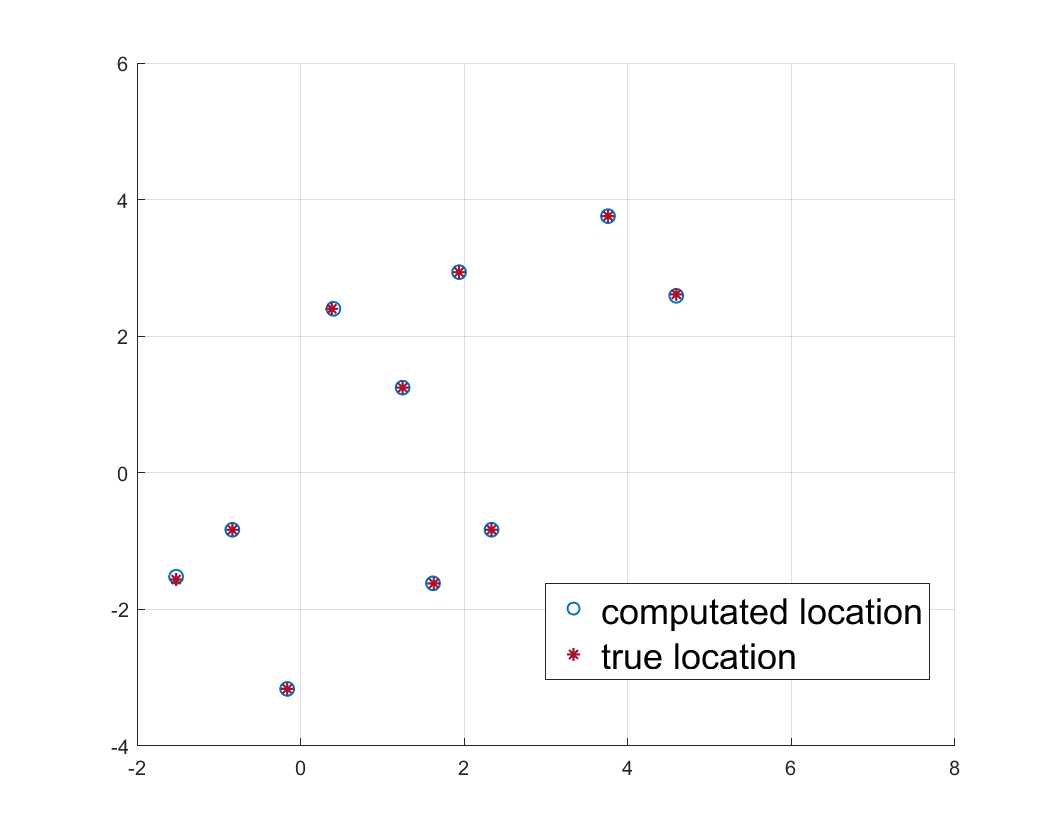}
	\caption{ Computed sensor location results.}
	\label{fg9}
\end{figure}

	\begin{figure*}[t!]
 \centering
\vspace{-10pt}
\hspace{-18pt}
\subfigure[$x_{11}(0)=3$]{
\begin{minipage}[c]{0.32\linewidth}
\centering
\includegraphics[width=6.5cm,height = 4.5cm]{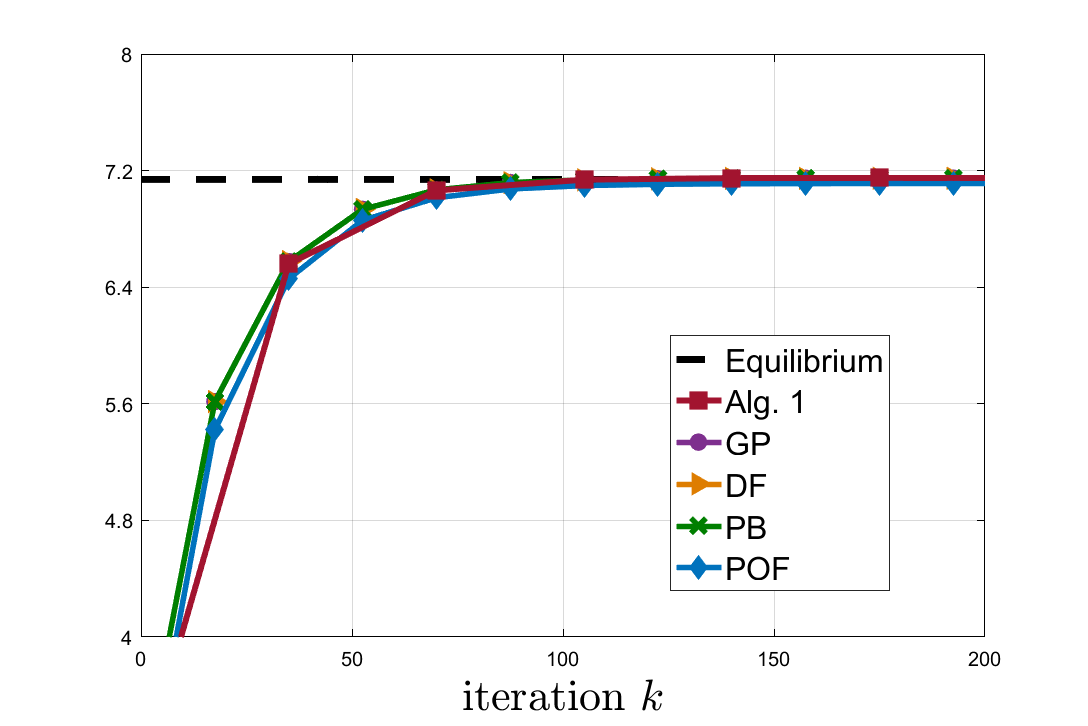}
\end{minipage}
}
\subfigure[$x_{11}(0)=-3$]{
\begin{minipage}[c]{0.32\linewidth}
\centering
\includegraphics[width=6.5cm,height = 4.5cm]{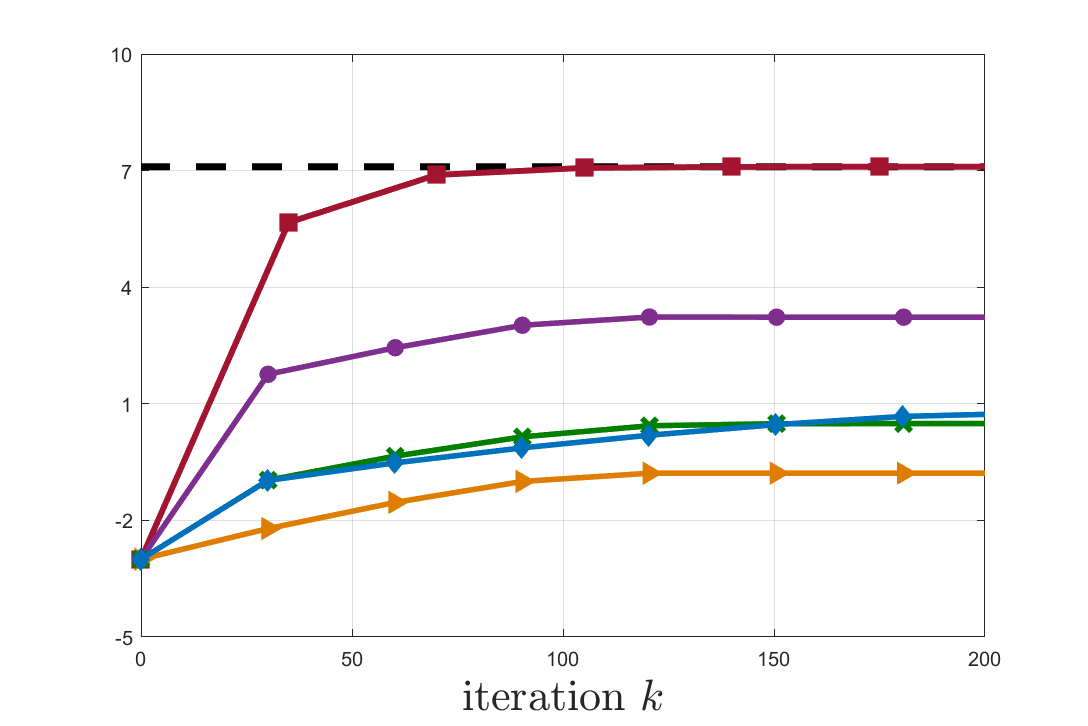}
\end{minipage}
}
\subfigure[$x_{11}(0)=-0.1$]{
\begin{minipage}[c]{0.32\linewidth}
\centering
\includegraphics[width=6.5cm,height = 4.5cm]{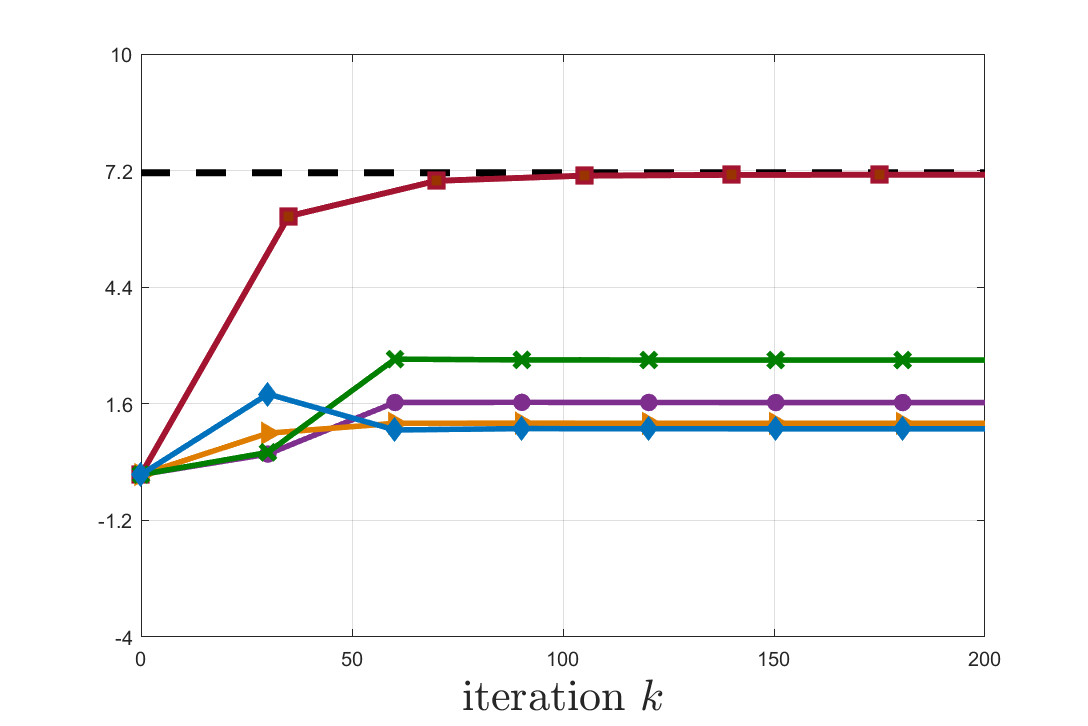}
\end{minipage}
} 
\caption{Comparison of convergence results   with different initial points. Alg. 1 is ours.}
\label{117}
 \end{figure*}
\begin{table*}[t]
	\centering
	\small
	\setlength\tabcolsep{13pt}
	\renewcommand\arraystretch{1.4}
		\caption{MLE of five methods with three initializations in three iterations.}  
		\begin{tabular}{lllllll}
			\hline
			\hline
			\specialrule{0em}{1pt}{1pt}
			\multicolumn{1}{l}{\multirow{2}{1.5cm}{Initialization}}
			&\multicolumn{1}{l}{\multirow{2}{1.8cm}{Iteration }}
			&  \multicolumn{5}{c}{Values of $ \operatorname{MLE} $}
			\\
			\cline{3-7}
			\multicolumn{1}{c}{}
			&
			\multicolumn{1}{c}{}
			 &  {\textbf{Alg. 1 (ours)}}&{proximal \cite{liu2020approximate}}&{PGD \cite{chen2021distributed}}&  {penalty \cite{facchinei2010penalty}}&{SGD \cite{mertikopoulos2019learning}} \\ \hline
			\specialrule{0em}{1pt}{1pt}
			\multirow{3}{1.5cm}{$ x_{11}^{0}=3 $} &\multicolumn{1}{l}{\multirow{1}{1.8cm}{5}}&\multirow{1}{1.5cm}{\textbf{1.2631}} &\multirow{1}{1.5cm}{1.5589}&\multirow{1}{1.5cm}{{1.0819}}&\multirow{1}{1.5cm}{1.9259}&\multirow{1}{1.5cm}{1.6590}\\
			{}& \multicolumn{1}{l}{50}&{\textbf{0.3009}}&  {0.1460} & {{0.1174}}&  {0.1625} & {0.2513}\\
   	{}& \multicolumn{1}{l}{200}&{\underline{\textbf{0.0021}}}&  {0.0343} & {0.0481}&  {0.0567} & {0.0529}\\
			\hline
			\specialrule{0em}{1pt}{1pt}
			\multirow{3}{1.5cm}{$ x_{11}^{0}=-3 $} &\multicolumn{1}{l}{\multirow{1}{1.8cm}{5}}&\multirow{1}{1.5cm}{\textbf{9.4408}} &\multirow{1}{1.5cm}{10.0225}&\multirow{1}{1.5cm}{9.9922}&\multirow{1}{1.5cm}{9.9855}&\multirow{1}{1.5cm}{9.9925}\\
			{}& \multicolumn{1}{l}{50}&{\textbf{3.3574}} & {4.8478}&  {8.8478}&  {7.5678} & {7.9568}\\
			{}& \multicolumn{1}{l}{200}&{\underline{\textbf{0.0379}}} & {3.8488}&  {7.2688}&  {6.5929} & {6.3470}\\
			\hline	
   \specialrule{0em}{1pt}{1pt}
			\multirow{3}{1.7cm}{$ x_{11}^{0}=-0.1 $} &\multicolumn{1}{l}{\multirow{1}{1.8cm}{5}}&\multirow{1}{1.5cm}{\textbf{5.5911}} &\multirow{1}{1.5cm}{7.0914}&\multirow{1}{1.5cm}{6.8472}&\multirow{1}{1.5cm}{7.0160}&\multirow{1}{1.5cm}{5.6742}\\
			{}& \multicolumn{1}{l}{50}&{\textbf{0.5977}}&  {6.6912} & {6.0461}&  {5.1330} & {5.8409}\\
			{}& \multicolumn{1}{l}{200}&{\underline{\textbf{0.0028}}}&  {5.4908} & {5.1005}&  {4.4875} & {6.1451}\\
			\hline	
   \hline	
	\end{tabular}
	\label{tab3}
\end{table*}
  
\subsection{Sensor localization}

In this section, we  apply 
our approach to solving a sensor localization problem.

\subsubsection{Experiment implementation}

\noindent\textbf{Model and data  pre-processing.} We consider a class of non-convex  games in sensor  localization with $N=10$ sensor nodes \cite{jia2013distributed,yang2018df}.
For $i\in\mathcal{I}$,  the position strategy set $\Omega_i$ is equipped with a unit square form 
$\Omega_i=\{x_i\in \mathbb{R}^{2}: x_{\operatorname{min}}\leq x_{il}\leq x_{\operatorname{max}}\}$ for $l=1,2$,  where $x_{\operatorname{min}}=-6$, $ x_{\operatorname{max}}=6$
and the payoff function  defined in (\ref{j1}) for  localization measurement  is accompanied with a deviation of personal estimation $\|D_ix_i-e_{i}\|^{2}$, i.e., 
\begin{equation*}
     J_{i}(x_{i},\boldsymbol{x}_{-i})=\sum_{j\in\mathcal{N}_{i}}( \left\|x_{i}-{x}_{j}\right\|^{2}-d_{i,j })^{2}+\|D_ix_i-e_{i}\|^{2}.
\end{equation*}
Take a ring graph as the senor network $
1\Leftrightarrow2\Leftrightarrow\cdots\Leftrightarrow10\Leftrightarrow1 
$, so that there are two neighbors in  $\mathcal{N}_{i}$ for sensor $i$. 
The distance parameters $d_{i,j}$ are randomly chosen from a compact region $[5, 10]$, and the  coefficient
matrices $D_i\in\mathbb{R}^{2\times2}$ and $e_{i}\in\mathbb{R}^{2}$ are  randomly generated constants. 

{The goal of the sensor localization problem  is to estimate the position of all sensor nodes as accurately as possible. Thus, it is desirable to obtain the optimal location selection $\boldsymbol{x}^{*}$ from a global perspective.  In this view, } we reformulate this problem with a potential  game model. The potential function is given as 
\begin{equation*}
   \Psi(\Lambda(x_i,\boldsymbol{x}_{-i}))=\sum_{i=1}^{N} \sum_{j\in\mathcal{N}_{i}}( \left\|x_{i}-{x}_{j}\right\|^{2}-d_{i,j })^{2}+\|D_ix_i-e_{i}\|^{2}.
\end{equation*}
Then we make a canonical transformation  along the procedure in this paper to handle  non-convexity. As mentioned in (\ref{p11}),  $ \mathscr{E}^{+}$ is a polyhedron here due to a common $\sigma$ after canonical transformation.

\noindent\textbf{Methods.} We employ Algorithm 1 to solve this problem. Assign $\alpha_k=\frac{0.0637}{\sqrt{k}}$ as the step size. Take $\phi_{i}(x_i)={\!\sum\nolimits_{l=1}\limits^{2} (x_{i,l} \!- \!x_{\operatorname{min}}) \log (x_{i,l} \!- \!x_{\operatorname{min}})+\!(x_{\operatorname{max}}\!-\!x_{i,l}) \log (x_{\operatorname{max}}\!-\!x_{i,l})}$ and $\varphi(\sigma)=\frac{1}{2}\|\sigma\|^2_2$. 
For further illustration in this task, 
 we compare Algorithm 1 with 
 several existing methods for solving  multi-player models,  
 including projected gradient descent (PGD) \cite{chen2021distributed},  penalty-based
methods \cite{facchinei2010penalty}, stochastic gradient descent (SGD) \cite{mertikopoulos2019learning}, and gradient-proximal methods  \cite{liu2020approximate}.

\noindent\textbf{Evaluation measures.} Denote $\boldsymbol{x}=\{x_{1},x_{2}\cdots,x_{10}\}$ as  sensor locations computed by Algorithm 1 and $\boldsymbol{x}^{\Diamond}$ as   the true sensor location.
The
performance of this game setting can be evaluated by the \textit{mean localization error}:
\begin{equation*}
  \operatorname{MLE}=\frac{1}{N} \sqrt{\sum_{i=1}^{N}\|x_{i}-x_{i}^{\Diamond}\|^2}.
\end{equation*}


\subsubsection{Experiment results}

As discussed in Section \ref{headings} and \ref{d5}, the set $ \mathscr{E}^{+}$ is nonempty and the existence condition is satisfied. Thus, Algorithm 1 always converges to a global NE. 
In Fig. \ref{fg2}, we show  the trajectories of all sensors’ strategies in Algorithm 1 with respect to one certain dimension. This reveals that all sensors find their appropriate localization on account of the convergence, 
which actually serves as the desired global NE. The location result is shown in Fig. \ref{fg9}, where 
the true sensor locations are denoted by stars, and the computed sensor locations are denoted by circles. As can be seen from Fig. \ref{fg9}, the computed location results match the true locations. This indicates that our approach ensures the accuracy of sensor location and is able to 
find the global NE under this non-convex scenario.

Moreover, we compare Algorithm 1 with several methods mentioned above. We randomly generate   three different initial points as three cases and record the value of  $\operatorname{MLE}$ of several methods under $5$, $50$, 
 $200$ iterations respectively to measure the accuracy of the computed locations.
 All results are listed
in Table \ref{tab3}. In the first case { $x^{0}_{11}=3$},  all methods locate the sensor nodes accurately, and the value of $\operatorname{MLE}$ decreases  with the increase of iterations. {This is because the initial is near the global NE.}
Moreover, in other two cases {$x^{0}_{11}=-3$ and $x^{0}_{11}=-0.1$}, the advantage of Algorithm 1 is displayed. It can be seen from  Table \ref{tab3} that only Algorithm 1 still maintains  $\operatorname{MLE}$ in a tolerance error range, while other methods can not guarantee this. This is because Algorithm 1 is insusceptible wherever the initial point lies. 
In  Fig. \ref{117} (a)-(c), we further check these three cases
in view of a fixed player's decision. 
It follows from Fig. \ref{117} that only our algorithm achieves the target, while others fail with varied initial points. 



\section{Discussions}\label{discuss}
In this section, we give some discussions on the obtained results in this paper, based on the comparisons with related works.

Firstly, we innovatively derived the existence condition of global NE in such a significant class of non-convex multi-player games.  {This game setting  has broad applications in robust training \cite{deng2021local}, sensor localization \cite{jia2013distributed}, and mechanism desgn \cite{ruby2015centralized, qin2022benefits}. Nevertheless, } 
the obtained results cannot be achieved by other current approaches to non-convex game settings so far. With the rapid development and wide applications of adversarial systems and models in machine learning,  non-convex games are playing a more and more important role in learning tasks. As we know, there have been some theoretical studies and algorithm designs for non-convex games, but most of these conclusions were focused on the two-player min-max problems. { According to categories  of players' objectives,  the research status can be roughly divided into  Polyak-{\L}ojasiewicz cases \cite{nouiehed2019solving,fiez2021global},  strongly-concave cases \cite{lin2020gradient,rafique2021weakly}, and  general non-convex non-concave cases \cite{heusel2017gans,daskalakis2018limit}.
Such popular researches owe to the  success of GAN  and its  variants
\cite{goodfellow2014generative,daskalakis2018training}.
 }  { Recently, learning methods in artificial intelligence are developed towards  multi-agent ways, distributed manners, or federated frameworks \cite{yu2019multi,li2019interaction,fan2021fault,qin2022benefits}. This means that adversarial training is  gradually generalized to multiple agents, no longer restricted to two opponents.} There have been initial efforts for non-convex multi-player settings. {
For example, \cite{pang2011nonconvex,hao2020piecewise} proposed a best-response scheme for local NE  seeking, while
 \cite{liu2020approximate} 
designed a  gradient-proximal algorithm to find an approximate solution. } As these works mentioned, finding local NE or alternative approximations is challenging but acceptable enough. Up till now, revealing the existence of global optimums or equilibria in non-convex settings is still an open problem \cite{1177151,8643982,9398583}. {In this view, it is important to investigate the existence condition of global NE in this paper, and we employed conjugate transformation of duality theory and continuous mapping of variational inequality to derive this in Theorem \ref{t2}.}


{Secondly, we designed a continuous ODE to compute the global NE 
and deduced its discrete algorithm with its step-size designs and convergence rates in two typical cases.} 
Recall the existing game-theoretical algorithms for multiple players based on the first-order information  \cite{yi2019operator,chen2021distributed,facchinei2010penalty}. Most of these works depend on convexity assumptions, that is, strongly or strictly convex payoffs of each player or directly monotone pseudo gradients, which are the core to finding NE in game models. However, totally convexity is a luxury in reality. It is inevitable  that the above approaches for convex game models perform unsatisfactorily when confronting non-convex settings, since their terminus merely lies in local NE or some approximations. Therefore, considering so important a class of non-convex in this paper, we are accountable for designing novel algorithms to seek the global NE therein. We realized the goal via the design of continuous conjugate-based ODE. { This breaks the limits of traditional convexity assumptions in studying multi-player game models.} By using Lyapunov stability theory in nonlinear systems, we obtained the convergence  of the designed dynamics in Theorem \ref{l1}  and the exponential rate in Theorem \ref{t4}, which is a pretty good convergence performance for continuous ODE. In addition, the induced discrete scheme also reached good convergence results in Theorem \ref{t6} and Theorem \ref{t9} respectively under generalized monotone cases and potential cases. 

Lastly, we give further discussions on the duality theory and associated techniques in this paper. Actually, the applications of duality theory  in machine learning are not rare \cite{8770111}, and the canonical duality theory utilized in this paper has also been studied before in optimization problems \cite{gao2017canonical,liang2019topology}. Unfortunately, it is not straightforward to transplant this technique from optimization to game models. This is because players' decisions are coupled. 
When players are making decisions, they also need to take into account the changes in other players' decision variables. This phenomenon yields that we should handle all players' decision variables as a unified profile. Therefore, we transformed the original problem into a complementary one by means of duality theory, and assigned a sufficient feasible set for the dual variables. We finally overcame the above bottleneck {by virtue of } variational inequalities, and obtained the existence condition of global NE.
Besides, the convergence proofs of the proposed algorithms to seek global NE are also different from the optimization perspective. 
Unlike optimization, where there is a uniform objective function for all variables, multi-player game models need new measures to analyze the convergence of realization algorithms. 
To this end, we employed the Bregman divergence and some inequalities to  negotiate the obstacles, including the three-point identity, the Fenchel's inequality, and the Jensen's inequality (see more details by Lemmas \ref{d4}-\ref{d9} in the Supplementary Materials).
The adoption of these techniques in the theoretical analysis  may provide a new path in the future study of large-scale multi-player interactions and interference.

\section{Conclusions and Future Directions}\label{conclusion}
We have considered  a typical class of non-convex multi-player games, and discussed how to seek their global NE.  By virtue of canonical duality theory and  VIs, 
we have proposed a conjugate-based ODE for obtaining the solution of a transformed VI problem, which actually induces the global NE of the original non-convex game if the duality relation can  be checked. After providing theoretical convergence guarantees of the ODE, we have derived the discretization, as well as the step-size settings and
the corresponding convergence rates under two typical non-convex conditions.

Our exploration does not cease to advance; future works would be conducted for enhancement and elaboration on the basis of this paper. In terms of the convergence rate, proper accelerated approaches may be combined for promising results; In terms of the multi-player background, players' interaction may rely on a communication network in consideration of privacy and  security, which may suggest the necessity for distributed or decentralized protocol.

\bibliographystyle{IEEEtran}
\bibliography{example_paper.bib}


\begin{IEEEbiography}[{\includegraphics[width=1in,height=1.25in,clip,keepaspectratio]{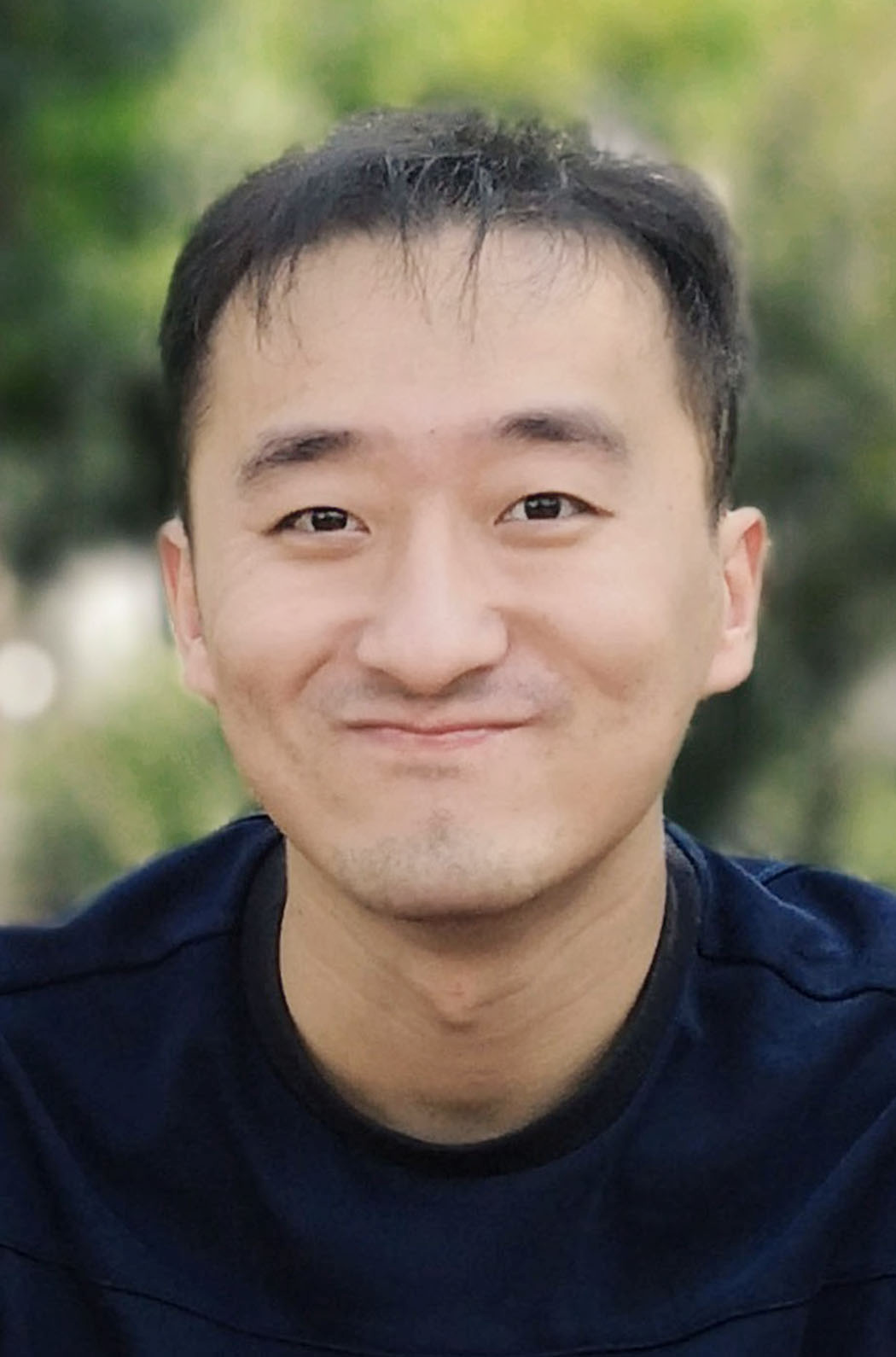}}]{Guanpu Chen}
received his B.Sc. in mathematics from the University of Science and Technology of China, Hefei, China, and Ph.D. in system science from the Chinese Academy of Sciences, Beijing, China. He is currently an algorithm scientist at JD Explore Academy, JD.com Inc. His research interest is in operational research and cybernetics, including multi-agent systems, distributed optimization, machine learning, algorithmic games, network games, and security games.
\end{IEEEbiography}

\begin{IEEEbiography}[{\includegraphics[width=1in,height=1.25in,clip,keepaspectratio]{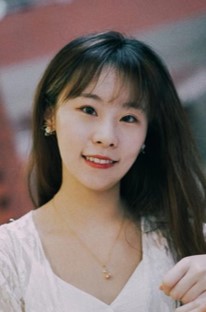}}]{Gehui Xu}
 received the B.Sc. degree in information and computing science from University of Mining and Technology of China, Beijing, China, in 2019. She is currently a Ph.D. candidate in Academy of Mathematics and Systems Science, Chinese Academy of Sciences, Beijing, China. Her research interests include game theory, distributed optimization, and machine learning. 
\end{IEEEbiography}

\begin{IEEEbiography}[{\includegraphics[width=1in,height=1.25in,clip,keepaspectratio]{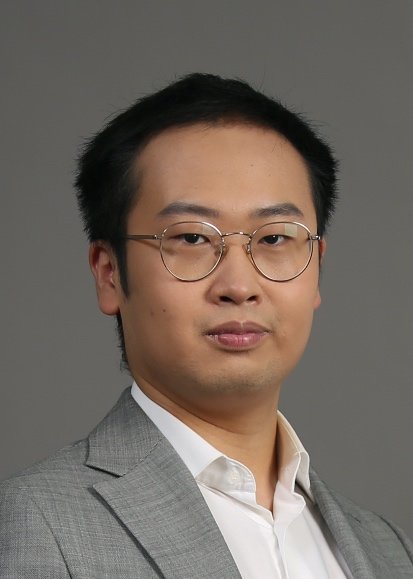}}]{Fengxiang He}
received his BSc in statistics from University of Science and Technology of China, MPhil and PhD in computer science from the University of Sydney. He is currently an algorithm scientist at JD Explore Academy, JD.com Inc, leading R\&D in trustworthy AI. His research interest is in the theory and practice of trustworthy AI, including deep learning theory and explainability, privacy-preserving ML, decentralized learning, and algorithmic game theory. He is the Area Chair of UAI, AISTATS, and ACML.
\end{IEEEbiography}

\begin{IEEEbiography}[{\includegraphics[width=1in,height=1.25in,clip,keepaspectratio]{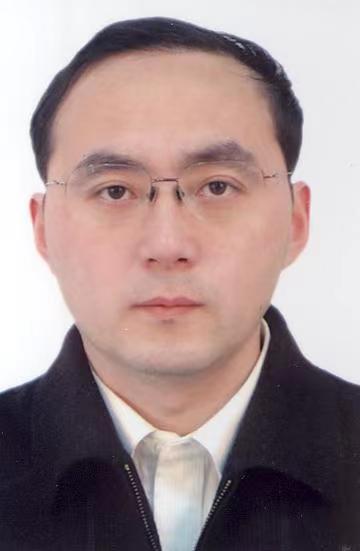}}]{Yiguang Hong} (F'16)
received his B.S. and M.S. degrees from Peking University, China, and  Ph.D. degree from the Chinese Academy of Sciences (CAS), China.
He is currently a professor of the Shanghai Institute of Intelligent Science and Technology, Tongji University, and also a professor of the Academy of Mathematics and Systems Science, CAS. 
His current research interests include nonlinear control, multiagent systems, distributed optimization/game, machine learning, and social networks.
Prof. Hong serves as Editor-in-Chief of Control Theory and Technology. 
He also serves or served as Associate Editors for many journals including the
IEEE Transactions on Automatic Control, IEEE Transactions on Control of Network
Systems, and IEEE Control Systems Magazine.
He is a recipient of the Guan Zhaozhi Award at the Chinese Control Conference,
Young Author Prize of the IFAC World Congress, Young Scientist Award of CAS, the
Youth Award for Science and Technology of China, and the National Natural Science
Prize of China. He is also a Fellow of IEEE.
\end{IEEEbiography}

\begin{IEEEbiography}[{\includegraphics[width=1in,height=1.25in,clip,keepaspectratio]{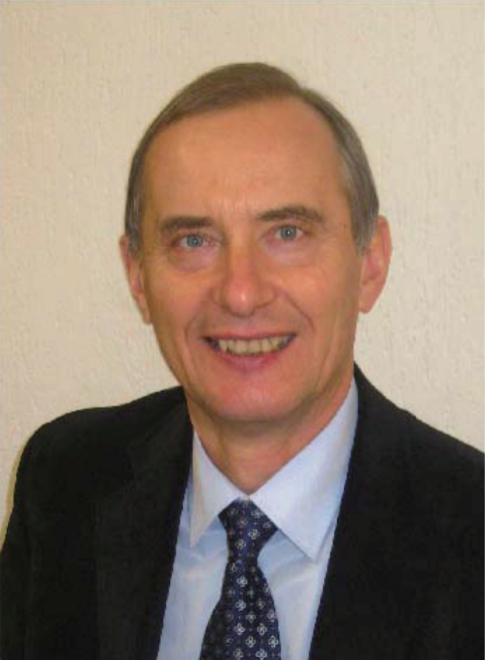}}]{Leszek Rutkowski}
(F'05) received the M.Sc., Ph.D., and D.Sc. degrees from the Wroc{ł}aw University of Technology, Wrocław, Poland, in 1977, 1980, and 1986, respectively, and the Honoris Causa degree from the AGH University of Science and Technology, Kraków, Poland, in 2014. He is with the Systems Research Institute of the Polish Academy of Sciences, Warsaw, Poland, and with the Institute of Computer Science, AGH University of Science and Technology, Krakow, Poland, in both places serving as a professor. He is an Honorary Professor of the Czestochowa University of Technology, Poland, and he also cooperates with the University of Social Sciences in Łódź, Poland. His research interests include machine learning, data stream mining, big data analysis, neural networks, stochastic optimization and control, agent systems, fuzzy systems, image processing, pattern classification, and expert systems. He has published seven monographs and more than 300 technical papers, including more than 40 in various series of IEEE Transactions. He is the president and founder of the Polish Neural Networks Society. He is on the editorial board of several most prestigious international journals. He is a recipient of the IEEE Transactions on Neural Networks  Outstanding Paper Award. He is a Full Member (Academician) of the Polish Academy of Sciences, elected in 2016, and a Member of the Academia Europaea, elected in 2022.

\end{IEEEbiography}

\begin{IEEEbiography}[{\includegraphics[width=1in,height=1.25in,clip,keepaspectratio]{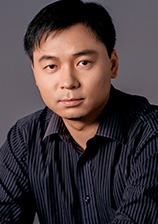}}]{Dacheng Tao}
(F'15) 
is currently Inaugural Director of JD Explore Academy and Senior Vice President of JD.com Inc. He is also Professor of Computer Science and ARC Laureate Fellow at the University of Sydney. He mainly applies statistics and mathematics to artificial intelligence and data science, and his research is detailed in one monograph and over 200 publications in prestigious journals and proceedings at leading conferences. He received the 2015 Australian Scopus-Eureka Prize, the 2018 IEEE ICDM Research Contributions Award, and the 2021 IEEE Computer Society McCluskey Technical Achievement Award. He is Fellow of the Australian Academy of Science, the Royal Society of NSW, AAAS, ACM, and IEEE.
\end{IEEEbiography}

 \clearpage
 \onecolumn
\begin{center}
	\huge 	Supplementary Materials
	\end{center}

\noindent{In the Supplementary Materials, we provide several detailed parts omitted from the main text. The necessary preliminaries include canonical duality theory, variational inequality, Bregman divergence, and some important inequalities. All the proofs of lemmas and theorems are presented in detail. }

\setcounter{equation}{0}
\setcounter{section}{0}
\renewcommand\thesection{S.\arabic{section}}
\renewcommand\theequation{s.\arabic{equation}}
\renewcommand\thesubsection{S.\arabic{section}.\arabic{subsection}}

\setcounter{mythm}{0}   
\renewcommand{\themythm}{S\arabic{mythm}}
\setcounter{lemma}{0}   
\renewcommand{\thelemma}{S\arabic{lemma}}

\section{Canonical duality theory}


We begin the supplementary of this paper with the following  fundamental concepts of  canonical duality theory.
A differentiable function $\Psi: \Theta\rightarrow \mathbb{R}$ is said to be a {canonical function} if  its derivative $\nabla\Psi : \Theta\rightarrow \Theta^{*}$ is a one-to-one mapping. 	
%
Besides, if $\Psi$ is a convex canonical function, its 
conjugate function $\Psi^{*}: \Theta^{*} \rightarrow \mathbb{R} $ can be uniquely defined by the Legendre transformation, that is,
\begin{align*}
	\Psi^{*}\left(\sigma\right)=\left\{\xi^{T} \sigma-\Psi(\xi) \mid \sigma=\nabla \Psi(\xi)\right\}, 
\end{align*}
where 	$\sigma\in \Theta^{*} $
is a canonical dual variable.
On this basis, 	there are corresponding canonical duality relations holding on $\Theta\times\Theta^{*}$: 
\begin{align*}
	\sigma&=\nabla \Psi(\xi),\\
	\Leftrightarrow\quad\quad \;\xi&=\nabla \Psi^{*}\left(\sigma\right),\\ \Leftrightarrow \quad \xi^{T} \sigma&= \Psi(\xi)+\Psi^{*}\left(\sigma\right).
\end{align*} 	
Here, $(\xi,\sigma)$ is called the Legendre canonical duality pair
on $\Theta\times\Theta^{*}$.

\section{Proof of Lemma \ref{t1}}\label{sl1}
In this section, we give the proof of
Lemma \ref{t1}, which investigates the relationship of stationary points between (\ref{ee2}) and  (\ref{f1}). Here we reclaim Lemma \ref{t1}  for convenience.

\noindent\textbf{Lemma 1}
There exists a profile $ \boldsymbol{x}^{\Diamond} $ as a Nash stationary point of (\ref{f1}) 
if $\boldsymbol{\sigma}^{\Diamond}\in \boldsymbol{\Theta}^{*}$ and for $ i\in \mathcal{I}$, 
$ ({x_{i}^{\Diamond}},{\sigma_{i}^{\Diamond}})$
is a stationary point of  complementarity function  $ \Gamma_{i} (x_{i},\sigma_{i},\boldsymbol{x}_{-i}^{\Diamond})$.

\noindent\textbf{Proof.} {For a given strategy profile $\boldsymbol{x}^{\Diamond}$,  if there exists  $\boldsymbol{\sigma}^{\Diamond}\in \boldsymbol{\Theta}^{*}$ such that for all  $ i\in \mathcal{I}$, $ ({x_{i}^{\Diamond}},{\sigma_{i}^{\Diamond}})$
	is a stationary point of  complementarity function $ \Gamma_{i} (x_{i},\sigma_{i},\boldsymbol{x}_{-i}^{\Diamond})$,
	then 
	it satisfies the following first order conditions:}
\begin{subequations}\label{er2}
	\begin{align} 
		&\mathbf{0}_{n} \in  \sigma_{i}^{\Diamond T}\nabla_{x_{i}} \Lambda_{i}(x_{i}^{\Diamond},\boldsymbol{x}_{-i}^{\Diamond})
		+\mathcal{N}_{\Omega_{i}}(x_{i}^{\Diamond}),\\
		&\mathbf{0}_{q_i} \in -\Lambda_{i}(x_{i}^{\Diamond},\boldsymbol{x}_{-i}^{\Diamond})+\nabla \Psi_{i}^{*}(\sigma_{i}^{\Diamond})+ \mathcal{N}_{\Theta^*_{i}}(\sigma_{i}^{\Diamond}),
	\end{align}
\end{subequations}
where $\mathcal{N}_{\Omega_{i}}(x_{i}^{\Diamond})$ is the normal cone at point $x_{i}^{\Diamond}$ on set $\Omega_{i}$, with a similar definition for the normal cone $\mathcal{N}_{\Theta^*_{i}}(\sigma_{i}^{\Diamond})$.
Following the definition of the convex canonical function $\Psi_i$, we can learn that its derivative $\nabla\Psi_i : \Theta_i\rightarrow \Theta^{*}_i$ is a one-to-one mapping from $\Theta_i$ to its range $\Theta^{*}_i$. Thus,  for given $\xi_{i}^{\Diamond} \in \Theta_i$ 
with $\xi_{i}^{\Diamond}=\Lambda_{i}(x_{i}^{\Diamond},\boldsymbol{x}_{-i}^{\Diamond})$, there exists a unique $\sigma_{i}^{\Diamond} \in \Theta^{*}_i$ such that 
\begin{align*}
	\sigma_{i}^{\Diamond}=\nabla \Psi_{i}(\xi_{i}^{\Diamond}).
\end{align*}
Meanwhile, given this Legendre canonical duality pair $(\xi_{i}^{\Diamond}, \sigma_{i}^{\Diamond})$ on $\Theta_i \times \Theta_i^*$,
the duality relation  holds that
\begin{align*}
	\sigma_{i}^{\Diamond}=\nabla \Psi_{i}(\xi_{i}^{\Diamond})\; \Longleftrightarrow \;\xi_{i}^{\Diamond}=\nabla \Psi_{i}^{*}(\sigma_{i}^{\Diamond}). 
\end{align*}
With all this in mind, 
(\ref{er2}b) can be transformed into
\begin{equation}\label{ss}
	\Lambda_{i}(x_{i}^{\Diamond},\boldsymbol{x}_{-i}^{\Diamond})=\nabla \Psi_{i}^{*}(\sigma_{i}^{\Diamond}).
\end{equation}
Using the the duality relation again, \eqref{ss} is equivalent to
\begin{equation}\label{s55}
	\sigma_{i}^{\Diamond}=\nabla \Psi_{i}(
	\Lambda_{i}(x_{i}^{\Diamond},\boldsymbol{x}_{-i}^{\Diamond})).
\end{equation}
By substituting \eqref{s55} into (\ref{er2}a), we have 
\begin{equation}\label{sss}
	\mathbf{0}_{n} \in  \nabla \Psi_{i}(\Lambda_{i}(x_{i}^{\Diamond},\boldsymbol{x}_{-i}^{\Diamond}))^{T} \nabla_{x_{i}}\Lambda_{i}(x_{i}^{\Diamond},\boldsymbol{x}_{-i}^{\Diamond})+\mathcal{N}_{\Omega_{i}}(x_{i}^{\Diamond}).
\end{equation}
According to the chain rule, 
\begin{equation*}
	\nabla \Psi_{i}(\Lambda_{i}(x_{i}^{\Diamond},\boldsymbol{x}_{-i}^{\Diamond}))^{T} \nabla_{x_{i}}\Lambda_{i}(x_{i}^{\Diamond},\boldsymbol{x}_{-i}^{\Diamond})=\nabla_{x_{i}} J_{i}(x_{i}^{\Diamond}, \boldsymbol{x}_{-i}^{\Diamond}).
\end{equation*}
Therefore, \eqref{sss} is equivalent to
\begin{equation}\label{ft4}
	\mathbf{0}_{n} \in \nabla_{x_{i}} J_{i}(x_{i}^{\Diamond}, \boldsymbol{x}_{-i}^{\Diamond})+\mathcal{N}_{\Omega_{i}}({x_{i}^{\Diamond}}). 
\end{equation}
Since (\ref{ft4}) is true for any player $i\in\mathcal{I}$, the profile $\boldsymbol{x}^{\Diamond}$ satisfies the  Nash stationary condition, which completes the proof.
\hfill $\square$

\section{Variational inequality}
{In this section, we introduce some concepts and properties related to the variational inequality (VI).}
Recall  the following notations according to problem (\ref{ee2}) 
\begin{equation*}
	\boldsymbol{z}=\operatorname{col}\{\boldsymbol{x},\boldsymbol{\sigma}\},\quad \boldsymbol{\Xi}=\boldsymbol{\Omega} \times \mathscr{E}^{+} \subset \mathbb{R}^{nN+q}.
\end{equation*}
For the conjugate gradient of canonical function $\Psi_i$ for $i\in\mathcal{I}$, denote  
\begin{equation*}
	\nabla \Psi^{*}\left(\boldsymbol{\sigma}\right)= \operatorname{col}\{\nabla \Psi_{i}^{*}\left(\sigma_{i}\right) \}_{i=1}^{N}.
\end{equation*} 
Also, denote the profile of all $\Lambda_i$ by
\begin{equation*}
	\Lambda\left(\boldsymbol{x}\right)=\operatorname{col}\left\{
	\Lambda_{i}\left(x_{i},\boldsymbol{x}_{-i}\right)
	\right\}_{i=1}^{N},
\end{equation*} 
and {the augmented partial derivative profile as }
\begin{equation*}
	G(\boldsymbol{x},\boldsymbol{\sigma}) =	\operatorname{col}\left\{
	\sigma_{i}^{T}\nabla_{x_{i}} \Lambda_{i}\left(x_{i},\boldsymbol{x}_{-i}\right)\right\}_{i=1}^{N}.
\end{equation*} 
In this way,
the pseudo-gradient of (\ref{ee2}) can be rewritten as
\begin{equation}\label{compact-pseudo}
	F(\boldsymbol{z})
	\triangleq\left[\begin{array}{c}
		G(\boldsymbol{x},\boldsymbol{\sigma}) \\
		-\Lambda\left(\boldsymbol{x}\right)+\nabla \Psi^{*}\left(\boldsymbol{\sigma}\right)
	\end{array}\right]=\left[\begin{array}{c}
		\operatorname{col}\{
		\sum\nolimits_{k=1}\nolimits^{q_i}[\sigma_{i}]_{k} \nabla_{x_{i}} \Lambda_{i, k}(x_{i},\boldsymbol{x}_{-i}) \}_{i=1}^{N}  \\
		\operatorname{col}\left\{
		-\Lambda_{i}\left(x_{i},\boldsymbol{x}_{-i}\right)+\nabla \Psi_{i}^{*}\left(\sigma_{i}\right)
		\right\}_{i=1}^{N}
	\end{array}\right].
\end{equation}
To proceed,  the accordingly introduced variational inequality (VI)  problem $\operatorname{VI}(\boldsymbol{\Xi},F)$ 
is defined as
\begin{equation}\label{e2er}
	\text{to find} \quad  \boldsymbol{z}\in  \boldsymbol{\Xi} \quad  \text{subject to} \quad  (\boldsymbol{z}^{\prime}-\boldsymbol{z})^{T}F(\boldsymbol{z}) \geq 0, \quad \forall \boldsymbol{z}^{\prime}\in \boldsymbol{\Xi}.    
\end{equation}
The solution of this VI problem is denoted by $\operatorname{SOL}(\boldsymbol{\Xi},F)$. 
Moreover,  since $F(\boldsymbol{z})$  is a continuous mapping and $\boldsymbol{\Xi}$ is a closed set,
we have the following result referring to \cite[Page 2-3]{facchinei2003finite}.
\begin{lemma}\label{v111} The solution set 
	$\operatorname{SOL}(\boldsymbol{\Xi},F)$ of $\operatorname{VI}(\boldsymbol{\Xi},F)$ in \eqref{e2er} is closed. Moreover,  
	any  profile  $\boldsymbol{z} \in \operatorname{SOL}(\boldsymbol{\Xi},F)$ if and only if 
	\begin{equation*}\label{e39}
		\mathbf{0}_{nN+q} \in  F(\boldsymbol{z})
		+\mathcal{N}_{\boldsymbol{\Xi}}(\boldsymbol{z}).
	\end{equation*}
\end{lemma}

\section{Proof of Theorem \ref{t2}}\label{sd}
In this section, we give the  proof of Theorem \ref{t2} by the preparation aforementioned about the canonical duality theory and the property in VI problems. 
We reproduce Theorem \ref{t2} here for convenience.

\noindent\textbf{Theorem 1}
		There exists $\boldsymbol{x}^{\Diamond}$ as the global NE		of the non-convex multi-player game (\ref{f1}) if 
$(\boldsymbol{x}^{\Diamond},\boldsymbol{\sigma}^{\Diamond})$ is a solution to $\operatorname{VI}(\boldsymbol{\Xi},F)$ with  $\sigma_{i}^{\Diamond}=\nabla \Psi_{i}\left(\xi_{i}\right) \mid_{\xi_{i}=\Lambda_{i}\left(x_{i}^{\Diamond}, \boldsymbol{x}_{-i}^{\Diamond}\right)}$ for $i\in\mathcal{I}$.

\noindent\textbf{Proof.}
Under Assumption 1, if 
there exists $\boldsymbol{\sigma}^{\Diamond}\in \mathscr{E}^{+}$ such that $\boldsymbol{z}^{\Diamond}=\operatorname{col}\{\boldsymbol{x}^{\Diamond},\boldsymbol{\sigma}^{\Diamond}\}$ is a solution to $\operatorname{VI}(\boldsymbol{\Xi},F)$, 
then it follows from Lemma \ref{v111} that 
\begin{equation}\label{e37}
	\mathbf{0}_{nN+q} \in  F(\boldsymbol{z}^{\Diamond})
	+\mathcal{N}_{\boldsymbol{\Xi}}(\boldsymbol{z}^{\Diamond}),
\end{equation}
which implies that for $i\in\mathcal{I}$,
\begin{align*} 
	&\mathbf{0}_{n} \in  \sigma_{i}^{\Diamond T}\nabla_{x_{i}} \Lambda_{i}(x_{i}^{\Diamond},\boldsymbol{x}_{-i}^{\Diamond})
	+\mathcal{N}_{\Omega_{i}}(x_{i}^{\Diamond}),\\
	&\mathbf{0}_{q_i} \in -\Lambda_{i}(x_{i}^{\Diamond},\boldsymbol{x}_{-i}^{\Diamond})+\nabla \Psi_{i}^{*}(\sigma_{i}^{\Diamond})+ \mathcal{N}_{\mathscr{E}_{i}^{+}}(\sigma_{i}^{\Diamond}),
\end{align*}
or  equivalently described as
%
%
%
\begin{equation}\label{e346578}
	\begin{aligned}
		( \sigma_{i}^{\Diamond T}\nabla_{x_{i}} \Lambda_{i}(x_{i}^{\Diamond},\boldsymbol{x}_{-i}^{\Diamond}))^{T} (x_{i}-x_{i}^{\Diamond})&\geq0, \quad \forall x_{i}\in \Omega_i,\\
		(-\Lambda_{i}(x_{i}^{\Diamond},\boldsymbol{x}_{-i}^{\Diamond})+\nabla \Psi_{i}^{*}(\sigma_{i}^{\Diamond}))^{T} (\sigma_{i}-\sigma_{i}^{\Diamond})&\geq0, \quad\forall \sigma_{i}\in \mathscr{E}_{i}^{+}. 
	\end{aligned}
\end{equation}
Moreover, if $\sigma_{i}^{\Diamond}=\nabla \Psi_{i}\left(\xi_{i}\right) \mid_{\xi_{i}=\Lambda_{i}\left(x_{i}^{\Diamond}, \boldsymbol{x}_{-i}^{\Diamond}\right)}$ for $i\in\mathcal{I}$, then 
the canonical duality relation hold on $\Theta_i\times\mathscr{E}_{ i}^{+}$ for $i\in\mathcal{I}$. This indicates that  the solution to $\operatorname{VI}(\boldsymbol{\Xi},F)$ is a stationary point profile of (\ref{ee2}) on $\Theta_i\times\Theta_i^{*}$.

Thus, 
similar to the chain rules employed in Lemma \ref{t1}, we can further derive that
\begin{align*}
	( \nabla_{x_{i}} J_{i}(x_{i}^{\Diamond}, \boldsymbol{x}_{-i}^{\Diamond}))^{T}(x_{i}-x_{i}^{\Diamond})\geq0,\quad \forall x_{i}\in \Omega_i.
\end{align*}
Moreover, when  $ \sigma_{i}\in\mathscr{E}_{ i}^{+}  $,  the Hessian matrix satisfies
\begin{align*}
	\nabla_{x_i}^{2} \Gamma_{i}(x_{i},\sigma_{i},\boldsymbol{x}_{-i})=\sum\nolimits_{k=1}\nolimits^{q_i}[\sigma_{i}]_{k} \nabla^{2}_{x_{i}} \Lambda_{i, k}(x_{i},\boldsymbol{x}_{-i}) \succeq \kappa_{x}\boldsymbol{I}_n,
\end{align*} which indicates 
the convexity of $\Gamma_{i}(x_{i},\sigma_{i},\boldsymbol{x}_{-i})$ with respect to $x_{i}$. 
Besides, due to the  convexity of $\Psi_i$, its Legendre conjugate $\Psi_i^*$ is also convex \cite{yang2000canonical}.  Therefore, the total complementary function  $\Gamma_{i}(x_{i},\sigma_{i},\boldsymbol{x}_{-i})$ is  concave in canonical dual  variable $\sigma_{i}$.

In this light, we can obtain the  globally optimality of $(\boldsymbol{x}^{\Diamond}, \boldsymbol{\sigma}^{\Diamond})$ on $\boldsymbol{\Omega}\times \mathscr{E}^{+} $, i.e.,
for $i\in\mathcal{I}$,
\begin{equation*}
	\Gamma_{i}(x_{i}^{\Diamond},  \sigma_i,\boldsymbol{x}_{-i}^{\Diamond})\leq\Gamma_{i}(x_{i}^{\Diamond},  \sigma_i^{\Diamond},\boldsymbol{x}_{-i}^{\Diamond}) \leq \Gamma_{i}(x_{i},  \sigma_i^{\Diamond},\boldsymbol{x}_{-i}^{\Diamond}),\quad \forall x_{i}\in \Omega_i,\; \sigma_{i}\in \mathscr{E}_{i}^{+}. 
\end{equation*}
The inequality relation above tells that 
\begin{equation*}
	J_{i}(x_{i}^{\Diamond}, \boldsymbol{x}_{-i}^{\Diamond})\leq J_{i}(x_{i}, \boldsymbol{x}_{-i}^{\Diamond}), \quad \forall x_{i}\in \Omega_i,\quad \forall i\in\mathcal{I}.
\end{equation*}
This confirms that $\boldsymbol{x}^{\Diamond}$ is the global NE of (\ref{f1}), which completes the proof. \hfill $\square$

\section{Convergence analysis of the conjugate-based ODE}\label{aee}

{In this section, we aim to provide proofs for Lemma \ref{l2}, Theorems \ref{l1} and \ref{t4}.}
We first provide some  preliminaries that are necessary in the convergence analysis of ODE (\ref{e21}), which are also widely-accepted concepts in convex analysis.
Take 
$h(z) : \Xi\rightarrow  \mathbb{R}$ as a differentiable    $\omega$-strongly convex  function   on a closed convex set
$\Xi$, which satisfies
\begin{equation*}
	h\left(\theta z+(1-\omega) z^{\prime}\right) \leq \theta h(z)+(1-\theta) h\left(z^{\prime}\right)-\frac{ \omega}{2}\, \theta(1-\theta)\left\|z^{\prime}-z\right\|^{2},\quad\forall z,z^{\prime}\in  \Xi,\,  \theta\in[0,1].
\end{equation*}
Additionally, 
$h$ admits a Lipschitz continuous gradient if there exists a constant $L> 0 $, such that 
\begin{equation*}
	\|\nabla h(z^{\prime})-\nabla h(z)\| \leq L\|z-z^{\prime}\|, \quad \forall z, z^{\prime} \in \Xi,
\end{equation*}
which  is equivalent to
\begin{equation*}
	h\left(z^{\prime}\right)-h(z) \leq\left(z^{\prime}-z\right)^{\mathrm{T}} \nabla h(z)+\frac{L}{2}\left\|z-z^{\prime}\right\|^{2}, \quad \forall z, z^{\prime} \in \Xi.
\end{equation*}
On the other hand, following the duality theory \cite{nemirovskij1983problem},  the  conjugate function of $h$ defined on the dual space $\Xi^{*}$  is
\begin{equation*}
	h^{*}(s)=\operatorname{sup}_{z \in \Xi}\left\{z^{T} s-h(z)\right\},
\end{equation*}   
where  $s\in\Xi^{*} $ serves as a dual variable. Moreover, consider $h$ as a  differentiable and strongly convex function on a closed convex set $\Xi$.  Then according to  \cite{diakonikolas2019approximate}, 
$h^{*}$ is  also convex and differentiable on $\Xi^{*}$, and satisfies
\begin{equation*}
	h^{*}(s)=\operatorname{min}_{z \in \Xi}\left\{-z^{T} s+h(z)\right\}.
\end{equation*}   
Moreover,  the conjugate gradient $\nabla h^{*}(s)$ who maps  $\Xi^{*}$ to $\Xi$ satisfies 
\begin{equation*}
	\nabla h^{*}\left(s\right) = \operatorname{argmin}_{z \in \Xi}\left\{-z^{T} s+h(z)\right\}.
\end{equation*}
With these preliminaries at hand, we  investigate the convergence of  ODE (\ref{e21}).
For simplicity, let us denote the following compact forms  associated with the gradients therein
\begin{equation*}
	\begin{array}{ll}
		\nabla\boldsymbol{\phi}(\boldsymbol{x})\triangleq \operatorname{col}\left\{\nabla \phi_{i}\left(x_{i}\right)\right\}_{i=1}^{N}, &
		\nabla \boldsymbol{\varphi}(\boldsymbol{\sigma})=\operatorname{col}\{\nabla \varphi_i(\sigma_i) \}_{i=1}^{N};
	\end{array}
\end{equation*}
\begin{equation*}
	\begin{array}{ll}
		\nabla \boldsymbol{\phi}^{*}(\boldsymbol{y})=\operatorname{col}\{{\nabla\phi_{i}^*}\left(y_{i}\right)\}_{i=1}^{N}, & \nabla \boldsymbol{\varphi}^{*}(\boldsymbol{\nu})=\operatorname{col}\{\nabla\varphi_{i}^*\left(\nu_i)\right\}_{i=1}^{N}.
	\end{array}
\end{equation*}
Hence, together with the compact forms  $G(\boldsymbol{x},\boldsymbol{\sigma}) $,  $\Lambda\left(\boldsymbol{x}\right)$, and $\nabla \Psi^{*}\left(\boldsymbol{\sigma}\right)$   defined in (\ref{compact-pseudo}), 
ODE (\ref{e21}) can be compactly presented  by
\begin{equation}\label{e1}
	\left\{\begin{array}{l}\vspace{0.15cm}
		\dot{\boldsymbol{y}}=-G(\boldsymbol{x},\boldsymbol{\sigma}) +\nabla\boldsymbol{\phi}(\boldsymbol{x})-\boldsymbol{y},\\\vspace{0.15cm}
		\dot{\boldsymbol{\nu}}={\Lambda}\left(\boldsymbol{x}\right)-\nabla \Psi^{*}\left(\boldsymbol{\sigma}\right)+\nabla \boldsymbol{\varphi}(\boldsymbol{\sigma})-\boldsymbol{\nu},\\\vspace{0.15cm}
		\boldsymbol{x}=\nabla \boldsymbol{\phi}^{*}(\boldsymbol{y}),\\\vspace{0.15cm}
		\boldsymbol{\sigma}=\nabla \boldsymbol{\varphi}^{*}\left(\boldsymbol{\nu}\right).
	\end{array}\right.
\end{equation}
On this basis, 
we first show a  relationship between the equilibrium in ODE (\ref{e1}) (or ODE (\ref{e21})) and the global NE of game (\ref{f1}).   Rewrite Lemma \ref{l1} here for conveniency.

\noindent\textbf{Lemma 2}
Suppose that  
$(\boldsymbol{y}^{\Diamond},\boldsymbol{\nu}^{\Diamond},\boldsymbol{x}^{\Diamond},\boldsymbol{\sigma}^{\Diamond})$ is an equilibrium point of ODE (\ref{e21}). If $\sigma_{i}^{\Diamond}=\nabla \Psi_{i}\left(\xi_{i}\right) \mid_{\xi_{i}=\Lambda_{i}\left(x_{i}^{\Diamond}, \boldsymbol{x}_{-i}^{\Diamond}\right)}$ for $i\in\mathcal{I}$,
then $ \boldsymbol{x}^{\Diamond} $ is {the} global NE of  (\ref{f1}). 


\noindent\textbf{Proof.}  
If $(\boldsymbol{y}^{\Diamond},\boldsymbol{x}^{\Diamond},\boldsymbol{\nu}^{\Diamond},\boldsymbol{\sigma}^{\Diamond})$ is an equilibrium point of ODE (\ref{e1}),
we have
\begin{subequations}\label{e8}
	\begin{align}
		&\boldsymbol{0}_{nN} =-G(\boldsymbol{x}^{\Diamond},\boldsymbol{\sigma}^{\Diamond}) +\nabla\boldsymbol{\phi}(\boldsymbol{x}^{\Diamond})-\boldsymbol{y}^{\Diamond},\\
		&\boldsymbol{0}_{q}=\Lambda\left(\boldsymbol{x}^{\Diamond}\right)-\nabla \Psi^{*}\left(\boldsymbol{\sigma}^{\Diamond}\right)+\nabla \boldsymbol{\varphi}(\boldsymbol{\sigma}^{\Diamond})-\boldsymbol{\nu}^{\Diamond},\\
		&\boldsymbol{x}^{\Diamond}=\nabla \boldsymbol{\phi}^{*}(\boldsymbol{y}^{\Diamond}),\\	
		&\boldsymbol{\sigma}^{\Diamond}=\nabla \boldsymbol{\varphi}^{*}(\boldsymbol{\nu}^{\Diamond}).
	\end{align}			
\end{subequations}
It follows from 
$\boldsymbol{y}^{\Diamond}=-G(\boldsymbol{x}^{\Diamond},\boldsymbol{\sigma}^{\Diamond}) +\nabla\boldsymbol{\phi}(\boldsymbol{x}^{\Diamond})$  that  (\ref{e8}c) becomes
\begin{equation}\label{ert}
	\boldsymbol{x}^{\Diamond}=\nabla \boldsymbol{\phi}^{*}(-G(\boldsymbol{x}^{\Diamond},\boldsymbol{\sigma}^{\Diamond}) +\nabla\boldsymbol{\phi}(\boldsymbol{x}^{\Diamond}) ).
\end{equation}
For $i\in\mathcal{I}$, (\ref{ert}) is equivalent to
\begin{equation*}
	x_{i}=\nabla \phi_{i}^{*}(- \sigma_{i}^{T}\nabla_{x_{i}} \Lambda_{i}\left(x_{i},\boldsymbol{x}_{-i}\right)+\nabla \phi_i(x_i)).
\end{equation*}
Moreover,  by recalling
\begin{equation*}
	\nabla \phi_i^{*}(y_i)= \operatorname{argmin}_{x_i \in \Omega_i}\left\{-x_i^{T} y_i+\phi_i(x_i)\right\},\quad 
\end{equation*}
and taking $y_i$ as $- \sigma_{i}^{T}\nabla_{x_{i}} \Lambda_{i}\left(x_{i},\boldsymbol{x}_{-i}\right)+\nabla \phi_i(x_i)$, {we  obtain the associated first order condition, expressed as  the following compact form}
\begin{equation}\label{fe3}
	\mathbf{0}_{nN} \in  G(\boldsymbol{x}^{\Diamond},\boldsymbol{\sigma}^{\Diamond})
	+\mathcal{N}_{\boldsymbol{\Omega}}(\boldsymbol{x}^{\Diamond}).
\end{equation}
Similarly, it follows from (\ref{e8}b) and  (\ref{e8}d)  that
\begin{equation*}
	%
	\boldsymbol{\sigma}^{\Diamond}=\nabla \boldsymbol{\varphi}^{*}(\Lambda\left(\boldsymbol{x}^{\Diamond}\right)-\nabla \Psi^{*}\left(\boldsymbol{\sigma}^{\Diamond}\right)+\nabla \boldsymbol{\varphi}(\boldsymbol{\sigma}^{\Diamond})),
\end{equation*}
which yields
\begin{equation}\label{fe2}
	\boldsymbol{0}_{q}\in-\Lambda\left(\boldsymbol{x}^{\Diamond}\right)+\nabla \Psi^{*}\left(\boldsymbol{\sigma}^{\Diamond}\right)+\mathcal{N}_{\mathscr{E}^{+}}(\boldsymbol{\sigma}^{\Diamond}).
\end{equation}
Thus, combining (\ref{fe3}) and (\ref{fe2}), it follows from Lemma \ref{v111}
that $\boldsymbol{z}^{\Diamond}=\operatorname{col}\{\boldsymbol{x}^{\Diamond},\boldsymbol{\sigma}^{\Diamond}\}$ is a solution to   $\operatorname{VI}(\boldsymbol{\Xi},F)$.
Moreover, due to
Theorem \ref{t2},   the solution of  $\operatorname{VI}(\boldsymbol{\Xi},F)$ derives a global NE of game (\ref{f1}) if $\sigma_{i}^{\Diamond}=\nabla \Psi_{i}\left(\xi_{i}\right) \mid_{\xi_{i}=\Lambda_{i}\left(x_{i}^{\Diamond}, \boldsymbol{x}_{-i}^{\Diamond}\right)}$ for $i\in\mathcal{I}$, which completes the proof.
\hfill $\square$




%
%
%

Now, we are  in a position to prove   the convergence  of conjugate-based ODE (\ref{e21}). Reclaim Theorem \ref{l1} here for convenience.

\noindent\textbf{Theorem 2}
	If $\mathscr{E}_{ i}^{+}$ is nonempty for $i\in\mathcal{I}$,
then ODE (\ref{e21}) is bounded and convergent.
Moreover, if 
the convergent point $(\boldsymbol{y}^{\Diamond},\boldsymbol{\nu}^{\Diamond},\boldsymbol{x}^{\Diamond},\boldsymbol{\sigma}^{\Diamond})$  satisfies 
$\sigma_{i}^{\Diamond}=\nabla \Psi_{i}\left(\xi_{i}\right) \mid_{\xi_{i}=\Lambda_{i}\left(x_{i}^{\Diamond}, \boldsymbol{x}_{-i}^{\Diamond}\right)}$ for $i\in\mathcal{I}$, 
then $\boldsymbol{x}^{\Diamond}$ is the 
global NE of  (\ref{f1}).

\noindent\textbf{Proof.}  
(i) We first prove that the trajectory $( \boldsymbol{y}(t)$, $\boldsymbol{x}(t)$,  $\boldsymbol{\nu}(t) $, $\boldsymbol{\sigma}(t) )$ is bounded along ODE (\ref{e21}).  
Construct a Lyapunov
candidate function as
\begin{equation}\label{fe22}
	\begin{aligned}
		V_{1}=& \sum_{i=1}^{N} D_{\phi_{i}^{*}}(y_{i},y_{i}^{\Diamond})+D_{\varphi_{i}^{*}}(\nu_{i},\nu_{i}^{\Diamond}),
	\end{aligned}
\end{equation}
where  Bregman divergences therein are expressed detailedly as
\begin{equation*}
	D_{\phi_{i}^{*}}(y_{i},y_{i}^{\Diamond})=\phi_{i}^{*}(y_{i})-\phi_{i}^{*}(y_{i}^{\Diamond})-\nabla \phi_{i}^{*}(y_{i}^{\Diamond})^{T}(y_{i}-y_{i}^{\Diamond}),
\end{equation*} 
\begin{equation*} D_{\varphi_{i}^{*}}(\nu_{i},\nu_{i}^{\Diamond})=\varphi_{i}^{*}(\nu_{i})-\varphi_{i}^{*}(\nu_{i}^{\Diamond})-\nabla \varphi_{i}^{*}(\nu_{i}^{\Diamond})^{T}(\nu_{i}-\nu_{i}^{\Diamond}).
\end{equation*} 
Consider the term $ D_{\phi_{i}^{*}}(y_{i},y_{i}^{\Diamond})$  for  $i\in\mathcal{I}$.
Since $ x_{i}=\nabla \phi_{i}^{*}\left(y_{i}\right)$ and  $ x_{i}^{\Diamond}=\nabla \phi_{i}^{*}\left(y_{i}^{\Diamond}\right)$,  it follows from the expression of $\nabla \phi_{i}^{*}$  in (\ref{c1}) that	
\begin{equation}\label{b1}
	\begin{aligned}
		\phi_{i}^{*}(y_{i}) &=x_{i}^{T} y_{i}-\phi_{i}(x_{i}), \quad
		\phi_{i}^{*}(y_{i}^{\Diamond}) =x_{i}^{\Diamond T} y_{i}^{\Diamond}-\phi_{i}(x_{i}^{\Diamond}).
	\end{aligned}
\end{equation}
Thus, by  (\ref{b1}), we get
\begin{equation*}
	\begin{aligned}
		D_{\phi_{i}^{*}}(y_{i},y_{i}^{\Diamond}) &=\phi_{i}^{*}(y_{i})-\phi_{i}^{*}(y_{i}^{\Diamond})-\nabla \phi_{i}^{*}(y_{i}^{\Diamond})^{T}(y_{i}-y_{i}^{\Diamond})\\ 
		&=\phi_{i}(x_{i}^{\Diamond})-\phi_{i}(x_{i})-(x_{i}^{\Diamond}-x_{i})^{T} y_{i}\\
		&=\phi_{i}(x_{i}^{\Diamond})-\phi_{i}(x_{i})-(x_{i}^{\Diamond}-x_{i})^{T} \nabla \phi(x_{i})+(x_{i}^{\Diamond}-x_{i})^{T} \nabla \phi(x_{i})-(x_{i}^{\Diamond}-x_{i})^{T} y_{i}.
	\end{aligned}
\end{equation*} 
Since $\phi_{i}$ is $\mu_{x}$-strongly convex on $\Omega_i$, 
we can further derive
\begin{equation*}
	\begin{aligned}
		D_{\phi_{i}^{*}}(y_{i},y_{i}^{\Diamond}) 
		&\geq \frac{\mu_{x}}{2}\left\|{x}_i-x_i^{\Diamond}\right\|^{2} + (x_{i}^{\Diamond}-x_{i})^{T} (\nabla \phi(x_{i})-y_{i}),
	\end{aligned}
\end{equation*} 
which yields
\begin{align}\label{dfrd}
	\sum_{i=1}^{N} D_{\phi_{i}^{*}}(y_{i},y_{i}^{\Diamond}) \geq \frac{\mu_{x}}{2}\left\|\boldsymbol{x}-\boldsymbol{x}^{\Diamond}\right\|^{2}+\sum_{i=1}^{N}(x_{i}^{\Diamond}-x_{i})^{T}\left(\nabla \phi_{i}\left(x_{i}\right)-y_{i}\right).
\end{align}
In fact, recall 
$\nabla \phi_{i}^{*}\left(y_{i}\right)=\operatorname{argmin}_{x \in \Omega_{i}}\left\{-x^{T} y_{i}+\phi_{i}(x)\right\}$. Due to the  optimality of $\nabla \phi_{i}^{*}\left(y_{i}\right)$ and the convexity of $\phi_{i}$, we have
\begin{equation}\label{b3}
	\begin{aligned}
		&(\nabla \phi_{i}^{*}(y_{i}))^{T}\left(\nabla \phi_{i}\left(\nabla \phi_{i}^{*}\left(y_{i}\right)\right)-y_{i}\right)\leq (\nabla \phi_{i}^{*}(y_{i}^{\Diamond}))^{T} \left(\nabla \phi_{i}\left(\nabla \phi_{i}^{*}\left(y_{i}\right)\right)-y_{i}\right).
	\end{aligned}
\end{equation}
{Furthermore, in consideration of  $ x_{i}=\nabla \phi_{i}^{*}\left(y_{i}\right)$  and 
	$ x_{i}^{\Diamond}=\nabla \phi_{i}^{*}\left(y_{i}^{\Diamond}\right)$ again,} \eqref{b3} indicates 
\begin{equation}\label{b30}
	\begin{aligned}
		0& \leq   \nabla \phi_{i}^{*}\left(y_{i}^{\Diamond}\right)^{T}\left(\nabla \phi_{i}\left(\nabla \phi_{i}^{*}\left(y_{i}\right)\right)-y_{i}\right)-\nabla \phi_{i}^{*}\left(y_{i}\right)^{T}\left(\nabla \phi_{i}\left(\nabla \phi_{i}^{*}\left(y_{i}\right)\right)-y_{i}\right)\\
		&= 
		(x_{i}^{\Diamond}-x_{i})^{T}\left(\nabla \phi_{i}\left(x_{i}\right)-y_{i}\right).
	\end{aligned}
\end{equation}
Thus, (\ref{dfrd}) becomes
\begin{equation*}\label{b33}
	\sum_{i=1}^{N} D_{\phi_{i}^{*}}(y_{i},y_{i}^{\Diamond}) \geq \frac{\mu_{x}}{2}\left\|\boldsymbol{x}-\boldsymbol{x}^{\Diamond}\right\|^{2}.
\end{equation*}
The analogous analysis of the term $ D_{\varphi_{i}^{*}}(\nu_{i},\nu_{i}^{\Diamond})$ in \eqref{fe22} can be carried on,
which  also indicates that
\begin{equation*}
	\begin{aligned}
		\sum_{i=1}^{N} D_{\varphi_{i}^{*}}(\nu_{i},\nu_{i}^{\Diamond}) \geq \frac{\mu_{\sigma}}{2}\left\|\boldsymbol{\sigma}-\boldsymbol{\sigma}^{\Diamond}\right\|^{2}+\sum_{i=1}^{N}(\sigma_{i}^{\Diamond}-\sigma_{i})^{T}\left(\nabla \varphi_{i}\left(\sigma_{i}\right)-\nu_{i}\right).
	\end{aligned}
\end{equation*}
Besides, recall 
$\nabla \varphi_{i}^{*}\left(\nu_{i}\right)=\operatorname{argmin}_{\sigma_i \in \mathscr{E}_{ i}^{+}}\{-\sigma_i^{T} \nu_i+\varphi_i(\sigma_i)\}$. Based on the  convexity of $\varphi_{i}$  and the
optimality of $\nabla \varphi_{i}^{*}\left(\nu_{i}\right)$, we obtain
\begin{equation}\label{b37}
	\begin{aligned}
		0  \leq(\sigma_{i}^{\Diamond}-\sigma_{i})^{T}\left(\nabla \varphi_{i}\left(\sigma_{i}\right)-\nu_{i}\right),
	\end{aligned}
\end{equation}
which similarly leads to
\begin{equation*}
	\begin{aligned}
		\sum_{i=1}^{N} D_{\varphi_{i}^{*}}(\nu_{i},\nu_{i}^{\Diamond}) \geq \frac{\mu_{\sigma}}{2}\left\|\boldsymbol{\sigma}-\boldsymbol{\sigma}^{\Diamond}\right\|^{2}.
	\end{aligned}
\end{equation*}
{As a result, we obtain the lower bound of \eqref{fe22} that}
\begin{align*}
	V_{1}
	\geq  {\mu}(\|\boldsymbol{x}-\boldsymbol{x}^{\Diamond}\|^{2}+\|\boldsymbol{\sigma}-\boldsymbol{\sigma}^{\Diamond}\|^{2})
	\geq 0,
\end{align*}
where  $\mu=\min\{\mu_{x}/2,\mu_{\sigma}/2 \}$.	
This means that $V_{1}$ is positive semi-definite, and $V_{1}=0$ if and only if $\boldsymbol{x}=\boldsymbol{x}^{\Diamond}$ and $\boldsymbol{\sigma}=\boldsymbol{\sigma}^{\Diamond}$. Moreover,  $V_{1}$ is radially unbounded in $\boldsymbol{x}(t)$ and $\boldsymbol{\sigma}(t)$.

Next, 
we investigate the derivative of $V_{1}$ along ODE (\ref{e21}), that is, 
\begin{equation*}
	\begin{aligned}
		\frac{d}{d t} V_1(t)&=\frac{d}{d t}  \sum_{i=1}^{N} D_{\phi_{i}^{*}}(y_{i},y_{i}^{\Diamond})+D_{\varphi_{i}^{*}}(\nu_{i},\nu_{i}^{\Diamond})\\
		&=\frac{d}{d t}\sum_{i=1}^{N}
		(\phi_{i}^{*}(y_i)-\phi_{i}^{*}(y_i^{\Diamond})-(y_i-y_i^{\Diamond})^{T}\nabla \phi_{i}^{*}\left(y_i^{*}\right))+\frac{d}{d t}\sum_{i=1}^{N}
		(\varphi_{i}^{*}(\nu_i)-\varphi_{i}^{*}(\nu_i^{\Diamond})-(\nu_i-\nu_i^{\Diamond})^{T}\nabla \varphi_{i}^{*}\left(\nu_i^{*}\right)   )  
		\\
		&=\sum_{i=1}^{N}(\nabla 
		\phi^{*}_i(y_{i})-\nabla \phi^{*}_i(y_{i}^{\Diamond}))^{T} \dot{y_i}(t)+\sum_{i=1}^{N}(\nabla 
		\varphi^{*}_i(\nu_{i})-\nabla \varphi^{*}_i(\nu_{i}^{\Diamond}))^{T} \dot{\nu_i}(t) \\
		&=\sum_{i=1}^{N}(x_i-x_i^{\Diamond})^{T} \dot{y_i}(t)+\sum_{i=1}^{N}(\sigma_i-\sigma_i^{\Diamond}))^{T} \dot{\nu_i}(t). \\ 
	\end{aligned}
\end{equation*}
Here we employ the compact form defined in (\ref{e1}) for a more concise statement below and derive 
\begin{equation}\label{vv1}
	\begin{aligned}
		\frac{d}{d t} V_1(t)
		=(\boldsymbol{x}-\boldsymbol{x}^{\Diamond})^{T}(-\boldsymbol{G}(\boldsymbol{x},\boldsymbol{\sigma} )+\nabla\boldsymbol{\phi}(\boldsymbol{x})-\boldsymbol{y} )+\left(\boldsymbol{\sigma}-\boldsymbol{\sigma}^{\Diamond}\right)^{T}(\Lambda\left(\boldsymbol{x}\right)-\nabla \Psi^{*}\left(\boldsymbol{\sigma}\right)+\nabla \boldsymbol{\varphi}(\boldsymbol{\sigma})-\boldsymbol{\nu}  ). 
	\end{aligned}
\end{equation}
Meanwhile, 	by rearranging the terms in  (\ref{vv1}), we have
\begin{equation}\label{r4}	
	\begin{aligned}
		\dot {V}_{1}&=
		-(\boldsymbol{x}-\boldsymbol{x}^{\Diamond})^{T}\boldsymbol{G}(\boldsymbol{x},\boldsymbol{\sigma} )-\left(\boldsymbol{\sigma}-\boldsymbol{\sigma}^{\Diamond}\right)^{T}(-\Lambda\left(\boldsymbol{x}\right)+\nabla \Psi^{*}\left(\boldsymbol{\sigma}\right))+(\boldsymbol{x}-\boldsymbol{x}^{\Diamond})^{T}(\nabla\boldsymbol{\phi}(\boldsymbol{x})-\boldsymbol{y})+\left(\boldsymbol{\sigma}-\boldsymbol{\sigma}^{\Diamond}\right)^{T}\left(\nabla \boldsymbol{\varphi}(\boldsymbol{\sigma})-\boldsymbol{\nu}\right)  \\
		&=-(\boldsymbol{z}-\boldsymbol{z}^{\Diamond})^{T} F(\boldsymbol{z})	+(\boldsymbol{x}-\boldsymbol{x}^{\Diamond})^{T}(\nabla\boldsymbol{\phi}(\boldsymbol{x})-\boldsymbol{y})+(\boldsymbol{\sigma}-\boldsymbol{\sigma}^{\Diamond})^{T}\left(\nabla \boldsymbol{\varphi}(\boldsymbol{\sigma})-\boldsymbol{\nu}\right),
	\end{aligned}
\end{equation}
where $\boldsymbol{z}=\operatorname{col}\{\boldsymbol{x},\boldsymbol{\sigma}\}$ and $F(\boldsymbol{z})=\operatorname{col}\{G(\boldsymbol{x},\boldsymbol{\sigma}), \; -\Lambda\left(\boldsymbol{x}\right)+\nabla \Psi^{*}\left(\boldsymbol{\sigma}\right)\}$ are defined in (\ref{compact-pseudo}).
Notice that \eqref{b30} and \eqref{b37} actually reveals that
\begin{equation}\label{r41}	
	(\boldsymbol{x}-\boldsymbol{x}^{\Diamond})^{T}(\nabla\boldsymbol{\phi}(\boldsymbol{x})-\boldsymbol{y})\leq 0,\quad (\boldsymbol{\sigma}-\boldsymbol{\sigma}^{\Diamond})^{T}\left(\nabla \boldsymbol{\varphi}(\boldsymbol{\sigma})-\boldsymbol{\nu}\right)\leq 0.
\end{equation}
Because $\boldsymbol{z}^{\Diamond}$ is a solution to  $\operatorname{VI}(\boldsymbol{\Xi}, F) $, we realize
\begin{equation}\label{e56}
	\left(\boldsymbol{z}-\boldsymbol{z}^{\Diamond}\right)^{T} F(\boldsymbol{z}^{\Diamond})\geq 0.
\end{equation}
Thus,  (\ref{r4}) yields the further scaling that
\begin{equation}\label{v4}
	\begin{aligned}
		\dot {V}_{1}	&=-(\boldsymbol{z}-\boldsymbol{z}^{\Diamond})^{T} F(\boldsymbol{z})	+(\boldsymbol{x}-\boldsymbol{x}^{\Diamond})^{T}(\nabla\boldsymbol{\phi}(\boldsymbol{x})-\boldsymbol{y})+(\boldsymbol{\sigma}-\boldsymbol{\sigma}^{\Diamond})^{T}\left(\nabla \boldsymbol{\varphi}(\boldsymbol{\sigma})-\boldsymbol{\nu}\right)\\
		&\leq -(\boldsymbol{z}-\boldsymbol{z}^{\Diamond})^{T} F(\boldsymbol{z})\\
		&=-(\boldsymbol{z}-\boldsymbol{z}^{\Diamond})^{T} (F(\boldsymbol{z})-F(\boldsymbol{z}^{\Diamond}))-(\boldsymbol{z}-\boldsymbol{z}^{\Diamond})^{T} F(\boldsymbol{z}^{\Diamond})\\
		&\leq-(\boldsymbol{z}-\boldsymbol{z}^{\Diamond})^{T} (F(\boldsymbol{z})-F(\boldsymbol{z}^{\Diamond})),
	\end{aligned}
\end{equation}
where the first inequality is due to \eqref{r41} and the second  inequality is due to (\ref{e56}).
Now, we consider the term
$(\boldsymbol{z}-\boldsymbol{z}^{\Diamond})^{T} (F(\boldsymbol{z})-F(\boldsymbol{z}^{\Diamond}))$ with details.
\begin{align*}
&\quad(\boldsymbol{z}-\boldsymbol{z}^{\Diamond})^{T}(F(\boldsymbol{z})-F(\boldsymbol{z}^{\Diamond}))\\
	&= (\boldsymbol{x}-\boldsymbol{x}^{\Diamond})^{T}(\boldsymbol{G}(\boldsymbol{x},\boldsymbol{\sigma} )-\boldsymbol{G}(\boldsymbol{x}^{\Diamond},\boldsymbol{\sigma}^{\Diamond}))+\left(\boldsymbol{\sigma}-\boldsymbol{\sigma}^{\Diamond}\right)^{T}(-\Lambda\left(\boldsymbol{x}\right)+\nabla \Psi^{*}\left(\boldsymbol{\sigma}\right)-(-\Lambda(\boldsymbol{x}^{\Diamond})+\nabla \Psi^{*}(\boldsymbol{\sigma}^{\Diamond})) )\\
	&=(\boldsymbol{x}-\boldsymbol{x}^{\Diamond})^{T}(\boldsymbol{G}(\boldsymbol{x},\boldsymbol{\sigma} )-\boldsymbol{G}(\boldsymbol{x}^{\Diamond},\boldsymbol{\sigma}^{\Diamond}))-\left(\boldsymbol{\sigma}-\boldsymbol{\sigma}^{\Diamond}\right)^{T}(\Lambda(\boldsymbol{x})-\Lambda(\boldsymbol{x}^{\Diamond}))+(\boldsymbol{\sigma}-\boldsymbol{\sigma}^{\Diamond})^{T}(\nabla \Psi^{*}(\boldsymbol{\sigma})-\nabla \Psi^{*}(\boldsymbol{\sigma}^{\Diamond})).
\end{align*}
Due to the  convexity of $\Psi_i$ for $i\in\mathcal{I}$, the Legendre conjugate $\Psi_i^*$ is also convex \cite{yang2000canonical}, which indicates
\begin{align*}
	(\boldsymbol{\sigma}-\boldsymbol{\sigma}^{\Diamond})^{T}(\nabla \Psi^{*}(\boldsymbol{\sigma})-\nabla \Psi^{*}(\boldsymbol{\sigma}^{\Diamond}))\geq 0.
\end{align*}
{Hence,  
	\begin{equation}\label{v5}
		\begin{aligned}
			&\quad(\boldsymbol{z}-\boldsymbol{z}^{\Diamond})^{T}(F(\boldsymbol{z})-F(\boldsymbol{z}^{\Diamond}))\\
			&\geq(\boldsymbol{x}-\boldsymbol{x}^{\Diamond})^{T}(\boldsymbol{G}(\boldsymbol{x},\boldsymbol{\sigma} )-\boldsymbol{G}(\boldsymbol{x}^{\Diamond},\boldsymbol{\sigma}^{\Diamond}))-\left(\boldsymbol{\sigma}-\boldsymbol{\sigma}^{\Diamond}\right)^{T}(\Lambda(\boldsymbol{x})-\Lambda(\boldsymbol{x}^{\Diamond})).
		\end{aligned}
	\end{equation}
	Expanding the expression in  (\ref{v5}), 
	\begin{equation}\label{v52}
		\begin{aligned}
			&\quad(\boldsymbol{x}-\boldsymbol{x}^{\Diamond})^{T}(\boldsymbol{G}(\boldsymbol{x},\boldsymbol{\sigma} )-\boldsymbol{G}(\boldsymbol{x}^{\Diamond},\boldsymbol{\sigma}^{\Diamond}))-\left(\boldsymbol{\sigma}-\boldsymbol{\sigma}^{\Diamond}\right)^{T}(\Lambda(\boldsymbol{x})-\Lambda(\boldsymbol{x}^{\Diamond}))\\
			&=\sum\nolimits_{i=1}\nolimits^{N} (x_i-x_i^{\Diamond})^{T}\left( \sum\nolimits_{k=1}\nolimits^{q_i}[\sigma_{i}]_{k} \nabla_{x_{i}} \Lambda_{i, k}(x_{i},\boldsymbol{x}_{-i})-\sum\nolimits_{k=1}\nolimits^{q_i}[\sigma_{i}^{\Diamond}]_{k} \nabla_{x_{i}} \Lambda_{i, k}(x_{i}^{\Diamond},\boldsymbol{x}_{-i}^{\Diamond})\right)\\
			&\quad\quad-\sum\nolimits_{i=1}\nolimits^{N} \sum\nolimits_{k=1}\nolimits^{q_i}([\sigma_{i}]_{k}-[\sigma_i^{\Diamond}]_{k})\left(\Lambda_{i,k}\left(x_{i},\boldsymbol{x}_{-i}\right)-\Lambda_{i,k}(x_{i}^{\Diamond},\boldsymbol{x}^{\Diamond}_{-i})\right).
		\end{aligned}
	\end{equation}
	Rearranging (\ref{v52}), we have
	\begin{equation}\label{v53}
		\begin{aligned}
			&\quad(\boldsymbol{x}-\boldsymbol{x}^{\Diamond})^{T}(\boldsymbol{G}(\boldsymbol{x},\boldsymbol{\sigma} )-\boldsymbol{G}(\boldsymbol{x}^{\Diamond},\boldsymbol{\sigma}^{\Diamond}))-\left(\boldsymbol{\sigma}-\boldsymbol{\sigma}^{\Diamond}\right)^{T}(\Lambda(\boldsymbol{x})-\Lambda(\boldsymbol{x}^{\Diamond}))\\&=\sum\nolimits_{i=1}\nolimits^{N} \sum\nolimits_{k=1}\nolimits^{q_i} \left( [\sigma_{i}]_{k} (x_i-x_i^{\Diamond})^{T} \nabla_{x_{i}} \Lambda_{i, k}(x_{i},\boldsymbol{x}_{-i})-[\sigma_{i}^{\Diamond}]_{k} (x_i-x_i^{\Diamond})^{T}\nabla_{x_{i}} \Lambda_{i, k}(x_{i}^{\Diamond},\boldsymbol{x}_{-i}^{\Diamond})\right)\\
			&\quad\quad-\sum\nolimits_{i=1}\nolimits^{N} \sum\nolimits_{k=1}\nolimits^{q_i}([\sigma_{i}]_{k}-[\sigma_i^{\Diamond}]_{k})\left(\Lambda_{i,k}\left(x_{i},\boldsymbol{x}_{-i}\right)-\Lambda_{i,k}(x_{i}^{\Diamond},\boldsymbol{x}^{\Diamond}_{-i})\right).
		\end{aligned}
	\end{equation}
	By merging terms in (\ref{v53}), we have
	\begin{equation}\label{v54}
		\begin{aligned}
			&\quad(\boldsymbol{x}-\boldsymbol{x}^{\Diamond})^{T}(\boldsymbol{G}(\boldsymbol{x},\boldsymbol{\sigma} )-\boldsymbol{G}(\boldsymbol{x}^{\Diamond},\boldsymbol{\sigma}^{\Diamond}))-\left(\boldsymbol{\sigma}-\boldsymbol{\sigma}^{\Diamond}\right)^{T}(\Lambda(\boldsymbol{x})-\Lambda(\boldsymbol{x}^{\Diamond}))\\&=\sum\nolimits_{i=1}\nolimits^{N}\! \sum\nolimits_{k=1}\nolimits^{q_i} \!
			\left((x_i-x_i^{\Diamond})^{T}( [\sigma_{i}]_{k}\nabla_{x_{i}} \Lambda_{i, k}(x_{i},\boldsymbol{x}_{-i})) \!+ \! [\sigma_{i}]_{k}\Lambda_{i,k}(x_{i}^{\Diamond},\boldsymbol{x}^{\Diamond}_{-i})\!-\![\sigma_{i}]_{k}\Lambda_{i,k}(x_{i},\boldsymbol{x}_{-i})\right) \\
			&\quad\quad\!+\!\!\sum\nolimits_{i=1}\nolimits^{N} \!\!\sum\nolimits_{k=1}\nolimits^{q_i}  \! \!\!\left((x_i^{\Diamond}\!-\!x_i)^{T}\! ([\sigma_{i}^{\Diamond}\!]_{k}\!\nabla_{x_{i}} \!\Lambda_{i, k}\!(x_{i}^{\Diamond},\boldsymbol{x}_{-i}^{\Diamond}))\!+\![\sigma_{i}^{\Diamond}\!]_{k}\! \Lambda_{i,k}\!(x_{i},\boldsymbol{x}_{-i})\! \!-\![\sigma_{i}^{\Diamond}]_{k}\!\Lambda_{i,k}(x_{i}^{\Diamond},\boldsymbol{x}^{\Diamond}_{-i}) \right).	
		\end{aligned}
\end{equation}}
Recalling  the definition in (\ref{s23}) that for $i\in\mathcal{I}$,
\begin{align*}
	\sigma_{i},\sigma_{i}^{\Diamond}\in
	\mathscr{E}_{ i}^{+}=\{\sigma_{i}\in\Theta_{i}^{*} :\; \sum\nolimits_{k=1}\nolimits^{q_i}[\sigma_{i}]_{k} \nabla^{2}_{x_{i}} \Lambda_{i, k}(x_{i},\boldsymbol{x}_{-i}) \succeq
	\kappa_{x}\boldsymbol{I}_{n}
	\}.
\end{align*}	
Hence, (\ref{v54}) satisfies
\begin{align*}
	&\sum\nolimits_{i=1}\nolimits^{N}\! \sum\nolimits_{k=1}\nolimits^{q_i} \!
	\left((x_i-x_i^{\Diamond})^{T}( [\sigma_{i}]_{k}\nabla_{x_{i}} \Lambda_{i, k}(x_{i},\boldsymbol{x}_{-i})) \!+ \! [\sigma_{i}]_{k}\Lambda_{i,k}(x_{i}^{\Diamond},\boldsymbol{x}^{\Diamond}_{-i})\!-\![\sigma_{i}]_{k}\Lambda_{i,k}(x_{i},\boldsymbol{x}_{-i})\right) \\
	&\quad\quad\!+\!\!\sum\nolimits_{i=1}\nolimits^{N} \!\!\sum\nolimits_{k=1}\nolimits^{q_i}  \! \!\!\left((x_i^{\Diamond}\!-\!x_i)^{T}\! ([\sigma_{i}^{\Diamond}\!]_{k}\!\nabla_{x_{i}} \!\Lambda_{i, k}\!(x_{i}^{\Diamond},\boldsymbol{x}_{-i}^{\Diamond}))\!+\![\sigma_{i}^{\Diamond}\!]_{k}\! \Lambda_{i,k}\!(x_{i},\boldsymbol{x}_{-i})\! \!-\![\sigma_{i}^{\Diamond}]_{k}\!\Lambda_{i,k}(x_{i}^{\Diamond},\boldsymbol{x}^{\Diamond}_{-i}) \right)\\
	\geq & 	\kappa_x  \|\boldsymbol{x}-\boldsymbol{x}^{\Diamond}\|^{2}.
\end{align*}
which
further yields 
\begin{equation*}
	(\boldsymbol{z}-\boldsymbol{z}^{\Diamond})^{T}(F(\boldsymbol{z})-F(\boldsymbol{z}^{\Diamond}))\geq 	\kappa_x  \|\boldsymbol{x}-\boldsymbol{x}^{\Diamond}\|^{2}.
\end{equation*} 
In this view, we can  accordingly get
\begin{equation}\label{g36}
	\dot {V}_{1}\leq-\kappa_x  \|\boldsymbol{x}-\boldsymbol{x}^{\Diamond}\|^{2} \leq 0.
\end{equation} 
{Since ${V}_{1}$ is radially unbounded in $\boldsymbol{x}(t)$ and $\boldsymbol{\sigma}(t)$, this implies that the trajectories of $\boldsymbol{x}(t)$ and $\boldsymbol{\sigma}(t)$ are bounded along the conjugate-based ODE (\ref{e21}). }

Secondly,  we show that $\boldsymbol{y}(t)$ and $\boldsymbol{\nu}(t)$ are bounded. Take another  Lyapunov candidate function as
$	V_{2}=\frac{1}{2}\|\boldsymbol{y}\|^{2},$
which is radially unbounded in $\boldsymbol{y}$. Along the trajectories of (\ref{e1}), the derivative of $V_2$  satisfies
\begin{equation*}
	\dot {V}_{2}\leq  \boldsymbol{y}^{T}(-\boldsymbol{G}(\boldsymbol{x},\boldsymbol{\sigma} )+\nabla\Phi(\boldsymbol{x}))-\|\boldsymbol{y}\|^2.
\end{equation*}
It is clear that 
\begin{align*}
	\dot {V}_{2}\leq -\|\boldsymbol{y}\|^{2}+p_{1}\|\boldsymbol{y}\|=-2 V_{2}+p_{1} \sqrt{2 V_{2}},
\end{align*}
with a positive constant $p_1$, which is because  $ \boldsymbol{x} $, $ \boldsymbol{\sigma} $ have been proved to be bounded.
Analogously, take a third  Lyapunov candidate function as
$	V_{3}=\frac{1}{2}\|\boldsymbol{\nu}\|^{2},$
which  is radially unbounded in $\boldsymbol{\sigma}$. Along the trajectories of (\ref{e1}), the derivative of $V_3$  satisfies
\begin{align*}
	\dot {V}_{3}&\leq  \boldsymbol{\nu}^{T}(\Lambda\left(\boldsymbol{x}\right)-\nabla \Psi^{*}\left(\boldsymbol{\sigma}\right)+\nabla \varphi(\boldsymbol{\sigma})-\boldsymbol{\nu} )-\|\boldsymbol{\nu}\|^2\\
	&\leq -\|\boldsymbol{\nu}\|^{2}+p_{2}\|\boldsymbol{\nu}\|=-2 V_{3}+p_{2} \sqrt{2 V_{3}},
\end{align*}
with a positive constant $p_2$.
Hence, it can be easily verified that $V_2$ and $V_3$  are bounded, so are $ \boldsymbol{y}(t) $ and $ \boldsymbol{\nu}(t) $. 

(ii) {	Now let us investigate the set
	\begin{align*}
		Q \triangleq\left\{(\boldsymbol{x}, \boldsymbol{y}, \boldsymbol{\sigma}, \boldsymbol{\nu}): \frac{d}{d t} V_1=0\right\},
	\end{align*}
	and take set $ I_{v} $ as its largest invariant subset. It follows from the invariance principle \cite[Theorem 2.41]{chellaboina2008nonlinear}
	that $ (\boldsymbol{x}, \boldsymbol{y}, \boldsymbol{\sigma}, \boldsymbol{\nu})\rightarrow I_{v}$ as $ t \rightarrow \infty $, and $ I_{v} $ is a positive invariant set. 
	Then it follows from the derivation in (\ref{g36}) that 
	\begin{align*}	
		I_{v} \subseteq\left\{(\boldsymbol{x}, \boldsymbol{y}, \boldsymbol{\sigma}, \boldsymbol{\nu}): \boldsymbol{x}=\boldsymbol{x}^{\Diamond}  \right\}.
	\end{align*}
	This indicates that any trajectory along ODE (\ref{e21}) results in  the convergence of variable $\boldsymbol{x}$, that is, $\boldsymbol{x}(t)\rightarrow\boldsymbol{x}^{\Diamond} $ as $t\rightarrow\infty$. 
	Moreover, if $\sigma_{i}^{\Diamond}=\nabla \Psi_{i}\left(\xi_{i}\right) \mid_{\xi_{i}=\Lambda_{i}\left(x_{i}^{\Diamond}, \boldsymbol{x}_{-i}^{\Diamond}\right)}$ for $i\in\mathcal{I}$, then the convergent point $\boldsymbol{x}^{\Diamond}$ is indeed a global NE.
	So far, we have accomplished the proof.} 	
%
\hfill $\square$


Based on the  proof of Theorem \ref{l1},  we further show the convergence rate of ODE (\ref{e21}).   Here, we also reproduce Theorem \ref{t4} below for convenience:

\noindent\textbf{Theorem 3}
If $\mathscr{E}_{ i}^{+}$ is nonempty and $ \Psi_{i}\left(\cdot\right)$ is $\frac{1}{\kappa_{\sigma}}$-smooth	for $i\in\mathcal{I}$,  
then 
(\ref{e21}) converges at an exponential rate, i.e.,
\begin{align*}
	\|\boldsymbol{z}(t)-\boldsymbol{z}^{\Diamond}\|\leq \sqrt{\frac{{\tau}}{{\mu}}}\|\boldsymbol{z}(0)\|\operatorname{exp}(- \frac{{\kappa}}{2{\tau}}\,t),
\end{align*}
where  
$\mu=\min\{{\mu_{x}}/{2},{\mu_{\sigma}}/{2} \}$, 
${\kappa}=\min\{\kappa_{\sigma}, \kappa_{x} \}$, ${\tau}=\max\{{L_{x}}/{2\mu_{x}},{L_{\sigma}}/{2\mu_{\sigma}} \}$.

\noindent\textbf{Proof.}  Take the same Lyapunov function as in Theorem \ref{l1}:
\begin{align*}
	V_{1}=& \sum_{i=1}^{N} D_{\phi_{i}^{*}}(y_{i},y_{i}^{\Diamond})+D_{\varphi_{i}^{*}}(\nu_{i},\nu_{i}^{\Diamond}).
\end{align*}
Recalling the analysis in Theorem \ref{l1},  we have
\begin{align}\label{v7}
	V_{1}\geq  {\mu}\left(\left\|\boldsymbol{x}-\boldsymbol{x}^{\Diamond}\right\|^{2}+\left\|\boldsymbol{\sigma}-\boldsymbol{\sigma}^{\Diamond}\right\|^{2}\right)={\mu}\left\|\boldsymbol{z}-\boldsymbol{z}^{\Diamond}\right\|^{2},
\end{align}
where $\mu=\min\{\mu_{x}/2,\mu_{\sigma}/2 \}$.
Based on the standard duality relations, 
the $\mu_{x}$-strong  convexity of generating function $\phi_{i}$ on $\Omega_i$ implies that its conjugate gradient  $\nabla\phi_{i}^{*}$ is continuously differentiable on $\mathbb{R}^{n}$
with $1/\mu_{x}$-Lipschitz continuous gradient \cite{ben2001ordered}. 	
Thus, 
\begin{equation}\label{dde}
	\begin{aligned}
		\sum_{i=1}^{N} D_{\phi_{i}^{*}}(y_{i},y_{i}^{\Diamond})&=\sum_{i=1}^{N} \phi_{i}^{*}(y_{i})-\phi_{i}^{*}(y_{i}^{\Diamond})-\nabla \phi_{i}^{*}(y_{i}^{\Diamond})^{T}(y_{i}-y_{i}^{\Diamond})\\
		&\leq \frac{1}{2\mu_{x}} \sum_{i=1}^{N} \|y_{i}-y_{i}^{\Diamond} \|^{2}.
	\end{aligned}
\end{equation}
Moreover, with the  duality relation 
$ \nabla \phi_{i}(x_{i})=y_{i}$ and 
$ \nabla \phi_{i}(x_{i}^{\Diamond})=y_{i}^{\Diamond}$,  \eqref{dde} yields
\begin{equation}\label{dde1}
	\begin{aligned}
		\sum_{i=1}^{N} D_{\phi_{i}^{*}}(y_{i},y_{i}^{\Diamond})
		&\leq \frac{1}{2\mu_{x}} \sum_{i=1}^{N} \|y_{i}-y_{i}^{\Diamond} \|^{2}\\
		&=\frac{1}{2\mu_{x}} \sum_{i=1}^{N} \|\nabla \phi_{i}(x_{i})-\nabla \phi_{i}(x_{i}^{\Diamond}) \|^{2}\\
		&\leq \frac{L_{x}}{2\mu_{x}} \|\boldsymbol{x}-\boldsymbol{x}^{\Diamond} \|^{2},
	\end{aligned}
\end{equation}

where  the last inequality is due  to the $L_{x}$-Lipschitz continuity  of generating function $\phi_{i}$.
Analogously, 
$\nabla\varphi_{i}^{*}$ is $1/\mu_{\sigma}$-Lipschitz  because of the $\mu_{\sigma}$-strongly convexity of generating function $\varphi_{i}$ on $\mathscr{E}_{i}^{+}$. 
With $ \nabla \varphi_{i}(\sigma_{i})=\nu_{i}$ and 
$ \nabla \varphi_{i}(\sigma_{i}^{\Diamond})=\nu_{i}^{\Diamond}$,
we also have
\begin{align*}
	\sum_{i=1}^{N} D_{\varphi_{i}^{*}}(\nu_{i},\nu_{i}^{\Diamond})&=\sum_{i=1}^{N}\varphi_{i}^{*}(\nu_{i})-\varphi_{i}^{*}(\nu_{i}^{\Diamond})-\nabla \varphi_{i}^{*}(\nu_{i}^{\Diamond})^{T}(\nu_{i}-\nu_{i}^{\Diamond})\\
	&\leq \frac{1}{2\mu_{\sigma}} \sum_{i=1}^{N} \|\nu_{i}-\nu_{i}^{\Diamond} \|^{2}\\
	&=\frac{1}{2\mu_{\sigma}} \sum_{i=1}^{N} \|\nabla \varphi_{i}(\sigma_{i})-\nabla \varphi_{i}(\sigma_{i}^{\Diamond}) \|^{2}\\
	&\leq \frac{L_{\sigma}}{2\mu_{\sigma}} \|\boldsymbol{\sigma}-\boldsymbol{\sigma}^{\Diamond} \|^{2},
\end{align*}
where the last inequality is due  to the $L_{\sigma}$-Lipschitz continuity  of generating function $\varphi_{i}$.
Therefore, 
\begin{align*}
	V_{1}\leq \sum_{i=1}^{N} D_{\phi_{i}^{*}}(y_{i},y_{i}^{\Diamond})+D_{\varphi_{i}^{*}}(\nu_{i},\nu_{i}^{\Diamond}) \leq {\tau} \|\boldsymbol{z}-\boldsymbol{z}^{\Diamond} \|^{2},  
\end{align*}
where 
${\tau}=\max\{{L_{x}}/{2\mu_{x}},{L_{\sigma}}/{2\mu_{\sigma}} \}$.
Moreover, 	
following the proof of Theorem \ref{l1},  the derivate of  $ {V}_{1}$  satisfies
\begin{align*}
	\dot {V}_{1}&\leq
	-(\boldsymbol{z}-\boldsymbol{z}^{\Diamond})^{T} (F(\boldsymbol{z})-F(\boldsymbol{z}^{\Diamond})).
\end{align*}
{Due to  the definition of set $\mathscr{E}_{ i}^{+}$ in (\ref{s23}) and 
	the  $\kappa_{\sigma}$-strongly convexity of $ \Psi^{*}_i$ 
	for $i\in\mathcal{I}$,  we have
	\begin{align*}
		&\quad(\boldsymbol{z}-\boldsymbol{z}^{\Diamond})^{T}(F(\boldsymbol{z})-F(\boldsymbol{z}^{\Diamond}))\\
		&= (\boldsymbol{x}-\boldsymbol{x}^{\Diamond})^{T}(\boldsymbol{G}(\boldsymbol{x},\boldsymbol{\sigma} )-\boldsymbol{G}(\boldsymbol{x}^{\Diamond},\boldsymbol{\sigma}^{\Diamond}))+\left(\boldsymbol{\sigma}-\boldsymbol{\sigma}^{\Diamond}\right)^{T}(-\Lambda\left(\boldsymbol{x}\right)+\nabla \Psi^{*}\left(\boldsymbol{\sigma}\right)-(-\Lambda\left(\boldsymbol{x}^{\Diamond}\right)+\nabla \Psi^{*}\left(\boldsymbol{\sigma}^{\Diamond}\right)) )\\
		&=(\boldsymbol{x}-\boldsymbol{x}^{\Diamond})^{T}(\boldsymbol{G}(\boldsymbol{x},\boldsymbol{\sigma} )-\boldsymbol{G}(\boldsymbol{x}^{\Diamond},\boldsymbol{\sigma}^{\Diamond}))-\left(\boldsymbol{\sigma}-\boldsymbol{\sigma}^{\Diamond}\right)^{T}(\Lambda\left(\boldsymbol{x}\right)-\Lambda(\boldsymbol{x}^{\Diamond}))+(\boldsymbol{\sigma}-\boldsymbol{\sigma}^{\Diamond})^{T}(\nabla \Psi^{*}(\boldsymbol{\sigma})-\nabla \Psi^{*}(\boldsymbol{\sigma}^{\Diamond}))\\
		&\geq \kappa_{x}\| \boldsymbol{x}-\boldsymbol{x}^{\Diamond}  \|^{2}+\kappa_{\sigma}\| \boldsymbol{\sigma}-\boldsymbol{\sigma}^{\Diamond}  \|^{2}\\
		&\geq \kappa \|\boldsymbol{z}-\boldsymbol{z}^{\Diamond}\|^{2},
	\end{align*}
	where $\kappa=\min\{\kappa_{\sigma}, \kappa_{x} \}$.}
Therefore,
\begin{equation}\label{a1}
	\begin{aligned}
		\dot {V}_{1}
		&\leq -\kappa \|\boldsymbol{z}-\boldsymbol{z}^{\Diamond}\|^{2}.
	\end{aligned}
\end{equation}
It follows from (\ref{v7}) and (\ref{a1}) that 
\begin{align*}
	\dot {V}_{1}\leq -\kappa \|\boldsymbol{z}-\boldsymbol{z}^{\Diamond}\|^{2}\leq -\frac{\kappa}{\tau} {V}_{1},
\end{align*}
which actually yields the exponential convergence rate.
In other words,
\begin{align*}
	\mu\|\boldsymbol{z}(t)-\boldsymbol{z}^{\Diamond}\|^{2}
	&\leq {V}_{1}(\boldsymbol{z}(t))\\&\leq {V}_{1}(\boldsymbol{z}(0))\operatorname{exp}(- \frac{{\kappa}}{{\tau}}\,t)\\&
	\leq \tau \|\boldsymbol{z}(0)\|^2\operatorname{exp}(- \frac{{\kappa}}{{\tau}}\,t).
\end{align*}
Thus, we can also obtain 
\begin{align*}
	\|\boldsymbol{z}(t)-\boldsymbol{z}^{\Diamond}\|\leq \sqrt{\frac{\tau}{\mu}}\|\boldsymbol{z}(0)\|\operatorname{exp}(- \frac{{\kappa}}{2{\tau}}\,t),
\end{align*}
which implies this conclusion. \hfill $\square$

\section{Bregman divergence and some inequalities}


After the analysis of ODE (\ref{e21}), in this section, we turn to investigate  the discrete algorithm 1 induced from ODE (\ref{e21}). Also, 
we  provide some auxiliary results  that are needed in the following contents. 

First of all, the Bregman divergence associated to a  generating function $h: \Xi\rightarrow \mathbb{R}$ is defined as
\begin{align*}
	D_{h}(z^{\prime}, z)=h(z^{\prime})-h(z)-\left(z^{\prime}-z\right)^{T}\nabla h(z), \;\forall  z,z^{\prime} \in \Xi. 
\end{align*}
In what follows, 
%
%
we give
basic bounds on the Bregman divergence. Firstly, the basic ingredient for these bounds is a generalization of the (Euclidean) law of cosines,  which is known in the literature as the  ``three-point identity'' \cite{chen1993convergence}:
\begin{lemma}\label{di}
	Let the  continuously differentiable generating function $h$ be  $\omega$-strongly convex  on set $\Xi$. For $z, z^{\prime},z^{+} $ in $\Xi$, there holds
	\begin{equation}\label{fgy}	
		\begin{aligned}
			D_{h}(z^{\prime}, z^{+})+D_{h}(z^{+}, z) 
			&=D_{h}(z^{\prime}, z)+\langle z^{\prime}-z^{+}, \nabla h(z)-\nabla h(z^{+})\rangle.
		\end{aligned}
	\end{equation}
\end{lemma}
\textbf{Proof.} It follows from the definition of the Bregman divergence that 
\begin{align*}
	D_{h}(z^{\prime}, z^{+})&=h(z^{\prime})-h\left(z^{+}\right)-\left(z^{\prime}-z^{+}\right)^{T}\nabla h(z^{+}), \\
	D_{h}(z^{+}, z)&=h(z^{+})-h(z)-\left(z^{+}-z\right)^{T}\nabla h(z), \\
	D_{h}(z^{\prime}, z)&=h(z^{\prime})-h\left(z\right)-\left(z^{\prime}-z\right)^{T}\nabla h(z).
\end{align*}
This lemma is thus true by adding the first two equalities and subtracting the last one. \hfill$\square$
%

Secondly, with the identity above, we have the following  upper  bound of a Bregman divergence. 
\begin{lemma}\label{d4}
	Let the  continuously differentiable generating function $h$ be  $\omega$-strongly convex  on set $\Xi$.  For $z, z^{\prime}$ in $\Xi$, and  $z^{+}=\Pi_{\Xi}^{h}(g)=\operatorname{argmin}_{z \in \Xi}\left\{-z^{T} g+h(z)\right\}$, there holds
	\begin{equation}\label{e78}
		\begin{aligned}
			D_h(z^{\prime}, z^{+}) & \leq D_{h}(z^{\prime}, z)-D_{h}(z^{+}, z)+(g-\nabla h(z))^{T} (z^{+}-z^{\prime}).
		\end{aligned}
	\end{equation}
\end{lemma}
\textbf{Proof.} 
Based on the three-point identity \eqref{fgy}, we  obtain
\begin{equation*}\label{fgy1}	
	D_h(z^{\prime}, z^{+})+D_{h}(z^{+}, z)= D_{h}(z^{\prime}, z)+(z^{+}-z^{\prime})^{T}( \nabla h(z^{+})-\nabla h(z)).
\end{equation*}
Rearranging these terms gives 
\begin{equation}\label{fgy3}
	D_h(z^{\prime}, z^{+}) = D_{h}(z^{\prime}, z)-D_{h}(z^{+}, z)+(z^{+}-z^{\prime})^{T}( \nabla h(z^{+})-\nabla h(z)).
\end{equation}
Moreover, with the fact that $z^{+}=\Pi_{\Xi}^{h}(g)=\operatorname{argmin}_{z \in \Xi}\left\{-z^{T} g+h(z)\right\} $, we learn from the optimality of $z^{+}$ and the convexity of $h$ that
\begin{equation*}
	(-g+\nabla h(z^{+}))^{T}z^{+}\leq (-g+\nabla h(z^{+}))^{T}z^{\prime},
\end{equation*}
which implies
\begin{equation}\label{fgy4}
	(z^{+}-z^{\prime})^{T} \nabla h(z^{+})\leq (z^{+}-z^{\prime})^{T}g.
\end{equation}
Thus, (\ref{e78}) holds by plugging (\ref{fgy4}) into  (\ref{fgy3}). 
\hfill$\square$

Before the end of this section, we introduce two classic inequalities in the following.


\begin{lemma}[Fenchel's inequality]\label{d7}
	Take $f$ as a continuous function on set $C$. Then the  Fenchel conjugate $f^{*}$  in dual space $C^{*}$ is $	f^{*}(b)=\operatorname{sup}_{a \in C}\left\{a^{T} b-f(a)\right\}$, which results in the following  inequality
	%
	%
	%
	\begin{equation*}
		a^{T}b\leq  f (a)+ f^{*}(b).
	\end{equation*}
\end{lemma}

\begin{lemma}[Jessen's inequality]\label{d9}
	Take $f $ as a convex function on a convex set $U$,  then 
	\begin{equation*}
		f(\sum_{l=1}^{k}\gamma_l x_l)\leq \sum_{l=1}^{k} \gamma_l f(x_l),
	\end{equation*}
	where $x_1,\cdots, x_k\in U$ and $\gamma_1, \cdots, \gamma_k >0$ with $\gamma_1+\cdots+\gamma_k=1$. 
\end{lemma}

%


\section{Proof of Theorem  \ref{t6}}\label{st4}

With help of the basis mentioned above, we  show the convergence analysis of Algorithm 1 on a class of $N$-player generalized monotone games. Here, we reproduce  Theorem \ref{t6} below for convenience:

\noindent\textbf{Theorem 4 } If $\mathscr{E}_{ i}^{+}$ is nonempty and 	 $\Pi_{\Theta_{ i}}^{\Psi_i}(\cdot)$ is $\kappa_{\sigma}$-strongly monotone, 	then Algorithm \ref{nplayer} converges at a rate of $\mathcal{O}({1}/{k})$ with step size $\alpha_k=\frac{2}{\kappa(k+1)}$, i.e.,
\begin{align*}
	\|\boldsymbol{x}^k-\boldsymbol{x}^{\Diamond} \|^2+\|\boldsymbol{\sigma}^k-\boldsymbol{\sigma}^{\Diamond} \|^2
	\leq 
	\frac{1}{k+1}\frac{{M}_1}{{\mu}^2{\kappa}^2},
\end{align*}
where  ${\mu}=\min\{\frac{\mu_{x}}{2},\frac{\mu_{\sigma}}{2} \}$,  ${\kappa}=\min\{\kappa_{\sigma}, \kappa_{x} \}$, and ${M}_1$ is a positive constant.

\noindent\textbf{Proof.} Take the collection of the Bregman divergence  as
\begin{equation}\label{dfr}
	\Delta(\boldsymbol{z}^{\Diamond},\boldsymbol{z}^{k+1})\triangleq\sum_{i=1}^{N} D_{\phi_{i}}(x_{i}^{\Diamond},x_{i}^{k+1})+D_{\varphi_{i}}(\sigma_{i}^{\Diamond},\sigma_{i}^{k+1}),
\end{equation}
where 
\begin{equation*}
	D_{\phi_{i}}(x_{i}^{\Diamond},x_{i}^{k+1})=\phi_{i}(x_{i}^{\Diamond})-\phi_{i}(x_{i}^{k+1})-\nabla \phi_{i}(x_{i}^{k+1})^{T}(x_{i}^{\Diamond}-x_{i}^{k+1}),
\end{equation*} 
\begin{equation*} D_{\varphi_{i}}(\sigma_{i}^{\Diamond},\sigma_{i}^{k+1})=\varphi_{i}(\sigma_{i}^{\Diamond})-\varphi_{i}(\sigma_{i}^{k+1})-\nabla \varphi_{i}(\sigma_{i}^{k+1})^{T}(\sigma_{i}^{\Diamond}-\sigma_{i}^{k+1}).
\end{equation*}
Because   $\phi_{i}$  is $\mu_{x}$-strongly convex  and $\varphi_{i}$ is $\mu_{\sigma}$-strongly convex for $i\in \mathcal{I}$, 
we obtain that
\begin{equation}\label{dfry}
	\begin{aligned}
		\Delta(\boldsymbol{z}^{\Diamond},\boldsymbol{z}^{k+1})
		&\geq 
		\frac{\mu_{x}}{2}\sum_{i=1}^{N}  \left\|{x}_i^{k+1}-{x}_i^{\Diamond}\right\|^{2}+\frac{\mu_{\sigma}}{2}\sum_{i=1}^{N}\left\|{\sigma}_i^{k+1}-{\sigma}_i^{\Diamond}\right\|^{2}\\
		&\geq
		\mu \|\boldsymbol{z}^{k+1}-\boldsymbol{z}^{\Diamond}\|^2,
	\end{aligned}
\end{equation}
where $\mu=\min\{{\mu_{x}}/{2},{\mu_{\sigma}}/{2} \}$. 
Then, consider  the term $ D_{\phi_{i}}(x_{i}^{\Diamond},x_{i}^{k+1})$ in \eqref{dfr}. 
By employing three-point identity in Lemma \ref{di}, we obtain that
\begin{equation}\label{ertd}
	D_{\phi_{i}}(x_{i}^{\Diamond},x_{i}^{k+1})=D_{\phi_{i}}(x_{i}^{\Diamond},x_{i}^{k})-D_{\phi_{i}}(x_{i}^{k+1},x_{i}^{k})+(\nabla \phi_{i}(x_{i}^{k+1})-\nabla \phi_{i}(x_{i}^{k}))^{T}(x_{i}^{k+1}-x_{i}^{\Diamond}).
\end{equation}
Denote
\begin{equation*}
	g_i=\nabla \phi_{i}(x_i^{k})-\alpha_k \sigma_{i}^{kT}\nabla_{x_{i}} \Lambda_{i}(x_{i}^{k},\boldsymbol{x}_{-i}^{k}).
\end{equation*} 
%
%
According to Algorithm 1,
\begin{equation*}
	x_{i}^{k+1} =\Pi_{\Omega_{i}}^{\phi_i}(g_i  )=\operatorname{argmin}_{x \in \Omega_{i}}\left\{-x^{T} g_i+\phi_{i}(x)\right\}, 
\end{equation*}
which implies that
\begin{equation*}\label{b36}
	\begin{aligned}
		0& \leq \left(\nabla \phi_{i}\left(x_{i}^{k+1} \right)-g_{i}\right)^{T}x_{i}^{\Diamond} -\left( \left(\nabla \phi_{i}\left( x_{i}^{k+1}  \right)\right)-g_{i}\right)^{T}x_{i}^{k+1}\\
		&= 
		\left(\nabla \phi_{i}\left(x_{i}^{k+1}\right)-g_{i}\right)^{T} (x_{i}^{\Diamond}-x_{i}^{k+1}).
	\end{aligned}
\end{equation*}
In addition,
\begin{align*}
	(\nabla \phi_{i}(x_{i}^{k+1})^{T}(x_{i}^{k+1}-x_{i}^{\Diamond})\leq (\nabla \phi_{i}(x_i^{k})-\alpha_k \sigma_{i}^{kT}\nabla_{x_{i}} \Lambda_{i}(x_{i}^{k},\boldsymbol{x}_{-i}^{k}))^{T}(x_{i}^{k+1}-x_{i}^{\Diamond}).
\end{align*}
Then \eqref{ertd} becomes 
\begin{equation}\label{erd}
	D_{\phi_{i}}(x_{i}^{\Diamond},x_{i}^{k+1})\leq D_{\phi_{i}}(x_{i}^{\Diamond},x_{i}^{k})-D_{\phi_{i}}(x_{i}^{k+1},x_{i}^{k})-\alpha_k (\sigma_{i}^{kT}\nabla_{x_{i}} \Lambda_{i}(x_{i}^{k},\boldsymbol{x}_{-i}^{k}))^{T}(x_{i}^{k+1}-x_{i}^{\Diamond}).
\end{equation}
Similarly, as for the term $ D_{\varphi_{i}}(\sigma_{i}^{\Diamond},\sigma_{i}^{k+1})$ in \eqref{dfr}, we get
\begin{equation}\label{erd1}
	D_{\varphi_{i}}(\sigma_{i}^{\Diamond},\sigma_{i}^{k+1})\leq D_{\varphi_{i}}(\sigma_{i}^{\Diamond},\sigma_{i}^{k})-D_{\varphi_{i}}(\sigma_{i}^{k+1},\sigma_{i}^{k})-\alpha_k (-\Lambda_{i}\left(x_{i}^{k},\boldsymbol{x}_{-i}^{k}\right)+\xi_i^{k})^{T}(\sigma_{i}^{k+1}-\sigma_{i}^{\Diamond}),
\end{equation}
where  $\xi_i^{k}=\Pi_{\Theta_i}^{\Psi_{i}}(\sigma_i^{k})  $. To proceed, 
combining  \eqref{erd} and (\ref{erd1}) gives 
\begin{align*}
	\Delta(\boldsymbol{z}^{\Diamond},\boldsymbol{z}^{k+1})&=\sum_{i=1}^{N} D_{\phi_{i}}(x_{i}^{\Diamond},x_{i}^{k+1})+D_{\varphi_{i}}(\sigma_{i}^{\Diamond},\sigma_{i}^{k+1})\\
	&\leq 
	\sum_{i=1}^{N}  D_{\phi_{i}}(x_{i}^{\Diamond},x_{i}^{k})-D_{\phi_{i}}(x_{i}^{k+1},x_{i}^{k})-\alpha_k (\sigma_{i}^{kT}\nabla_{x_{i}} \Lambda_{i}(x_{i}^{k},\boldsymbol{x}_{-i}^{k}))^{T}(x_{i}^{k+1}-x_{i}^{\Diamond})\\
	&\quad\;+\sum_{i=1}^{N} D_{\varphi_{i}}(\sigma_{i}^{\Diamond},\sigma_{i}^{k})-D_{\varphi_{i}}(\sigma_{i}^{k+1},\sigma_{i}^{k})-\alpha_k (-\Lambda_{i}\left(x_{i}^{k},\boldsymbol{x}_{-i}^{k}\right)+\xi_i^{k})^{T}(\sigma_{i}^{k+1}-\sigma_{i}^{\Diamond}),\\
	&= \Delta(\boldsymbol{z}^{\Diamond},\boldsymbol{z}^{k}) - \alpha_{k} F(\boldsymbol{z}^{k})^{T}(\boldsymbol{z}^{k+1}-\boldsymbol{z}^{\Diamond})-\Delta(\boldsymbol{z}^{k+1},\boldsymbol{z}^{k}).
\end{align*}
Hence, 
\begin{equation*}\label{fr1}
	\begin{aligned}
		\Delta(\boldsymbol{z}^{\Diamond},\boldsymbol{z}^{k+1})
		&\leq \Delta(\boldsymbol{z}^{\Diamond},\boldsymbol{z}^{k}) - \alpha_{k} F(\boldsymbol{z}^{k})^{T}(\boldsymbol{z}^{k+1}-\boldsymbol{z}^{\Diamond})-\Delta(\boldsymbol{z}^{k+1},\boldsymbol{z}^{k})\\
		&= \Delta(\boldsymbol{z}^{\Diamond},\boldsymbol{z}^{k}) -\! \alpha_{k} F(\boldsymbol{z}^{k})^{T}(\boldsymbol{z}^{k}-\boldsymbol{z}^{\Diamond})+ \alpha_{k} F(\boldsymbol{z}^{k})^{T}(\boldsymbol{z}^{k}-\boldsymbol{z}^{k+1})\!-\Delta(\boldsymbol{z}^{k+1},\boldsymbol{z}^{k})\\ 
		& \leq \Delta(\boldsymbol{z}^{\Diamond},\boldsymbol{z}^{k}) -\! \alpha_{k} F(\boldsymbol{z}^{k})^{T}(\boldsymbol{z}^{k}-\boldsymbol{z}^{\Diamond})+ \alpha_{k} F(\boldsymbol{z}^{k})^{T}(\boldsymbol{z}^{k}-\boldsymbol{z}^{k+1})\!-\mu\|\boldsymbol{z}^{k}-\boldsymbol{z}^{k+1} \|^2,
	\end{aligned}
\end{equation*}
where the last inequality is due to the similar property in  (\ref{dfry}). 
On this basis, by additionally employing  Fenchel's inequality  and subsituting $f$ in Lemma \ref{d7} with $\frac{1}{2}\|\cdot\|$, 
we derive that
\begin{align*}
	\alpha_{k} F(\boldsymbol{z}^{k})^{T}(\boldsymbol{z}^{k}-\boldsymbol{z}^{k+1})
	&\leq \frac{(2\mu)}{2}\|\boldsymbol{z}^{k}-\boldsymbol{z}^{k+1} \|^2+ \frac{(2\mu)^{-1}}{2} \alpha_{k}^{2}\| F(\boldsymbol{z}^{k})^{T}\|^2_{*}\\
	&=\mu\|\boldsymbol{z}^{k}-\boldsymbol{z}^{k+1} \|^2+ \frac{1}{4\mu} \alpha_{k}^{2}\| F(\boldsymbol{z}^{k})^{T}\|^2_{*}\\
	&=\mu\|\boldsymbol{z}^{k}-\boldsymbol{z}^{k+1} \|^2+ \frac{1}{4\mu} \alpha_{k}^{2}\| F(\boldsymbol{z}^{k})^{T}\|^2,
\end{align*}
where the last equality follows from the fact that
the conjugate norm  
of $\ell_2$ norm is also  $\ell_2$ norm itself.

Hence, we can make further scaling so that
\begin{equation}\label{fr}
	\begin{aligned}
		\Delta(\boldsymbol{z}^{\Diamond},\boldsymbol{z}^{k+1})
		& \leq \Delta(\boldsymbol{z}^{\Diamond},\boldsymbol{z}^{k}) -\! \alpha_{k} F(\boldsymbol{z}^{k})^{T}(\boldsymbol{z}^{k}-\boldsymbol{z}^{\Diamond})+ \alpha_{k} F(\boldsymbol{z}^{k})^{T}(\boldsymbol{z}^{k}-\boldsymbol{z}^{k+1})\!-\mu\|\boldsymbol{z}^{k}-\boldsymbol{z}^{k+1} \|^2\\
		& \leq \Delta(\boldsymbol{z}^{\Diamond},\boldsymbol{z}^{k}) -\! \alpha_{k} F(\boldsymbol{z}^{k})^{T}(\boldsymbol{z}^{k}-\boldsymbol{z}^{\Diamond})+ \mu\|\boldsymbol{z}^{k}-\boldsymbol{z}^{k+1} \|^2+ \frac{\alpha_{k}^{2}}{4\mu} \| F(\boldsymbol{z}^{k})^{T}\|^2 -\mu\|\boldsymbol{z}^{k}-\boldsymbol{z}^{k+1} \|^2\\
		& = \Delta(\boldsymbol{z}^{\Diamond},\boldsymbol{z}^{k})-\! \alpha_{k} F(\boldsymbol{z}^{k})^{T}(\boldsymbol{z}^{k}-\boldsymbol{z}^{\Diamond})+ \frac{\alpha_{k}^2}{4\mu}\|F(\boldsymbol{z}^{k}) \|^2\\
		& = \Delta(\boldsymbol{z}^{\Diamond},\boldsymbol{z}^{k})-\! \alpha_{k} (F(\boldsymbol{z}^{k})-F(\boldsymbol{z}^{\Diamond}))^{T}(\boldsymbol{z}^{k}-\boldsymbol{z}^{\Diamond})-\alpha_{k}F(\boldsymbol{z}^{\Diamond})^{T}(\boldsymbol{z}^{k}-\boldsymbol{z}^{\Diamond}) + \frac{\alpha_{k}^2}{4\mu}\|F(\boldsymbol{z}^{k}) \|^2\\
		& \leq \Delta(\boldsymbol{z}^{\Diamond},\boldsymbol{z}^{k})-\! \alpha_{k} (F(\boldsymbol{z}^{k})-F(\boldsymbol{z}^{\Diamond}))^{T}(\boldsymbol{z}^{k}-\boldsymbol{z}^{\Diamond}) + \frac{\alpha_{k}^2}{4\mu}\|F(\boldsymbol{z}^{k}) \|^2,
	\end{aligned}
\end{equation}
where the last inequality is true because $\boldsymbol{z}^{\Diamond}$ is a solution to $\operatorname{VI}(\boldsymbol{\Xi}, F) $. 
%
Moreover, under Assumption 1, and with  $\kappa_{\sigma}$-strongly monotonicity of operator  $\Pi_{\Theta_{ i}}^{\Psi_i}(\cdot)$. Hence, there holds the inequality.
\begin{equation*}
	(F(\boldsymbol{z}^{k})-F(\boldsymbol{z}^{\Diamond}))^{T}(\boldsymbol{z}^{k}-\boldsymbol{z}^{\Diamond})\geq \kappa \|\boldsymbol{z}^{k}-\boldsymbol{z}^{\Diamond}\|^2,
\end{equation*}
where $\kappa=\min\{2\kappa_x,\kappa_\sigma\}$.
Then, it derives that
\begin{equation*}
	\begin{aligned}
		\Delta(\boldsymbol{z}^{\Diamond},\boldsymbol{z}^{k+1})
		\leq \Delta(\boldsymbol{z}^{\Diamond},\boldsymbol{z}^{k})-\! \alpha_{k}\kappa \|\boldsymbol{z}^{k}-\boldsymbol{z}^{\Diamond}\|^2 + \frac{\alpha_{k}^2}{4\mu}\|F(\boldsymbol{z}^{k}) \|^2.
	\end{aligned}
\end{equation*}

{Denote $ \eta_k= \kappa\alpha_k $ with $\eta_{0}=1$.
	We can verify that 
	\begin{equation*}
		\frac{1-\eta_{k+1}}{\eta_{k+1}^{2}} \leq \frac{1}{\eta_{k}^{2}}, \quad \forall k \geq 0.
	\end{equation*} 
}
Then, with the subsitute above,
\begin{equation}\label{frr}
	\begin{aligned}
		\Delta(\boldsymbol{z}^{\Diamond},\boldsymbol{z}^{k+1})\leq  \Delta(\boldsymbol{z}^{\Diamond},\boldsymbol{z}^{k})-\! {\eta_k} \|\boldsymbol{z}^{k}-\boldsymbol{z}^{\Diamond}\|^2 + \frac{\eta_{k}^2}{4\kappa^2\mu}\|F(\boldsymbol{z}^{k}) \|^2.
	\end{aligned}
\end{equation}
On the one hand, recalling the property of the Bregman divergence \cite{nedic2014stochastic}, we have
\begin{equation*}
	\Delta(\boldsymbol{z}^{\Diamond},\boldsymbol{z})\leq \frac{1}{2}\|\boldsymbol{z}-\boldsymbol{z}^{\Diamond}\|^2 \leq \|\boldsymbol{z}-\boldsymbol{z}^{\Diamond}\|^2, \quad \forall \boldsymbol{z}, \boldsymbol{z}^{\Diamond} \in \boldsymbol{\Xi}.
\end{equation*}
{On the other hand, under Assumption 1, since the stationary points are with finite values, the set $ \mathscr{E}_{ i}^{+} $ for $i\in\mathcal{I}$ can be regarded as  bounded without loss of generality. Then together with the compactness of  the feasible set $ \Omega_{ i} $ for $i\in\mathcal{I}$, there exists a finite constant $M_1>0$ such that $\|F(\boldsymbol{z})\|^2\leq M_1$.}
On this basis, we obtain 
\begin{equation}\label{frr3}
	\begin{aligned}
		\Delta(\boldsymbol{z}^{\Diamond},\boldsymbol{z}^{k+1})&\leq  \Delta(\boldsymbol{z}^{\Diamond},\boldsymbol{z}^{k})-\! \eta_k \Delta(\boldsymbol{z}^{\Diamond},\boldsymbol{z}^{k}) + \frac{\eta_{k}^2}{4\kappa^2\mu}\|F(\boldsymbol{z}^{k}) \|^2\\
		&\leq (1-\eta_{k})\Delta(\boldsymbol{z}^{\Diamond},\boldsymbol{z}^{k}) + \frac{\eta_{k}^2}{4\kappa^2\mu}\|F(\boldsymbol{z}^{k}) \|^2\\
		&\leq (1-\eta_{k})\Delta(\boldsymbol{z}^{\Diamond},\boldsymbol{z}^{k}) + \frac{\eta_{k}^2}{4\kappa^2\mu}{M}_1.
	\end{aligned}
\end{equation}

Multiplying both sides of the relation above by $1/\eta_{k}^2$,
and recalling the property $\frac{1-\eta_{k+1}}{\eta_{k+1}^{2}} \leq \frac{1}{\eta_{k}^{2}}$, we have
\begin{equation*}\label{frr1}
	\begin{aligned}
		\frac{1}{\eta_{k}^{2}}\Delta(\boldsymbol{z}^{\Diamond},\boldsymbol{z}^{k+1})&\leq \frac{1-\eta_{k}}{\eta_{k}^{2}} \Delta(\boldsymbol{z}^{\Diamond},\boldsymbol{z}^{k}) + \frac{{M}_1}{4\kappa^2\mu}\\
		&\leq \frac{1}{\eta_{k-1}^{2}} \Delta(\boldsymbol{z}^{\Diamond},\boldsymbol{z}^{k}) + \frac{{M}_1}{4\kappa^2\mu}.
	\end{aligned}
\end{equation*}
Hence, take the sum of these  inequalities over $k, \cdots , 1$ with $\eta_{0}=1$, that is,
\begin{equation*}\label{frr2}
	\begin{aligned}
		\frac{1}{\eta_{k}^{2}}\Delta(\boldsymbol{z}^{\Diamond},\boldsymbol{z}^{k+1})&\leq \Delta(\boldsymbol{z}^{\Diamond},\boldsymbol{z}^{1}) + k\frac{{M}_1}{4\kappa^2\mu}.
	\end{aligned}
\end{equation*}
Additionally, by taking $k=1$ and $\eta_{0}=1$ in (\ref{frr3}),  $\Delta(\boldsymbol{z}^{\Diamond},\boldsymbol{z}^{1})\leq \frac{\eta_{k}^2{M}_1}{4\kappa^2\mu}$.
Therefore, recalling the step size setting  $ \eta_k= \kappa\alpha_k= {2}/{(k+1)}$, for all $k\geq1$, we get
\begin{equation}\label{frr4}
	\begin{aligned}
		\Delta(\boldsymbol{z}^{\Diamond},\boldsymbol{z}^{k+1})\leq\eta_{k}^{2} (k+1) \frac{{M}_1}{4\kappa^2\mu}=\frac{1}{k+1}\frac{{M}_1}{{\mu}{\kappa}^2}.
	\end{aligned}
\end{equation}
Recall 
\begin{equation*}
	\Delta(\boldsymbol{z}^{\Diamond},\boldsymbol{z}^{k+1})\geq\mu \|\boldsymbol{z}^{k+1}-\boldsymbol{z}^{\Diamond}\|^2=\mu(\|\boldsymbol{x}^k-\boldsymbol{x}^{\Diamond} \|^2+\|\boldsymbol{\sigma}^k-\boldsymbol{\sigma}^{\Diamond} \|^2).
\end{equation*}
Therefore, we are finally rewarded by
\begin{equation*}
	\|\boldsymbol{x}^k-\boldsymbol{x}^{\Diamond} \|^2+\|\boldsymbol{\sigma}^k-\boldsymbol{\sigma}^{\Diamond} \|^2
	\leq 
	\frac{1}{k+1}\frac{{M}_1}{{\mu}^2{\kappa}^2},
\end{equation*}
which completes the proof.
\hfill $\square$

\section{Proof of Theorem \ref{t9}}\label{st5}

As mentioned in $\S$5, 
the  nonconvex $N$-player potential game in (\ref{p2}) is
endowed with a unified complementary function in (\ref{p11}), that is,
\begin{equation*}
	\Gamma(x_{i},\sigma,\boldsymbol{x}_{-i})=\sigma^{T} \Lambda\left(x_{i},\boldsymbol{x}_{-i}\right)-\Psi^{*}\left(\sigma\right).
\end{equation*}
Thus, we can employ
the gradient information of this unified complementary function in algorithm iterations, so as to reduce the computational cost  in Algorithm 1. Accordingly, we can rewrite Algorithm 1 for potential games in the following Algorithm \ref{nplayerp}.
\begin{algorithm}[H]
	\caption{}
	\label{nplayerp}
	\renewcommand{\algorithmicrequire}{\textbf{Input:}}
	\renewcommand{\algorithmicensure}{\textbf{Initialize:}}
	\begin{algorithmic}[1]
		\REQUIRE Step size  $ \{\alpha_{k} \} $, proper generating functions $\phi_{i}$ on $\Omega_{i}$ and  $\varphi$ on $\mathscr{E}^{+}$.  
		\ENSURE Set $\sigma^{0}\in \mathscr{E}^{+}, x_{i}^{0}\in \Omega_{i}, \,i\in\{1,\dots,N\}  $, 
		\FOR{$k = 1,2,\cdots$}
		\STATE 
		compute the unified  conjugate of $\Psi$:\;
		$\xi^{k}=\Pi_{\Theta}^{\Psi}(\sigma^{k})$\\
		\STATE  update the unified  canonical dual variable:\;\vspace{0.1cm}
		
		$\quad\sigma^{k+1} = \Pi_{\mathscr{E}^{+}}^{\varphi}(\nabla \varphi(\sigma^{k})+\alpha_{k}(\Lambda\left(x_{i}^{k},\boldsymbol{x}_{-i}^{k}\right)-\xi^{k}))
		$
		\FOR{$i = 1,\cdots N$}
		\STATE update the decision  variable of player $i$:\;\vspace{0.1cm}
		
		$	x_{i}^{k+1} =\Pi_{\Omega_{i}}^{\phi_i}(\nabla \phi_{i}(x_i^{k})-\alpha_k \sigma^{kT}\nabla_{x_{i}} \Lambda\left(x_{i}^{k},\boldsymbol{x}_{-i}^{k}\right)  )
		$
		\ENDFOR
		\ENDFOR
	\end{algorithmic}
\end{algorithm}
Similarly, define $
\boldsymbol{z}=\operatorname{col}\left\{ \boldsymbol{x},\sigma\right\}
$, and  the simplified pseudo-gradient of  (\ref{p11}) as 
\begin{equation*}
	F(\boldsymbol{z})\triangleq\left[\begin{array}{c}
		\operatorname{col}\left\{
		\sigma^{T}\nabla_{x_{i}} \Lambda\left(x_{i},\boldsymbol{x}_{-i}\right)\right\}_{i=1}^{N} \\
		
		-\Lambda\left(x_{i},\boldsymbol{x}_{-i}\right)+\nabla \Psi^{*}\left(\sigma\right)
	\end{array}\right]\triangleq
	\left[\begin{array}{c}
		G(\boldsymbol{x},\sigma) \\
		-\Lambda\left(\boldsymbol{x}\right)+\nabla \Psi^{*}\left(\sigma\right)
	\end{array}\right].
\end{equation*}
Consider the weighted averaged iterates in course of $k$ iterates as
\begin{equation*}
	\hat{\boldsymbol{x}}^{k}= \frac{\sum_{j=1}^{k} \alpha_{j} \boldsymbol{x}^{j}}{\sum_{j=1}^{k} \alpha^{j}},
	\quad \hat{{\sigma}}^{k}= \frac{\sum_{j=1}^{k} \alpha_{j} \sigma^{j}}{\sum_{j=1}^{k} \alpha^{j}}.
\end{equation*}
Then we  show the convergence rate of Algorithm \ref{nplayerp} (or Algorithm 1 in potential games). 
We rewrite Theorem \ref{t9} below for convenience:

\noindent\textbf{Theorem 5}If $\mathscr{E}^{+}$ is nonempty and players' payoffs are subject to the potential function in
(\ref{p2}), then Algorithm 1 converges at a rate of $ \mathcal{O}(1/\sqrt{k}) $  with  step size $\alpha_k=\frac{2{\mu{d}}}{{M}_2\sqrt{k}}$,  i.e.,
\begin{align*}
	\Gamma(\hat{\boldsymbol{x}}^{k},\sigma^{\Diamond})-
	\Gamma(\boldsymbol{x}^{\Diamond},\hat{\sigma}^{k})
	\leq 
	\frac{1}{\sqrt{k}} \sqrt{\frac{{d}}{\mu}}{M}_2,
\end{align*}
where ${\mu}=\min\{\frac{\mu_{x}}{2},\frac{\mu_{\sigma}}{2} \}$,  
and ${d}$, ${M}_2$ are two positive constants.

\noindent\textbf{Proof}  Take another collection of the Bregman divergence  as
\begin{equation}\label{rt}
	\widetilde{\Delta}(\boldsymbol{z}^{\Diamond},\boldsymbol{z})\triangleq D_{\varphi}(\sigma^{\Diamond},\sigma)+\sum_{i=1}^{N} D_{\phi_{i}}(x_{i}^{\Diamond},x_{i}).
\end{equation}
Working as in the proof of Theorem \ref{t6}, we obtain the following inequality
by three-point identity and Fenchel's inequality:
\begin{equation}\label{rt1}
	\begin{aligned}
		\widetilde{\Delta}(\boldsymbol{z}^{\Diamond},\boldsymbol{z}^{k+1})		&\leq \widetilde{\Delta}(\boldsymbol{z}^{\Diamond},\boldsymbol{z}^{k}) - \alpha_{k} F(\boldsymbol{z}^{k})^{T}(\boldsymbol{z}^{k+1}-\boldsymbol{z}^{\Diamond})-\widetilde{\Delta}(\boldsymbol{z}^{k+1},\boldsymbol{z}^{k})\\
		&\leq \widetilde{\Delta}(\boldsymbol{z}^{\Diamond},\boldsymbol{z}^{k}) -\! \alpha_{k} F(\boldsymbol{z}^{k})^{T}(\boldsymbol{z}^{k}-\boldsymbol{z}^{\Diamond})+ \alpha_{k} F(\boldsymbol{z}^{k})^{T}(\boldsymbol{z}^{k}-\boldsymbol{z}^{k+1})\!-\widetilde{\Delta}(\boldsymbol{z}^{k+1},\boldsymbol{z}^{k})\\ 
		& \leq \widetilde{\Delta}(\boldsymbol{z}^{\Diamond},\boldsymbol{z}^{k}) -\! \alpha_{k} F(\boldsymbol{z}^{k})^{T}(\boldsymbol{z}^{k}-\boldsymbol{z}^{\Diamond})+ \alpha_{k} F(\boldsymbol{z}^{k})^{T}(\boldsymbol{z}^{k}-\boldsymbol{z}^{k+1})\!-\mu\|\boldsymbol{z}^{k}-\boldsymbol{z}^{k+1} \|^2\\
		& \leq \widetilde{\Delta}(\boldsymbol{z}^{\Diamond},\boldsymbol{z}^{k}) -\! \alpha_{k} F(\boldsymbol{z}^{k})^{T}(\boldsymbol{z}^{k}-\boldsymbol{z}^{\Diamond})+ \alpha_{k} F(\boldsymbol{z}^{k})^{T}(\boldsymbol{z}^{k}-\boldsymbol{z}^{k+1})\!-\mu\|\boldsymbol{z}^{k}-\boldsymbol{z}^{k+1} \|^2\\
		& \leq \widetilde{\Delta}(\boldsymbol{z}^{\Diamond},\boldsymbol{z}^{k})-\! \alpha_{k} F(\boldsymbol{z}^{k})^{T}(\boldsymbol{z}^{k}-\boldsymbol{z}^{\Diamond})+ \frac{\alpha_{k}^2}{4\mu}\|F(\boldsymbol{z}^{k}) \|^2.
	\end{aligned}
\end{equation}
Moreover,  according to $\sigma\in\mathscr{E}^{+}$ in (\ref{s23}),
\begin{align*}
	(\boldsymbol{x}^{\Diamond}-\boldsymbol{x}^{k})^{T}G(\boldsymbol{x}^{k},\sigma^{k})\leq \sigma^{kT}(\Lambda\left(\boldsymbol{x}^{\Diamond}\right)-\Lambda\left(\boldsymbol{x}^{k}\right)).
\end{align*}
As a result, 
\begin{equation}\label{rtyt}
	\begin{aligned}
		\left\langle F(\boldsymbol{z}^{k}), \boldsymbol{z}^{\Diamond}-\boldsymbol{z}^{k}\right\rangle &= 
		(\boldsymbol{x}^{\Diamond}-\boldsymbol{x}^{k})^{T}G(\boldsymbol{x}^{k},\sigma^{k})+(\sigma^{\Diamond}-\sigma^{k})^{T}(-\Lambda\left(\boldsymbol{x}^{k}\right)+\nabla \Psi^{*}\left(\sigma^{k}\right))\\
		&\leq \sigma^{kT} \Lambda\left(\boldsymbol{x}^{\Diamond}\right)-\Psi^{*}\left(\sigma^{k}\right)-(\sigma^{\Diamond T} \Lambda\left(\boldsymbol{x}^{kT}\right)-\Psi^{*}\left(\sigma^{\Diamond}\right))\\
		&= \Gamma(\boldsymbol{x}^{\Diamond},\sigma^{k})-\Gamma(\boldsymbol{x}^{k},\sigma^{\Diamond}).
	\end{aligned}
\end{equation}
By substituting  \eqref{rtyt} into \eqref{rt1} and rearranging the terms therein, we have 
\begin{align*}
	\alpha_{k}(\Gamma(\boldsymbol{x}^{k},\sigma^{\Diamond})-\Gamma(\boldsymbol{x}^{\Diamond},\sigma^{k}))&\leq\alpha_{k} F(\boldsymbol{z}^{k})^{T}(\boldsymbol{z}^{k}-\boldsymbol{z}^{\Diamond})\\
	&\leq \widetilde{\Delta}(\boldsymbol{z}^{\Diamond},\boldsymbol{z}^{k})-\widetilde{\Delta}(\boldsymbol{z}^{\Diamond},\boldsymbol{z}^{k+1})+ \frac{\alpha_{k}^2}{4\mu}\|F(\boldsymbol{z}^{k}) \|^2.
\end{align*}
Meanwhile, under Assumption 1, since the stationary points are with finite values, the set $ \mathscr{E}^{+} $ can be regarded as bounded without loss of generality. Together with the compactness of  $ \Omega_{ i} $ for $i\in\mathcal{I}$, there exists  finite constants $d>0$ and $M_2>0$ such that $\widetilde{\Delta}(\boldsymbol{z}^{\Diamond},\boldsymbol{z}^{1})\leq d$ and $\|F(\boldsymbol{z})\|^2\leq M_2$. 
Then it follows from the sum of the above inequalities over $1, \cdots , k$ that
\begin{align}\label{o1}
	\sum_{j=1}^{k} \alpha_{j}\left(
	\Gamma(\boldsymbol{x}^{j},\sigma^{\Diamond})-
	\Gamma(\boldsymbol{x}^{\Diamond},\sigma^{j})
	\right) \leq \widetilde{\Delta}(\boldsymbol{z}^{\Diamond},\boldsymbol{z}^{1})+\frac{\sum^{k}_{j=1}\alpha_{j}^2 {M}^2_{2}}{4\mu}.
\end{align}
For more intuitive presentation, we denote the weight by $\lambda_j=  \frac{\alpha_{j}}{\sum^{k}_{l=1}\alpha_{l}}$. Then (\ref{o1})  yields
\begin{align*}
\frac{\widetilde{\Delta}(\boldsymbol{z}^{\Diamond},\boldsymbol{z}^{1})+(4\mu)^{-1}{M}^2_{2}\sum^{k}_{j=1}\alpha_{j}^2 }{\sum^{k}_{j=1}\alpha_{j}}	
	&\geq\sum_{j=1}^{k} \frac{\alpha_{j}}{\sum^{k}_{l=1}\alpha_{l}} \left(
	\Gamma(\boldsymbol{x}^{j},\sigma^{\Diamond})-
	\Gamma(\boldsymbol{x}^{\Diamond},\sigma^{j})
	\right)\\
	&= \sum_{j=1}^{k}  \lambda_j \left(
	\Gamma(\boldsymbol{x}^{j},\sigma^{\Diamond})-
	\Gamma(\boldsymbol{x}^{\Diamond},\sigma^{j})
	\right)\\
	&\geq \Gamma(\sum_{j=1}^{k}  \lambda_j \boldsymbol{x}^{j},\sigma^{\Diamond})-
	\Gamma(\boldsymbol{x}^{\Diamond},\sum_{j=1}^{k}  \lambda_j \boldsymbol{\sigma}^{j})\\
	&= \Gamma(\hat{\boldsymbol{x}}^{k},\sigma^{\Diamond})-
	\Gamma(\boldsymbol{x}^{\Diamond},\hat{\sigma}^{k}),
\end{align*}
where the last inequality is true due to Jensen's inequality. 
Since the step size satisfies $\alpha_k=2\sqrt{\mu{d}}/{M}_2\sqrt{k}$, we finally derive that 
\begin{align*}
	\Gamma(\hat{\boldsymbol{x}}^{k},\sigma^{\Diamond})-
	\Gamma(\boldsymbol{x}^{\Diamond},\hat{\sigma}^{k})
	\leq \frac{1}{\sqrt{k}} \sqrt{\frac{{d}}{\mu}}{M}_2,
\end{align*}
which indicates the conclusion.
\hfill $\square$

\end{document}